\documentclass[11pt,a4paper]{article}
\usepackage{array}
\usepackage{float}
\usepackage{amsmath}
\usepackage{amsfonts}
\usepackage{amssymb}
\usepackage{latexsym}
\usepackage{epsfig}
\usepackage{graphicx}
\usepackage{mathrsfs}
\usepackage{fancyhdr}
\usepackage{sectsty}
\usepackage[longnamesfirst]{natbib}
\usepackage[labelsep=colon,font=it,labelfont=sc]{caption}
\usepackage{theorem}
\usepackage{url}
\usepackage{rotating}
\usepackage{multirow}
\usepackage{lscape}
\usepackage[truetex]{color}

{
\theorembodyfont{\upshape} 
\theoremheaderfont{\scshape} 

}
\newtheorem{assumption}{Assumption}

\newtheorem{corollary}{Corollary}

{\theorembodyfont{\upshape} 

}

\newtheorem{lemma}{{\sc Lemma}}

\newtheorem{proposition}{{\sc Proposition}}

\newenvironment{proof}[1][Proof]{\bigskip \noindent \textbf{#1:} }{\  \rule{0.5em}{0.5em}}

\theoremheaderfont{\bfseries}
\theorembodyfont{\upshape}
\sectionfont{\centering \sc}
\subsectionfont{\centering \sc}
\subsubsectionfont{\centering \sc}
\allsectionsfont{\centering \sc}
\allowdisplaybreaks

\bibpunct[:~]{(}{)}{;}{a}{,}{;}
\renewcommand{\cite}{\citet}
\begin{document}

\begin{center}
\bigskip \textsc{AN IDENTIFICATION AND TESTING STRATEGY FOR PROXY-SVARs WITH
WEAK PROXIES}

\renewcommand{\thefootnote}{}
\footnote{
\hspace{-7.2mm}
$^{a}$Department of Economics, University of Bologna, Italy.
$^{b}$Department of Economics, University of Exeter Business School, UK. 
Correspondence to: Giuseppe Cavaliere, Department of Economics, University of Bologna, Piazza Scaravilli 2, 
40126 Bologna, Italy; email: giuseppe.cavaliere@unibo.it. 
}
\addtocounter{footnote}{-1}
\renewcommand{\thefootnote}{\arabic{footnote}}%
{\normalsize \vspace{0.1cm} }

{\large \textsc{Giovanni Angelini}}$^{a}${\large \textsc{, Giuseppe Cavaliere%
}$^{a,b}$\textsc{, Luca Fanelli}}$^{a}$

\medskip

First draft: September{\small \ }2021{\small .\ }First revision: September
2022{\small .\ } \\[0pt]
Second revision: July 2023{\small .\ } This version: October 2023
\end{center}

\par
\begingroup
\leftskip =1 cm
\rightskip =1 cm
\small%

\begin{center}
\textsc{Abstract\vspace{-0.15cm}}
\end{center}

When proxies (external instruments) used to identify target structural
shocks are weak, inference in proxy-SVARs (SVAR-IVs) is nonstandard and the
construction of asymptotically valid confidence sets for the impulse
responses of interest requires weak-instrument robust methods. In the
presence of multiple target shocks, test inversion techniques require extra
restrictions on the proxy-SVAR parameters other those implied by the proxies
that may be difficult to interpret and test. We show that frequentist
asymptotic inference in these situations can be conducted through Minimum
Distance estimation and standard asymptotic methods if the proxy-SVAR can be
identified by using `strong' instruments for the \emph{non-target shocks};
i.e. the shocks which are not of primary interest in the analysis. The
suggested identification strategy hinges on a novel pre-test for the null of
instrument relevance based on bootstrap resampling which is not subject to
pre-testing issues. Specifically, the validity of post-test asymptotic
inferences remains unaffected by the test outcomes due to an asymptotic
independence result between the bootstrap and non-bootstrap statistics. The
test is robust to conditionally heteroskedastic and/or zero-censored
proxies, is computationally straightforward and applicable regardless of the
number of shocks being instrumented. Some illustrative examples show the
empirical usefulness of the suggested identification and testing strategy.

\bigskip

\bigskip


\vspace{0.15cm}


\vspace{0.15cm}

\par
\endgroup
\normalsize%
\pagebreak

\section{Introduction}

\textsc{Proxy-SVARs,} or SVAR-IVs, popularized by Stock (2008), Stock and
Watson (2012, 2018) and Mertens and Ravn (2013), have become standard tools
to track the dynamic causal effects produced by macroeconomic shocks on
variables of interest. In proxy-SVARs, the model is complemented with
`external' variables -- which we call `proxies', `instruments' or `external
variables' interchangeably; such variables carry information on the
structural shocks of interest, the \emph{target shocks}, and allow to
disregard the structural shocks not of primary interest in the analysis, the 
\emph{non-target shocks}. Recent contributions on frequentist inference in
proxy-SVARs include Montiel Olea, Stock and Watson (2021) and Jentsch and
Lunsford (2022); in the Bayesian framework, Arias, Rubio-Ramirez and
Waggoner (2021) and Giacomini, Kitagawa and Read (2022) discuss inference in
the case of set-identification.

Inference in proxy-SVARs depends on whether the proxies are strongly or
weakly correlated with the target shocks. If the connection between the
proxies and the target shocks is `local-to-zero', as in Staiger and Stock
(1997) and Stock and Yogo (2005), asymptotic inference is non-standard. In
such case, weak-proxy robust methods can be obtained by extending the logic
of Anderson-Rubin tests (Anderson and Rubin, 1949), see Montiel Olea {\emph{%
et al.}} (2021). Grid Moving Block Bootstrap Anderson-Rubin confidence sets
(`grid MBB AR') for normalized impulse response functions [IRFs] (Br\"{u}%
ggemann, Jentsch and Trenkler, 2016; Jentsch and Lunsford, 2019) can also be
applied in the special case where one proxy identifies one structural shock;
see Jentsch and Lunsford (2022).

When proxy-SVARs feature multiple target shocks, further inferential
difficulties arise. First, (point-)identification requires additional
restrictions,\ other than those provided by the instruments; see Mertens and
Ravn (2013), Angelini and Fanelli (2019), Arias \emph{et al.} (2021),
Montiel Olea {\emph{et al.}} (2021) and Giacomini {\emph{et al.}} (2022).
Second, in the frequentist setup the implementation of weak-instrument
robust inference as in Montiel Olea {\emph{et al.}} (2021) may imply a large
number of additional restrictions on the parameters of the proxy-SVAR
relative to those needed under strong proxies. These extra restrictions are
not always credible, and may be difficult to test; see Montiel Olea \emph{et
al.} (2021, Section A.7) and Section \ref%
{Section_supplementary_fiscal_proxy-SVAR} of our supplement.\footnote{%
From the perspective of Bayesian inference, one can in principle make the
usual argument that weak identification issues do not matter. For instance,
Caldara and Herbst (2019) discuss how it is still possible to obtain
numerical approximations of the exact finite-sample posterior distributions
of the parameters of proxy-SVARs when instruments are weak. Giacomini {\emph{%
et al.}} (2022) show that for set-identified proxy-SVARs with weak
instruments, the Bernstein-von Mises property fails for the estimation of
the upper and lower bonds of the identified set.} Fourth, the theory for the
grid bootstrap Anderson-Rubin confidence sets does not extend to cases where
multiple instruments identify multiple target shocks.

This paper is motivated by these inferential difficulties. In particular, we
design an identification and (frequentist) estimation strategy intended to
circumvent, when possible, the use of weak-instrument robust methods. The
idea we pursue is to identify the proxy-SVAR through an `indirect' approach,
where a vector of proxies (say, $w_{t}$), correlated with (all or some of)
the non-target shocks of the system and uncorrelated with the target shocks
(say, $z_{t}$), is used to infer the IRFs of interest indirectly. We call
this strategy `indirect identification strategy' or `indirect-MD' approach,
as opposed to the conventional `direct' approach based on instrumenting the
target shock(s) directly with the (potentially weak)\ proxies $z_{t}$. As
highlighted by our empirical illustrations, the indirect approach can prove
more useful to a practitioner than one might think.

The proxies $w_{t}$ contribute to defining a set of moment conditions upon
which we develop a novel Minimum Distance [MD] estimation approach (Newey
and McFadden, 1994). We derive novel necessary order conditions and
necessary and sufficient rank condition for the (local)\ identifiability of
the proxy-SVAR. If the proxies $w_{t}$ are strong for the non-target shocks
and the model is identified, asymptotically valid confidence intervals for
the IRFs of interest obtain in the usual way; i.e., either by the
delta-method or by bootstrap methods. Interestingly, the idea of using
instruments for the non-target shocks to identify and infer the effects of
structural shocks of interest was initially pursued via Bayesian methods in
Caldara and Kamps (2017), where two fiscal (target) shocks are recovered by
instrumenting the non-fiscal (non-target)\ shocks of the system. We defer to
Section \ref{Section_indirect_approach_MD} a detailed comparison of our
method with Caldara and Kamps (2017).

Key to the indirect identification strategy is the availability of strong
proxies for the non-target shocks. In particular, it is essential that the
investigator can screen `strong' from `weak' instruments, and that such
screening does not affect post-test inference. To do so, we further
contribute by designing a novel pre-test for strong against weak proxies
based on bootstrap resampling.

Inspired by the idea originally developed in Angelini, Cavaliere and Fanelli
(2022) for state-space models, we show that the bootstrap can be used to
infer the strength of instruments, other than building valid confidence
intervals for IRFs. In particular, we exploit the fact that under mild
requirements, the MBB estimator of the proxy-SVAR parameters is
asymptotically Gaussian when the instruments are strong while, under weak
proxies \`{a} la Staiger and Stock (1997), the distribution of MBB estimator
is random in the limit (in the sense of Cavaliere and Georgiev, 2020) and,
in particular, is non-Gaussian. This allows to show that a test for the null
of strong proxies can be designed as a normality test based on an
appropriate number of\ bootstrap repetitions; such test is consistent
against proxies which are weak in Staiger and Stock's (1997). An idea that
echoes this approach in the Bayesian setting can be found in Giacomini {%
\emph{et al.}} (2022), who suggest using non-normality of the posterior
distribution of a suitable function of proxy-SVAR parameters to diagnose the
presence of weak proxies. This idea is not pursued further in their paper.

Our suggested test has several important features. First, it controls size
under general conditions on VAR disturbances and proxies, including the case
of conditional heteroskedasticity and/or zero-censored proxies. Second, with
respect to extant tests such as Montiel Olea and Pflueger's (2013) effective
first-stage F-test for IV\ models with conditional heteroskedasticity,%
\footnote{%
See Montiel Olea \emph{et al.} (2021) for an overview on first-stage
regressions in proxy-SVARs\textbf{\ }or, alternatively, Lunsford (2016) for
tests based on regressing the proxy on the reduced-form residuals.} our test
can be applied in the presence of multiple structural shocks;\ as far as we
are aware, no test of strength for proxy-SVARs with multiple target shocks
has been formalized in the literature. Third, it is computationally
straightforward, as it boils down to running multivariate/univariate
normality tests on the MBB replications of bootstrap estimators of the
proxy-SVAR parameters. Fourth, it can be computed in the same way regardless
of the number of shocks being instrumented. Fifth, and most importantly, the
test does not affect second-stage inference, meaning that regardless of the
outcome of the test, post-test inferences are not affected. This property
marks an important difference relative to the literature on weak instrument
asymptotics, where the negative consequences of pretesting the strength of
proxies are well known and documented (see, \textit{inter alia}, Zivot,
Startz and Nelson, 1998; Hausman, Stock and Yogo, 2005; Andrews, Stock and
Sun, 2019; Montiel Olea {\emph{et al.}}, 2021).

The paper is organized as follows. In Section \ref%
{Section_Motivating_example} we motivate our approach with a simple
illustrative example. In Section \ref{Section_Model} we introduce the
proxy-SVAR and rationalize the suggested identification strategy. The
assumptions are summarized in Section \ref{Section_Assumptions}, while we
present our indirect-MD approach in Section \ref%
{Section_indirect_approach_MD}. Section \ref{Section_testing} deals with the
novel approach to testing for strong proxies. To illustrate the practical
implementation and relevance of our approach, we present in Section \ref%
{Section_empirical_illustrations} two illustrative examples that reconsider
models already estimated in the literature. Section \ref{Sez_Conclusions}
concludes. An accompanying supplement complements the paper along several
dimensions, including auxiliary lemmas and their proofs, the proofs the
propositions in the paper and an additional empirical illustration based on
a fiscal proxy-SVAR.

\section{Motivating example: a market (demand/supply) model}

\label{Section_Motivating_example}In this section we outline the main ideas
in the paper by considering a `toy' proxy-SVAR, where we omit the dynamics
without loss of generality.\textbf{\ }We consider a model that comprises a
demand and supply function for a good with associated structural shocks,
given by the equations%
\begin{equation}
\underset{Y_{t}}{\underbrace{\left( 
\begin{array}{c}
q_{t} \\ 
p_{t}%
\end{array}%
\right) }}=\underset{B}{\underbrace{\left( 
\begin{array}{cc}
\beta _{1,1} & \beta _{1,2} \\ 
\beta _{2,1} & \beta _{2,2}%
\end{array}%
\right) }}\underset{\varepsilon _{t}}{\underbrace{\left( 
\begin{array}{c}
\varepsilon _{d,t} \\ 
\varepsilon _{s,t}%
\end{array}%
\right) }}\equiv \left( 
\begin{array}{c}
\beta _{1,1}\varepsilon _{d,t}+\beta _{1,2}\varepsilon _{s,t} \\ 
\beta _{2,1}\varepsilon _{d,t}+\beta _{2,2}\varepsilon _{s,t}%
\end{array}%
\right)  \label{structrual_toy}
\end{equation}%
where $q_{t}$ and $p_{t}$ are quantity and price at time $t$, respectively.
The nonsingular matrix $B$ captures the instantaneous impact, on $%
Y_{t}:=(q_{t},p_{t})^{\prime }$, of the structural shocks $\varepsilon
_{d,t},\varepsilon _{s,t}$, which are assumed to have unit variance and to
be uncorrelated. We temporary (and conventionally) label $\varepsilon _{d,t}$
as the `demand shock' and $\varepsilon _{s,t}$ as the `supply shock', and
assume that the objective of the analysis is the identification and
estimation of the instantaneous impact of the \emph{demand} \emph{shock} on $%
Y_{t}$ through the `external variables' approach. Hence, $\varepsilon _{d,t}$
is the \emph{target }shock, $\varepsilon _{s,t}$ is the \emph{non-target}
shock, and the parameters of interest are the on-impact responses $\frac{%
\partial Y_{t}}{\partial \varepsilon _{d,t}}=B_{\bullet 1}:=(\beta
_{1,1},\beta _{2,1})^{\prime }$; here $B_{\bullet 1}$ denotes the first
column of $B$.

Since the two equations in (\ref{structrual_toy}) are essentially identical
for arbitrary parameter values, nothing distinguishes a demand shock from a
supply shock in the absence of further information/restrictions. The typical
`direct approach' to this partial identification problem is to consider an
instrument $z_{t}$ correlated with the demand shock, ${E}(z_{t}\varepsilon
_{d,t})=\phi \neq 0$ (relevance condition),\ and uncorrelated with the
supply shock, ${E}(z_{t}\varepsilon _{s,t})=0$ (exogeneity condition). Now,
consider the case where the investigator strongly suspects that $z_{t}$ is a
weak proxy (meaning that $\phi $ can be `small'), but they also know that
there exists an external variable $w_{t}$, correlated with the non-target
supply shock and uncorrelated with the demand shock; formally, ${E}%
(w_{t}\varepsilon _{s,t})=\lambda \neq 0$ and ${E}(w_{t}\varepsilon
_{d,t})=0 $. Then, the proxy $w_{t}$\ can be used to recover the parameters
of interest (i.e., $B_{\bullet 1}$) `indirectly'; i.e., by instrumenting the
non-target supply shock $\varepsilon _{s,t}$, rather than the target demand
shock $\varepsilon _{d,t}$. To show how, let $A:=B^{-1}$ and consider the
alternative representation of (\ref{structrual_toy}): 
\begin{equation*}
\underset{A}{\underbrace{\left( 
\begin{array}{cc}
\alpha _{1,1} & \alpha _{1,2} \\ 
\alpha _{2,1} & \alpha _{2,2}%
\end{array}%
\right) }}\underset{Y_{t}}{\underbrace{\left( 
\begin{array}{c}
q_{t} \\ 
p_{t}%
\end{array}%
\right) }}=\left( 
\begin{array}{c}
A_{1\bullet }Y_{t} \\ 
A_{2\bullet }Y_{t}%
\end{array}%
\right) \underset{\varepsilon _{t}}{=\underbrace{\left( 
\begin{array}{c}
\varepsilon _{d,t} \\ 
\varepsilon _{s,t}%
\end{array}%
\right) }},
\end{equation*}%
where $A_{1\bullet }:=(\alpha _{1,1},\alpha _{1,2})$ and $A_{2\bullet }$
denote the first row and the second row of $A$, respectively. Since $w_{t}$\
is correlated with $p_{t}$ but uncorrelated with $\varepsilon _{d,t}$, it is
seen that for $\alpha _{11}\neq 0$, $w_{t}$ can be used in the equation:%
\begin{equation*}
q_{t}=-\frac{\alpha _{1,2}}{\alpha _{1,1}}p_{t}+\frac{1}{\alpha _{1,1}}%
\varepsilon _{d,t}
\end{equation*}%
as an instrument for $p_{t}$ in order to estimate the parameters in $%
A_{1\bullet }$, that is, $\alpha _{1,1}$ and $\alpha _{1,2}$. This delivers
an `estimate' of the demand shock, $\hat{\varepsilon}_{d,t}=\hat{A}%
_{1\bullet }Y_{t}=\hat{\alpha}_{1,1}q_{t}+$\ $\hat{\alpha}_{1,2}p_{t}$ ($%
t=1,\ldots ,T$). Finally, since (\ref{structrual_toy}) and $A=B^{-1}$
jointly imply $B=\Sigma _{u}A^{\prime }$, it holds that%
\begin{equation}
B_{\bullet 1}=\Sigma _{u}A_{1\bullet }^{\prime }
\label{crucial_relationship_partial}
\end{equation}%
where $\Sigma _{u}:=${${E}$}$(Y_{t}Y_{t}^{\prime })$ can be estimated (e.g.,
by its sample analog, $\hat{\Sigma}_{u}:=T^{-1}\sum_{t=1}^{T}Y_{t}Y_{t}^{%
\prime }$) under mild requirements. Hence, an indirect plug-in estimator of
the parameters of interest $B_{\bullet 1}$ is given by $\hat{B}_{\bullet 1}:=%
\hat{\Sigma}_{u}\hat{A}_{1\bullet }^{\prime }$. If the instrument $w_{t}$ is
a `strong' proxy for the supply shock, in the sense formally defined in
Section \ref{Section_Assumptions}, standard asymptotic inference on $%
B_{\bullet 1}$ can then be performed using $\hat{B}_{\bullet 1}$.

This toy example shows that strong proxies for the non-target shocks,
provided they exist, can be used to infer the causal effects of the target
shocks indirectly, in a partial identification logic. Importantly, the
investigator can strategically exploit the fact that if the proxies $z_{t}$
available for the target shock are `weak', the use of weak-instrument robust
methods for the parameters of interest ($B_{\bullet 1}$ in this example) can
be circumvented if they can alternatively rely on strong proxies $w_{t}$ for
the non-target shocks.

In the following, we assume that there exist proxies $w_{t}$ for the
non-target shocks that might be alternatively used instead of the
(potentially weak) proxies $z_{t}$ available for the target structural
shocks. The strength of $w_{t}$ is a key ingredient of this strategy; hence,
in Section \ref{Section_testing} we present our novel pre-test of relevance,
which consistently detects proxies which are weak in the sense of Staiger
and Stock (1997). Since the test does not affect post-test inferences, if
the null of relevance is not rejected, inference based on $w_{t}$ can be
conducted by standard methods with no need for Bonferroni-type adjustments.
In contrast, should the null of relevance be rejected, the investigator can
rely on weak-instrument robust methods based either on the proxies\ $z_{t}$,
if the target shocks are instrumented,\ or on the proxies $w_{t}$ if the
non-target shocks are instrumented.

\section{Model and identification strategies}

\label{Section_Model}Consider the SVAR model:%
\begin{equation}
Y_{t}=\Pi X_{t}+u_{t}\text{, \ \ }u_{t}=B\varepsilon _{t}\text{ \ (}%
t=1,\ldots ,T\text{)}  \label{VAR-RF2}
\end{equation}%
where $Y_{t}$ is the $n\times 1$ vector of endogenous variables, $%
X_{t}:=(Y_{t-1}^{\prime },\ldots ,Y_{t-l}^{\prime })^{\prime }$ collects $l$
lags of the variables, $\Pi :=(\Pi _{1},\ldots ,\Pi _{l})$ is the $n\times
nl $ matrix containing the autoregressive (slope) parameters, and $u_{t}$ is
the $n\times 1$ vector of reduced form disturbances with covariance matrix $%
\Sigma _{u}:=${${E}$}$(u_{t}u_{t}^{\prime })$. Deterministic terms have been
excluded without loss of generality, and the initial values $Y_{0},\ldots
,Y_{1-l}$ are fixed in the statistical analysis. The system of equations $%
u_{t}=B\varepsilon _{t}$\ in (\ref{VAR-RF2}) defines the reduced form
disturbances $u_{t}$ in terms of the $n\times 1$ vector of structural
shocks, $\varepsilon _{t}$, through the nonsingular $n\times n$ matrix $B$
of on-impact coefficients. The structural shocks are normalized such that $%
\Sigma _{\varepsilon }:=${${E}$}$(\varepsilon _{t}\varepsilon _{t}^{\prime
})=I_{n}$.

We partition the structural shocks as $\varepsilon _{t}:=(\varepsilon
_{1,t}^{\prime },\varepsilon _{2,t}^{\prime })^{\prime }$, where $%
\varepsilon _{1,t}$ collects the $1\leq k<n$ \emph{target} structural
shocks, and $\varepsilon _{2,t}$ collects the remaining $n-k$ structural
shocks of the system. We have%
\begin{equation}
u_{t}=\left( 
\begin{array}{l}
u_{1,t} \\ 
u_{2,t}%
\end{array}%
\right) =\left( 
\begin{array}{cc}
B_{1,1} & B_{1,2} \\ 
B_{2,1} & B_{2,2}%
\end{array}%
\right) \left( 
\begin{array}{l}
\varepsilon _{1,t} \\ 
\varepsilon _{2,t}%
\end{array}%
\right) \equiv B_{\bullet 1}\varepsilon _{1,t}+\text{ }B_{\bullet
2}\varepsilon _{2,t}  \label{partition_B}
\end{equation}%
where $u_{1,t}$ and $u_{2,t}$ have the same dimensions as $\varepsilon
_{1,t} $ and $\varepsilon _{2,t}$, respectively, and $B_{\bullet
1}:=(B_{1,1}^{\prime },$ $B_{2,1}^{\prime })^{\prime }$ is the $n\times k$
matrix collecting the on-impact coefficients associated with the target
structural shocks ($B_{1,1} $ and $B_{2,1}$ are $k\times k$ and $(n-k)\times
k$ blocks, respectively). Finally, the $n\times (n-k)$ matrix $B_{\bullet 2}$
collects the instantaneous impact of the non-target shocks on the variables.
We are interested in the $h $ period ahead responses of the $i$-th variable
in $Y_{t}$ ($i=1,\ldots ,n$)\ to the $j$-th shock in $\varepsilon _{1,t}$ ($%
j=1,\ldots ,k$); as is standard, such responses can be computed from the
companion form representation as 
\begin{equation}
\gamma_{\bullet j}(h):=(S_{n}^{\prime }\mathcal{C}_{y}^{h}S_{n})B_{\bullet
1}e_{k,j},  \label{IRF_j}
\end{equation}%
where $\mathcal{C}_{y}$ is the VAR companion matrix, $S_{n}:=(I_{n}$ $%
,0_{n\times n(l-1)})$ is a selection matrix and $e_{k,j}$ is the $k\times 1$
vector containing `1' in the $j$-th position and zero elsewhere.\footnote{%
Notice that we focus on \emph{absolute} IRFs -- the quantities $\gamma
_{i,j}(h)$, $\gamma _{i,j}(h)$ being the $i$-th element of $\gamma_{\bullet
j}(h)$ in (\ref{IRF_j}) -- rather than on \emph{relative} IRFs, $\gamma
_{i,j}(h)/\gamma _{1,j}(0)$, which measure the response of $Y_{i,t}$ to the $%
j$-th shock in $\varepsilon _{1,t}$ that increases $Y_{1,t}$ by one unit
on-impact.}

The common, `\emph{direct}' approach to infer the parameters of interest in $%
B_{\bullet 1}$ and hence solve the partial identification problem arising
from the estimation of the IRFs in (\ref{IRF_j}) is to find $r\geq k$
observable proxies, collected in the vector $z_{t}$, correlated with the
target shocks $\varepsilon _{1,t}$\ and uncorrelated with $\varepsilon
_{2,t} $. Thus, $z_{t} $ is related to $\varepsilon _{1,t}$ by the linear
measurement system 
\begin{equation}
z_{t}=\Phi \varepsilon _{1,t}+\omega _{z,t}  \label{equation_link}
\end{equation}%
where the matrix $\Phi :=${${E}$}$(z_{t}\varepsilon _{1,t}^{\prime }) $
captures the link between the proxies $z_{t}$ and the target shocks $%
\varepsilon _{1,t}$; $\omega _{z,t}$ is a measurement error, assumed to be
uncorrelated with the structural shocks $\varepsilon _{t}$. By combining (%
\ref{equation_link})\ with (\ref{partition_B}) and taking expectations, one
obtains the moment conditions 
\begin{equation}
\Sigma _{u,z}=B_{\bullet 1}\Phi ^{\prime }  \label{moments_B_form}
\end{equation}%
where $\Sigma _{u,z}:=${${E}$}$(u_{t}z_{t}^{\prime })$ is the $n\times r$
covariance matrix between $u_{t}$ and $z_{t}$. Stock (2008), Stock and
Watson (2012, 2018)\ and Mertens and Ravn (2013) exploit the moment
conditions in (\ref{moments_B_form}) as starting point for the
identification of the IRFs in (\ref{IRF_j}).

Alternatively, as shown in the example in Section \ref%
{Section_Motivating_example}, the IRFs in (\ref{IRF_j}) can be identified by
and `\emph{indirect approach}', where a vector of proxies $w_{t}$ are used
to instrument the non-target shocks. Specifically, for $A=B^{-1}$, model (%
\ref{VAR-RF2}) can be expressed in the form:%
\begin{equation}
AY_{t}=\Upsilon X_{t}+\varepsilon _{t}\text{, \ \ }Au_{t}=\varepsilon _{t}%
\text{ \ \ (}t=1,\ldots ,T\text{)}  \label{SVAR_A}
\end{equation}%
where $\Upsilon :=A\Pi $ and $A$ summarizes the simultaneous relationships
that characterize the observed variables. The system of equations $%
Au_{t}=\varepsilon _{t}$ can then be partitioned as 
\begin{equation}
Au_{t}\equiv \left( 
\begin{array}{c}
A_{1\bullet }u_{t} \\ 
A_{2\bullet }u_{t}%
\end{array}%
\right) \equiv \left( 
\begin{array}{c}
A_{1,1}u_{1,t}+A_{1,2}u_{2,t} \\ 
A_{2,1}u_{1,t}+A_{2,2}u_{2,t}%
\end{array}%
\right) =\left( 
\begin{array}{l}
\varepsilon _{1,t} \\ 
\varepsilon _{2,t}%
\end{array}%
\right)  \label{partition_A}
\end{equation}%
where the $k\times n$ matrix $A_{1\bullet }:=(A_{1,1},A_{1,2})$ collects the
first $k$ rows of $A$, $A_{2\bullet }$ the remaining $n-k$ rows, and the VAR
disturbances $u_{1,t}$ and $u_{2,t}$ have the same dimension as $\varepsilon
_{1,t}$ and $\varepsilon _{2,t}$, respectively. Under identifying
restrictions on $A_{1\bullet }$ and $A_{2\bullet }$, the term $\varepsilon
_{1,t}$ in equation (\ref{partition_A}) can be interpreted as the structural
shocks of a simultaneous system of equations \`{a}~ la Leeper, Sims and Zha
(1996).

Using the SVAR representation (\ref{partition_A}), we can infer the
parameters in $A_{1\bullet }$ by exploiting the vector of external proxy
variables $w_{t}$, correlated with (all or some of) the non-target shocks $%
\varepsilon _{2,t}$ and uncorrelated with the target shocks $\varepsilon
_{1,t}$. In Section \ref{Section_indirect_approach_MD} we discuss in detail
how the parameters in $A_{1\bullet }$ can be identified by using $w_{t}$
through a MD approach; the estimation of $B_{\bullet 1}$ and the IRFs (\ref%
{IRF_j})\ follow indirectly, as in (\ref{crucial_relationship_partial}),
from the relation $B_{\bullet 1}=\Sigma _{u}A_{1\bullet }^{\prime }$.

The next section states the assumptions behind our estimation approach and
qualifies the concepts of strong/weak proxies we refer to throughout the
paper.

\section{Assumptions and asymptotics}

\label{Section_Assumptions}Our first two main assumptions pertain to the
reduced form VAR.

\begin{assumption}[Reduced form, stationarity]
The data generating process (DGP) for $Y_{t}$\ satisfies (\ref{VAR-RF2})
with a stable companion matrix $\mathcal{C}_{y}$, i.e. all eigenvalues of $%
\mathcal{C}_{y}$ lie inside the unit disk.
\end{assumption}

\begin{assumption}[Reduced form, VAR innovations]
The VAR disturbances satisfy the following conditions: \newline
(i) $\left\{ u_{t}\right\} $ is a strictly stationary weak white noise; 
\newline
(ii) {${E}$}$(u_{t}u_{t}^{\prime })=\Sigma _{u}<\infty $ is positive
definite;\newline
(iii) $u_{t}$ satisfies the $\alpha $-mixing conditions in Assumption 2.1 of
Br\"{u}ggemann {et al.} (2016); \newline
(iv) $u_{t}$ has absolutely summable cumulants up to order eight.
\end{assumption}

Assumption 1 features a typical maintained hypothesis of correct
specification and incorporates a stability condition which rules out the
presence of unit roots. Assumption 2 is as in Francq and Ra\"{\i}ssi (2006)
and Boubacar Mainnasara and Francq (2011). Assumption 2(ii) is a standard
unconditional homoskedasticity condition on VAR disturbances and proxies.
The $\alpha $-mixing conditions in Assumption 2(iii) cover a large class of
uncorrelated, but possibly dependent, variables, including the case of
conditionally heteroskedastic disturbances.\ Assumption 2(iv) is a technical
condition necessary to prove the consistency of the MBB in this setting, see
Br\"{u}ggemann \emph{et al.} (2016); see also Assumption 2.4 in Jentsch and
Lunsford (2022).\footnote{%
The MBB is similar in spirit to a standard residual-based bootstrap where
the VAR residuals are resampled with replacement. However, instead of
resampling one VAR residual at a time the MBB, which is robust against forms
of `weak dependence' that may arise under $\alpha $-mixing conditions,
resamples blocks of the VAR residuals/proxies in order to replicate their
serial dependence structure. We refer to Jentsch and Lunsford (2019, 2022)
and Mertens and Ravn (2019) for a comprehensive discussion of the merits of
the MBB relative to other bootstrap methods in proxy-SVARs. Section \ref%
{Section_Supplementary_MBB_algorithm} in the Supplement sketches the
essential steps behind the MBB algorithm.}

The next assumption refers to the structural form.

\begin{assumption}[Structural form]
Given the SVAR in (\ref{VAR-RF2}),\ the matrix $B$ is nonsingular and its
inverse is denoted by $A=B^{-1}$.
\end{assumption}

Assumption 3 establishes the nonsingularity of the matrix $B$, which implies
the conditions $rank[B_{\bullet 1}]=k$ in (\ref{partition_B}) and $%
rank[A_{1\bullet }]=k$ in (\ref{partition_A}).

The next assumption is crucial to our approach. Henceforth, with $\tilde{%
\varepsilon}_{2,t}$ we denote a subset of the vector of non-target shocks $%
\varepsilon _{2,t}$ containing $s\leq n-k$ elements. We assume, without loss
of generality, that $\tilde{\varepsilon}_{2,t}$ corresponds to the first $s$
elements of $\varepsilon _{2,t}$, and it is intended that $\varepsilon
_{2,t}\equiv \tilde{\varepsilon}_{2,t}$ when $s=n-k$.

\begin{assumption}[Proxies for the non-target shocks]
\label{Assn 4}There exist $s\leq n-k$ proxy variables, collected in the
vector $w_{t}$, such that the following linear measurement system holds: 
\begin{equation}
w_{t}=\Lambda \tilde{\varepsilon}_{2,t}+\omega _{w,t},
\label{strong_proxies}
\end{equation}%
where $\Lambda :=${${E}$}$(w_{t}\tilde{\varepsilon}_{2,t}^{\prime })$ is an $%
s\times s$ matrix of relevance parameters and $\omega _{w,t}$ is a
measurement error term, uncorrelated with $\varepsilon _{t}$.
\end{assumption}

Assumption 4 establishes the existence of $s$ external variables which are
correlated with $s$ non-target shocks with covariance matrix $\Lambda :=${${E%
}$}$(w_{t}\tilde{\varepsilon}_{2,t}^{\prime })$, and are uncorrelated with
the target structural shocks, {${E}$}$(w_{t}\varepsilon _{1,t}^{\prime })=0$.%
\footnote{%
In principle, Assumption 4 can be generalized to allow for more proxies than
instrumented non-target shocks; i.e., $\dim (w_{t})>\dim (\tilde{\varepsilon}%
_{2,t})=s$. Without loss of generality, we focus on the case where $\Lambda $
in (\ref{strong_proxies})\ is a square matrix.} Assumption 4 implies that $%
\Sigma _{u,w}:=${${E}$}$(u_{t}w_{t}^{\prime })=\tilde{B}_{\bullet 2}\Lambda
^{\prime }$, where $\tilde{B}_{\bullet 2}:=\frac{\partial Y_{t}}{\partial 
\tilde{\varepsilon}_{2,t}^{\prime }}$ collects the $s$ columns of $\tilde{B}%
_{\bullet 2}$ associated with the instantaneous effects of the shocks $%
\tilde{\varepsilon}_{2,t}$; obviously, $\tilde{B}_{\bullet 2}\equiv
B_{\bullet 2}$ when $s=n-k$ ($\tilde{\varepsilon}_{2,t}\equiv \varepsilon
_{2,t}$). The illustrations we present in Section \ref%
{Section_empirical_illustrations} and in the Supplement show that Assumption
4 holds in many problems of interest.

Assumption 4 postulates the existence of proxies for the non-target shocks
but does not allow for models where the correlation between the proxies $%
w_{t}$ and the instrumented shocks $\tilde{\varepsilon}_{2,t}$ is \emph{weak}%
, i.e. arbitrarily close to zero. Weak correlation between $w_{t}$ and $%
\tilde{\varepsilon}_{2,t}$ can be allowed as in Montiel Olea \emph{et al.}
(2021, Section 3.2) by considering \emph{sequences }of models such that {${E}
$}$(w_{t}\tilde{\varepsilon}_{2,t}^{\prime })=\Lambda _{T}$, where $\Lambda
_{T}\rightarrow \Lambda $, and $\Lambda $ of reduced rank is allowed. To
illustrate, set $s=1$, so that $w_{t}$, $\tilde{\varepsilon}_{2,t}$ and {${E}
$}$(w_{t}\tilde{\varepsilon}_{2,t})$ in (\ref{strong_proxies}) are all
scalars. Then, we can consider a sequence of models with {${E}$}$(w_{t}%
\tilde{\varepsilon}_{2,t})=\lambda _{T}\rightarrow \lambda \in \mathbb{R}$.
In Montiel Olea \emph{et al.} (2021), a `strong instrument' corresponds to $%
\lambda \neq 0$; see also Assumption 2.3 in Jentsch and Lunsford (2022). A
`weak instrument' in the sense of Staiger and Stock (1997) corresponds to $%
\lambda _{T}=cT^{-1/2}$, where $\left\vert c\right\vert <\infty $ is a
scalar location parameter; under this embedding, $\lambda _{T}\rightarrow 0$%
, with the case of an `irrelevant' proxy corresponding to $c=0$. If the
proxy is strong ($\lambda \neq 0$), the asymptotic distribution of the
estimator of the parameters $( \tilde{B}_{\bullet 2}$, $\lambda _{T}^{\prime
})^{\prime }$ (or of the impulse responses to the shock $\tilde{\varepsilon}%
_{2,t}$)\ is Gaussian (see Supplement, Section \ref%
{Section_supplementary_Lemmas}). On the contrary, this is not guaranteed
when $\lambda =0$. For instance, if $\lambda _{T}=cT^{-1/2}$, the asymptotic
distribution of the estimator of $(\tilde{B}_{\bullet 2}^{\prime }$, $%
\lambda _{T}^{\prime })^{\prime }$ is non-Gaussian and the parameter $c$
governs the extent of the departure from the Gaussian distribution (see
Supplement, Section \ref{Section_supplementary_Lemmas}).

To deal with the case of multiple shocks ($s>1$), the embedding above can be
extended by considering a sequence of models with {${E}$}$(w_{t}\tilde{%
\varepsilon}_{2,t}^{\prime })=\Lambda _{T},$ $T=1,2,\ldots $, with the case
of strong proxies corresponding to 
\begin{equation}
\Lambda _{T}\rightarrow \Lambda ,\text{ }rank[\Lambda ]=s.
\label{eq strong proxxx}
\end{equation}%
\ Weak instruments as in Staiger and Stock (1997) correspond to the case
where $\Lambda _{T}$ can be approximated by 
\begin{equation}
\Lambda _{T}=CT^{-1/2}\text{ },\text{ \ }\left\Vert C\right\Vert <\infty
\label{eq weak proxxx}
\end{equation}%
$C$ being an $s\times s$ matrix with finite norm.

\section{Indirect-MD estimation}

\label{Section_indirect_approach_MD}We now present our indirect-MD
estimation approach based on the SVAR representation (\ref{partition_A}) and
the availability of external (strong) proxies $w_{t}$ for the non-target
shocks. In this framework, given the estimator of the parameters in $%
A_{1\bullet }$ we described below, the IRFs in (\ref{IRF_j}) are recovered
by using (\ref{crucial_relationship_partial}).

The first $k$ equations of system (\ref{partition_A}) read%
\begin{equation}
A_{1\bullet }u_{t}\equiv A_{1,1}u_{1,t}+A_{1,2}u_{2,t}=\varepsilon _{1,t}.
\label{structural_sub_A_bis}
\end{equation}%
Taking the variance of both sides of (\ref{structural_sub_A_bis}), we obtain
the $\frac{1}{2}k(k+1)$ moment conditions%
\begin{equation}
A_{1\bullet }\Sigma _{u}A_{1\bullet }^{\prime }=I_{k}.
\label{moments_A1_variance2}
\end{equation}%
Post-multiplying (\ref{structural_sub_A_bis}) by $w_{t}^{\prime }$ and
taking expectations yield the additional $ks$ moment conditions%
\begin{equation}
A_{1\bullet }\Sigma _{u,w}=0_{k\times s}.  \label{moments_A1}
\end{equation}%
Taken together, (\ref{moments_A1_variance2}) and (\ref{moments_A1}) provide $%
m:=\frac{1}{2}k(k+1)+ks$ independent moment conditions that can be used to
estimate the parameters in $A_{1\bullet }$. The idea is simple: the moment
conditions (\ref{moments_A1_variance2})-(\ref{moments_A1}) define a set of
`distances' between reduced form and structural parameters, which can be
minimized once $\Sigma _{u}$ and $\Sigma _{u,w}$ are replaced with their
consistent estimates. When $k>1$, however, the proxies alone do not suffice
to point-identify the proxy-SVAR, and it is necessary to impose additional
parametric restrictions; see Mertens and Ravn (2013), Angelini and Fanelli
(2019),\ Montiel Olea \emph{et al.} (2021), Arias \emph{et al.} (2021) and
Giacomini \emph{et al.} (2022). Depending on the information/theory
available, the additional restrictions can involve the parameters in $%
A_{1\bullet }$ or those in $B_{\bullet 1}$, and can be sign- or
point-restrictions.\footnote{%
See Section \ref{Section_supplementary_restrictions_on_B1} in the Supplement
for cases where additional point-restrictions are placed on the parameters
in $B_{\bullet 1}$.} We rule out the case of sign-restrictions and, as in
Angelini and Fanelli (2019), focus on general (possibly non-homogeneous)
linear constraints on $A_{1\bullet }$, as given by%
\begin{equation}
{vec}(A_{1\bullet })=S_{A_{1}}\alpha +s_{A_{1}}  \label{restrictions_A1}
\end{equation}%
where $\alpha $ is the vector of (free) structural parameters in $%
A_{1\bullet } $, $S_{A_{1}}$ is a full-column rank selection matrix and $%
s_{A_{1}}$ is a known vector. Under (\ref{restrictions_A1}), we provide
below necessary and sufficient conditions for local identification of the
proxy-SVAR; we refer to Bacchiocchi and Kitagawa (2022) for a thorough
investigation of SVARs that attain local identification, but may fail to
attain global identification.

Let $\sigma ^{+}:=({vech}(\Sigma _{u})^{\prime },{vec}(\Sigma
_{u,w})^{\prime })^{\prime }$ be the $m\times 1$ vector of reduced form
parameters entering the moment conditions in (\ref{moments_A1_variance2})-(%
\ref{moments_A1}). Let $\hat{\sigma}_{T}^{+}:=({vech}(\hat{\Sigma}%
_{u})^{\prime }$, ${vec}(\hat{\Sigma}_{u,w})^{\prime })^{\prime }$ be the
estimator of $\sigma ^{+}$, and $\sigma _{0}^{+}$ the corresponding true
value. $\hat{\sigma}_{T}^{+}$ is easily obtained from $\hat{\Sigma}_{u,w}:=%
\frac{1}{T}\sum_{t=1}^{T}\hat{u}_{t}w_{t}^{\prime }$\ and $\hat{\Sigma}_{u}:=%
\frac{1}{T}\sum_{t=1}^{T}\hat{u}_{t}\hat{u}_{t}^{\prime }$, $\hat{u}_{t}$, $%
t=1,..,T$, being the VAR residuals. By Lemma \ref{Lemma S.1} in the
Supplement, $T^{1/2}(\hat{\sigma}_{T}^{+}-\sigma _{0}^{+})\overset{d}{%
\rightarrow }N(0_{a\times 1},V_{\sigma ^{+}})$, with $V_{\sigma ^{+}}$
positive definite asymptotic covariance matrix that can be estimated
consistently under fairly general conditions. The moment conditions (\ref%
{moments_A1_variance2})-(\ref{moments_A1}) and the restrictions in (\ref%
{restrictions_A1}) can be summarized by the distance function%
\begin{equation}
g(\sigma ^{+},\alpha ):=\left( 
\begin{array}{c}
{vech}(A_{1\bullet }\Sigma _{u}A_{1\bullet }^{\prime }-I_{k}) \\ 
{vec}(A_{1\bullet }\Sigma _{u,w})%
\end{array}%
\right)  \label{distance_g}
\end{equation}%
where $A_{1\bullet }$ depends on $\alpha $ through (\ref{restrictions_A1}).
At the true parameter values, $g(\sigma _{0}^{+},\alpha _{0})=0_{m\times 1}$%
. The MD estimator of $\alpha $ is defined as%
\begin{equation}
\hat{\alpha}_{T}:=\arg \min_{\alpha \in \mathcal{P}_{\alpha }}\hat{Q}%
_{T}(\alpha ),\text{ \ }\hat{Q}_{T}(\alpha ):=g_{T}(\hat{\sigma}%
_{T}^{+},\alpha )^{\prime }\hat{V}_{gg}(\bar{\alpha})^{-1}g_{T}(\hat{\sigma}%
_{T}^{+},\alpha )  \label{MD_A_model}
\end{equation}%
where $g_{T}(\cdot ,\cdot )$ denotes the function $g(\cdot ,\cdot )$ once $%
\sigma ^{+}$ is replaced with $\hat{\sigma}_{T}^{+}$, $\mathcal{P}_{\alpha }$
is the parameter space, $\hat{V}_{gg}(\alpha ):=$ $G_{\sigma ^{+}}(\hat{%
\sigma}_{T}^{+},\alpha )\hat{V}_{\sigma ^{+}}G_{\sigma ^{+}}(\hat{\sigma}%
_{T}^{+},\alpha )^{\prime }$, $\hat{V}_{\sigma ^{+}}$ is a consistent
estimator of $V_{\sigma ^{+}}$, and $G_{\sigma ^{+}}(\sigma ^{+},\alpha )$
is the $m\times m$ Jacobian matrix $G_{\sigma ^{+}}(\sigma ^{+},\alpha ):=%
\frac{\partial g(\sigma ^{+},\alpha )}{\partial \sigma ^{+\prime }}$.
Finally, $\bar{\alpha}$ (interior point of $\mathcal{P}_{\alpha }$)\ is some
preliminary estimate of $\alpha $; for example, $\bar{\alpha}$ might be the
MD estimate of $\alpha $ obtained in a first-step by replacing $\hat{V}_{gg}(%
\bar{\alpha})$ in (\ref{MD_A_model})\ with the identity matrix, in which
case $\hat{\alpha}_{T}$ from (\ref{MD_A_model})\ corresponds to a classical
two-step MD estimator (see Newey and McFadden, 1994). Note that, despite
under Assumption 4 it holds $\Sigma _{u,w}:=\tilde{B}_{\bullet 2}\Lambda
^{\prime }$ (see Section \ref{Section_Assumptions}), in (\ref{MD_A_model})
the investigator needs not take a stand on the restrictions that might
characterize $\Lambda $ and $\tilde{B}_{\bullet 2}$.\footnote{%
Gains in efficiency can be achieved if these matrices are subject to
constraints that are explicitly imposed in the minimization problem (\ref%
{MD_A_model}) via the matrix $\Sigma _{u,w}$. For instance, if $\Lambda $ is
known to be diagonal (meaning that each proxy variable in $w_{t}$ solely
instruments one structural shock in $\tilde{\varepsilon}_{2,t}$), one can
use a constrained estimator of the covariance matrix $\Sigma _{u,w}$ in (\ref%
{MD_A_model}). This can be done by using $\hat{\Sigma}_{u,w}:=\widehat{%
\tilde{B}}_{\bullet 2}^{\prime }\hat{\Lambda}$, where $\hat{\Lambda}$ and $%
\widehat{\tilde{B}}_{\bullet 2}$ are obtained in a previous step through the
CMD approach we discuss in Section \ref{Section_bs_strength}.}

The next proposition establishes the necessary and sufficient rank
condition, as well as the necessary order condition for local identification
of the proxy-SVAR identified by the proxies $w_{t}$. $\mathcal{N}_{\alpha
_{0}}$ denotes a neighborhood of $\alpha _{0}$ in $\mathcal{P}_{\alpha }$,
with $\alpha _{0}$ true value of the structural parameters in the matrix $%
A_{1\bullet }$, and $D_{k}^{+}$ the generalized Moore-Penrose inverse of the
duplication matrix $D_{k}$, see Supplement, Section \ref{Section_bs_notation}%
.

\begin{proposition}[Point-identification]
\label{Prop 1}Consider the proxy-SVAR obtained by combining the SVAR (\ref%
{VAR-RF2}) with the proxies $w_{t}$ in (\ref{strong_proxies}) for the $s\leq
n-k$ non-target structural shocks $\tilde{\varepsilon}_{2,t}$. Assume that
the parameters in $A_{1\bullet }$ satisfy the $m:=\frac{1}{2}k(k+1)+ks$
independent moment conditions (\ref{moments_A1_variance2}) and (\ref%
{moments_A1}) and, for $k>1$, are restricted as in (\ref{restrictions_A1}).
Under Assumptions 1--4 and sequences of models in which {${E}$}$(w_{t}\tilde{%
\varepsilon}_{2,t}^{\prime })=\Lambda _{T}\rightarrow \Lambda $: \newline
(i)\ a necessary and sufficient condition for identification is that%
\begin{equation}
rank\left[ G_{\alpha }(\sigma ^{+},\alpha )\right] =a
\label{rank_Jacobian_a}
\end{equation}%
holds in $\mathcal{N}_{\alpha _{0}}$, where $a= dim(\alpha)$ and 
\begin{equation*}
G_{\alpha }(\sigma ^{+},\alpha ):=\left( 
\begin{array}{c}
2D_{k}^{+}(A_{1\bullet }\Sigma _{u}\otimes I_{k}) \\ 
(\Lambda \tilde{B}_{2\bullet }\otimes I_{k})%
\end{array}%
\right) S_{A_{1}};
\end{equation*}%
(ii) a necessary order condition is $a\leq m$; when $k>1$, this implies that
at least $\frac{1}{2}k(k-1)$ additional restrictions must be imposed on the
proxy-SVAR parameters.
\end{proposition}

As it is typical for SVARs and proxy-SVARs, the identification result in
Proposition \ref{Prop 1} holds `up to sign', meaning that the rank condition
in (\ref{rank_Jacobian_a})\ is valid regardless of the sign normalizations
of the rows of the matrix $A_{1\bullet }$. The necessary order condition, $%
a\leq m$, simply states that when $s$ shocks are instrumented, the number of
moment conditions used to estimate the proxy-SVAR must be larger or at least
equal to the total number of unknown structural parameters. It is not
strictly necessary that $s=n-k$, meaning that identification can be achieved
also by instrumenting part of the non-target shocks, provided there are
enough uncontroversial restrictions on $A_{1\bullet }$ through (\ref%
{restrictions_A1}).

An important consequence of Proposition \ref{Prop 1} is stated in the next
corollary, which establishes that the necessary and sufficient rank
condition for the identification of the proxy-SVAR fails when the proxies
are weak in the sense of (\ref{eq weak proxxx}).

\begin{corollary}[Identification failure]
\label{cor 1}Under the assumptions of Proposition \ref{Prop 1}, the
necessary and sufficient rank condition for identification in (\ref%
{rank_Jacobian_a})\ fails if the proxies satisfy (\ref{eq weak proxxx}).
\end{corollary}

The next proposition summarizes the asymptotic properties of the MD
estimator $\hat{\alpha}_{T}$ derived from (\ref{MD_A_model}) under local
identification.

\begin{proposition}[Asymptotic properties]
\label{prop 2}Under the conditions of Proposition \ref{Prop 1}, let the true
value $\alpha _{0}$ be an interior of $\mathcal{P}_{\alpha }$ (assumed
compact). If the necessary and sufficient rank condition in (\ref%
{rank_Jacobian_a}) is satisfied, then $\hat{\alpha}_{T}$ of (\ref{MD_A_model}%
) has the following properties:\newline
(i) $\hat{\alpha}_{T}\overset{p}{\rightarrow }\alpha _{0}$; \newline
(ii) $T^{1/2}\left( \hat{\alpha}_{T}-\alpha _{0}\right) \overset{d}{%
\rightarrow }N(0_{a\times 1},V_{\alpha })$, $V_{\alpha }:=\left\{ G_{\alpha
}(\sigma _{0}^{+},\alpha _{0})^{\prime }V_{gg}(\bar{\alpha})^{-1}G_{\alpha
}(\sigma _{0}^{+},\alpha _{0})\right\} ^{-1}$ with $V_{gg}(\alpha
):=G_{\sigma ^{+}}(\sigma _{0}^{+},\alpha )V_{\sigma ^{+}}G_{\sigma
^{+}}(\sigma _{0}^{+},\alpha )^{\prime }$ and $G_{\alpha }(\sigma
^{+},\alpha )$ as in Proposition~\ref{Prop 1}.
\end{proposition}

Proposition \ref{prop 2} ensures that the MD estimator $\hat{\alpha}_{T}$ is
consistent and asymptotically Gaussian if the rank condition holds.
Inference on the IRFs (\ref{IRF_j}) can be based on standard asymptotic
methods by classical delta-method arguments. Conversely, by Corollary 1,
consistency and asymptotic normality is not guaranteed to hold if the
instruments satisfy the local-to-zero embedding (\ref{eq weak proxxx}). The
rank of the Jacobian matrix $G_{\alpha }(\sigma ^{+},\alpha )$ in
Proposition \ref{Prop 1} depends on the the covariance matrix $\Sigma
_{w,u}=\Lambda \tilde{B}_{\bullet 2}^{\prime }$, which in turn reflects the
strength of the proxies $w_{t}$. The pre-test of relevance we discuss in
Section \ref{Section_testing} is based on an estimator of the parameters in $%
\Lambda $ and $\tilde{B}_{\bullet 2}$.

We end this section by noticing that our indirect-MD\ method presents
several differences with respect to Caldara and Kamps's (2017) approach to
proxy-SVARs. Caldara and Kamps (2017) interpret the structural equations of
their fiscal proxy-SVAR, the analog of system (\ref{structural_sub_A_bis}),
as fiscal reaction functions whose unsystematic components correspond to the
fiscal shocks of interest. They then identify the implied fiscal multipliers
by a Bayesian penalty function approach. We differ from Caldara and Kamps
(2017) in the motivations behind our analysis, as well as in the frequentist
nature of our approach\footnote{%
See Section S.6 in the Supplement for a comparison between the suggested MD
approach and the `standard' IV approach.}. Caldara and Kamps's (2017) main
objective is the estimation of fiscal multipliers from policy (fiscal)
reaction functions using external instruments. In contrast, our primary
objective is to rationalize a strategy intended to circumvent, when
possible, the use of weak-instrument robust methods. Finally, as our
empirical application in Section \ref{Section_empirical_illustrations}
illustrates, our approach is not confined or limited to cases where the
estimated structural equations read as policy reaction functions.

\section{Testing instrument relevance}

\label{Section_testing}\label{Section_testing copy(1)}In this section we
present our pre-test for relevance of the proxies. Our test exploits the%
\emph{\ }different asymptotic properties of a bootstrap estimator of
proxy-SVAR parameters under the regularity conditions in Proposition \ref%
{prop 2} -- which imply that the strong proxy condition (\ref{eq strong
proxxx}) is verified -- and under the weak IV sequences of Staiger and Stock
(1997) in (\ref{eq weak proxxx}). The test works for general $\alpha $%
-mixing VAR disturbances and/or zero-censored proxies, and is
computationally invariant to the number of shocks being instrument.
Importantly, the outcomes of the test do not affect post-test inferences
because of an asymptotic independence result between bootstrap and
non-bootstrap statistics that we summarize in Proposition \ref{prop 7}
below. This implies that the asymptotic coverage of IRFs confidence
intervals constructed using our indirect approach remains unaffected if the
bootstrap pre-test does not reject the null hypothesis of relevance of the
proxies $w_{t}$. Similarly, the asymptotic coverage is not affected even if
the bootstrap pre-test does reject the relevance of $w_{t}$ and
weak-instrument robust methods (using either the proxies $z_{t}$, or the
proxies $w_{t}$) are employed.

We organize this section as follows. In Section \ref{Section_bs_strength} we
discuss the bootstrap estimator used to capture the strength of the proxies
and then derive its asymptotic distribution. In Section \ref%
{Section_bs_diagnostic_test} we explain the mechanics of the test. In
Section \ref{Section_MC_results} we summarize its finite sample performance
through simulation experiments. Finally, Section \ref{Sub_section_screening}
focuses on its key properties.

\subsection{Bootstrap estimator and asymptotic distribution}

\label{Section_bs_strength}As noticed in Section \ref%
{Section_indirect_approach_MD}, the covariance matrix $\Sigma _{w,u}:=${${E}$%
}$(w_{t}u_{t}^{\prime })=\Lambda \tilde{B}_{\bullet 2}^{\prime }$ is a key
ingredient of the Jacobian $G_{\alpha }(\sigma ^{+},\alpha )$, which
determines the asymptotic properties of the MD estimator $\hat{\alpha}_{T}$;
see Propositions 1 and 2. In this section, we analyze a bootstrap estimator
of the parameters in $\Lambda $ and $\tilde{B}_{\bullet 2}^{\prime }$; this
estimator will subsequently serve as a measure of the strength of the
proxies $w_{t}$.

Let $\Omega _{w}$ be the $s\times s$ matrix defined by $\Omega _{w}:=\Sigma
_{w,u}\Sigma _{u}^{-1}\Sigma _{u,w}$. By combining $\Sigma _{w,u}=\Lambda 
\tilde{B}_{\bullet 2}^{\prime }$ with the `standard' SVAR\ covariance
restrictions, $\Sigma _{u}=BB^{\prime }$, by simple algebra we obtain the
relation $\Omega _{w}=\Lambda \tilde{B}_{\bullet 2}^{\prime }(BB^{\prime
})^{-1}\tilde{B}_{\bullet 2}^{\prime }\Lambda ^{\prime }=\Lambda \Lambda
^{\prime }$. Hence, the link between the reduced form parameters in $\Omega
_{w},\Sigma _{w,u}$ and the proxy-SVAR\ parameters in the $(n+s)\times s$
matrix $(\tilde{B}_{\bullet 2}^{\prime }$ , $\Lambda ^{\prime })^{\prime }$
is summarized by the following set of moment conditions%
\begin{equation}
\Omega _{w}=\Lambda \Lambda ^{\prime }\text{ ,\ }\Sigma _{w,u}=\Lambda 
\tilde{B}_{\bullet 2}^{\prime }  \label{moment_conditions_crucial}
\end{equation}%
which capture the connection between the proxies $w_{t}$ and the non-target
shocks $\tilde{\varepsilon}_{2,t}$. We denote by $\theta :=(\beta
_{2}^{\prime },\lambda ^{\prime })^{\prime }$ the $q_{\theta }\times 1$
vector containing the (free)\ parameters in the matrix $(\tilde{B}_{\bullet
2}^{\prime }$, $\Lambda ^{\prime })^{\prime }$; here, $\beta _{2}$ collects
the non-zero on-impact coefficients in $\tilde{B}_{\bullet 2}$ and $\lambda $
the non-zero elements in $\Lambda $. While the parameters in $\theta $ are
not economically interesting on their own, the asymptotic distribution of
the estimator of $\theta $ is informative on the strength of the proxies $%
w_{t}$.

The moment conditions (\ref{moment_conditions_crucial}) can be summarized by
the distance function $d(\mu ,\theta ):=\mu -f(\theta )$, with $\mu :=({vech}%
(\Omega _{w})^{\prime },{vec}(\Sigma _{w,u})^{\prime })^{\prime }$ and $%
f(\theta )=({vech}(\Lambda \Lambda ^{\prime })^{\prime }$, ${vec}(\Lambda 
\tilde{B}_{\bullet 2}^{\prime })^{\prime })^{\prime }$. At the true
parameter values, $d(\mu _{0},\theta _{0})=0$. In order to estimate $\theta $
through a MD approach, one needs an estimator of the reduced form parameters 
$\mu $. This is given by $\hat{\mu}_{T}:=({vech}(\hat{\Omega}_{w})^{\prime
}, $ ${vec}(\hat{\Sigma}_{w,u})^{\prime })^{\prime }$, where $\hat{\Omega}%
_{w}:=\hat{\Sigma}_{u,w}\hat{\Sigma}_{u}^{-1}\hat{\Sigma}_{u,w}$, $\hat{%
\Sigma}_{u,w}:=T^{-1}\sum_{t=1}^{T}\hat{u}_{t}w_{t}^{\prime }$ and $\hat{%
\Sigma}_{u}:=T^{-1}\sum_{t=1}^{T}\hat{u}_{t}\hat{u}_{t}^{\prime }$. When the
proxy-SVAR\ is identified as in Proposition \ref{Prop 1}, $T^{1/2}(\hat{\mu}%
_{T}-\mu _{0})$ is asymptotically Gaussian with positive definite asymptotic
covariance matrix $V_{\mu }:=J_{\sigma ^{+}}V_{\sigma ^{+}}J_{\sigma
^{+}}^{\prime }$, $J_{\sigma ^{+}}$ being the full-row rank Jacobian matrix $%
J_{\sigma ^{+}}:=\frac{\partial \mu }{\partial \sigma ^{+\prime }}$, see
Lemma \ref{Lemma S.2} in the Supplement, and $\hat{V}_{\mu }:=\hat{J}%
_{\sigma ^{+}}\hat{V}_{\sigma ^{+}}\hat{J}_{\sigma ^{+}}^{\prime }$ is a
consistent estimator of $V_{\mu }$.\footnote{%
In the `sandwich' expression $\hat{V}_{\mu }:=\hat{J}_{\sigma ^{+}}\hat{V}%
_{\sigma ^{+}}\hat{J}_{\sigma ^{+}}^{\prime },$ $\hat{V}_{\sigma ^{+}}$ is a
consistent estimator of $V_{\sigma ^{+}}$, see Supplement, Section \ref%
{Section_supplementary_Lemmas},\ and $\hat{J}_{\sigma ^{+}}$ is obtained
from the expression of $J_{\sigma ^{+}}$ in Lemma \ref{Lemma S.2} by
replacing $\Sigma _{w,u}$ and $\Sigma _{u}$ with the estimators $\hat{\Sigma}%
_{u,w}$ and $\hat{\Sigma}_{u}^{-1}$, respectively.} Conversely, by Lemma \ref%
{Lemma S.3} in the Supplement, $T^{1/2}(\hat{\mu}_{T}-\mu _{0})$ is not
asymptotically Gaussian when the proxies $w_{t}$ satisfy the local-to-zero
condition (\ref{eq weak proxxx}). Then, a classical MD (CMD) estimator of $%
\theta $ can defined as%
\begin{equation}
\hat{\theta}_{T}:=\arg \min_{\theta \in \mathcal{P}_{\theta }}\hat{Q}%
_{T}(\theta ),\text{ \ \ }\hat{Q}_{T}(\theta ):=d_{T}(\hat{\mu}_{T},\theta
)^{\prime }\hat{V}_{\mu }^{-1}d_{T}(\hat{\mu}_{T},\theta )
\label{CMD_problem}
\end{equation}%
where $d_{T}(\cdot ,\cdot )$ denotes the function $d(\cdot ,\cdot )$ once $%
\mu $ is replaced with $\hat{\mu}_{T}$, and $\mathcal{P}_{\theta }$ is the
parameter space.\footnote{\label{Footnote_restrictionsB2}For $s>1$, the
estimation problem (\ref{CMD_problem})\ requires that at least $(1/2)s(s-1)$
restrictions are placed on $\tilde{B}_{\bullet 2}^{\prime }$ and/or on $%
\Lambda $; see Proposition 1 in Angelini and Fanelli (2019) and the proof of
Lemma \ref{Lemma S.4} in the Supplement.} Lemma \ref{Lemma S.4} in the
Supplement shows that under the conditions of Proposition \ref{Prop 1}, $%
T^{1/2}(\hat{\theta}_{T}-\theta _{0})\overset{d}{\rightarrow }N(0,V_{\theta
})$, where $\theta _{0}:=(\beta _{2,0}^{\prime },\lambda _{0}^{\prime
})^{\prime }$ is the true value of $\theta $, $J_{\theta }$ is the
full-column rank Jacobian matrix $J_{\theta }:=\frac{\partial f(\theta )}{%
\partial \theta ^{\prime }}$, and $V_{\theta }:=\left( J_{\theta }^{\prime
}V_{\mu }^{-1}J_{\theta }\right) ^{-1}$. Hence, $\Gamma
_{T}:=T^{1/2}V_{\theta }^{-1/2}(\hat{\theta}_{T}-\theta _{0})$ is
asymptotically standard normal, and $\hat{V}_{\theta }:=(\hat{J}_{\theta
}^{\prime }\hat{V}_{\mu }^{-1}\hat{J}_{\theta })^{-1}$ is a consistent
estimator of $V_{\theta }$. In contrast, Lemma \ref{Lemma S.5} shows that,
asymptotically, $\Gamma _{T}$ is non-Gaussian when the instruments satisfy
the local-to-zero embedding in (\ref{eq weak proxxx}); its asymptotic
distribution is explicitly derived in the proof of Lemma \ref{Lemma S.5}.

The bootstrap counterpart of $\hat{\theta}_{T}$ (henceforth, MBB-CMD), given
by 
\begin{equation}
\hat{\theta}_{T}^{\ast }:=\arg \min_{\theta \in \mathcal{P}_{\theta }}\hat{Q}%
_{T}^{\ast }(\theta )\text{ \ , \ }\hat{Q}_{T}^{\ast }(\theta ):=d(\hat{\mu}%
_{T}^{\ast },\theta )^{\prime }\hat{V}_{\mu }^{-1}d(\hat{\mu}_{T}^{\ast
},\theta )  \label{CMD-MBB}
\end{equation}%
where $\hat{\mu}_{T}^{\ast }:=({vech}(\hat{\Omega}_{w}^{\ast })^{\prime },{%
vec}(\hat{\Sigma}_{w,u}^{\ast })^{\prime })^{\prime }$ is the bootstrap
analog of $\hat{\mu}_{T}$, is also affected by the strength of the proxies.
Specifically, Proposition \ref{prop 3} below shows that when the proxies are
strong in the sense of condition (\ref{eq strong proxxx}), the asymptotic
distribution of $\Gamma _{T}^{\ast }:=T^{1/2}\hat{V}_{\theta }^{-1/2}(\hat{%
\theta}_{T}^{\ast }-\hat{\theta}_{T})$, conditional on the data, is
asymptotically Gaussian.\footnote{%
As remarked in the Supplement, see Sections \ref%
{Section_supplementary_Lemmas} and \ref{Section_Supplementary_MBB_algorithm}%
, the asymptotic validity of the MBB requires that $\ell ^{3}/T\rightarrow 0$%
, where $\ell $ is the block length parameter behind resampling, see Jentsch
and Lunsford (2019, 2022). It is maintained that this condition holds in
Proposition \ref{prop 3} as well as in all cases in which the MBB is
involved. In the Monte Carlo experiments considered in Section \ref%
{Section_MC_results} and in the empirical illustrations considered in
Section \ref{Section_empirical_illustrations} and Section \ref%
{Section_supplementary_fiscal_proxy-SVAR}, $\ell $ is chosen as in Jentsch
and Lunsford (2019) and Mertens and Ravn (2019).} This result is consistent
with Theorem 4.1 in Jentsch and Lunsford (2022) on MBB consistency in
proxy-SVARs. In contrast, we show in Proposition \ref{prop 4} that under the
weak proxies embedding (\ref{eq weak proxxx}), the limiting distribution of $%
\Gamma _{T}^{\ast }$, conditional on the data, is random and non-Gaussian
(see equations (\ref{relationship_boot}) and (\ref{eq weak convergence in
distribution}) in the Supplement; see also Cavaliere and Georgiev, 2020, for
details on weak convergence\ in distribution).

\begin{proposition}[Bootstrap asymptotic distribution, strong proxies]
\label{prop 3}Con\-sider the CMD estimator $\hat{\theta}_{T}$ obtained from (%
\ref{CMD_problem})\ and its MBB counterpart $\hat{\theta}_{T}^{\ast }$
derived from (\ref{CMD-MBB}). Under the conditions of Proposition \ref{Prop
1}, if the necessary and sufficient rank condition for identification in (%
\ref{rank_Jacobian_a}) is satisfied\footnote{%
As is standard, with `$X_{T}^{\ast }\overset{d^{\ast }}{\rightarrow }_{p}X$'
we denote convergence of $X_{T}^{\ast }$ in conditional distribution to $X$,
in probability, as defined in the Supplement, Section \ref%
{Section_bs_notation}.}, $\Gamma _{T}^{\ast }:=T^{1/2}\hat{V}_{\theta
}^{-1/2}(\hat{\theta}_{T}^{\ast }-\hat{\theta}_{T})\overset{d^{\ast }}{%
\rightarrow }_{p}N(0_{q_{\theta }\times 1},I_{q_{\theta }})$.
\end{proposition}

\begin{proposition}[Bootstrap asymptotic distribution, weak proxies]
\label{prop 4} Consider the CMD estimator $\hat{\theta}_{T}$ obtained from (%
\ref{CMD_problem})\ and its MBB counterpart $\hat{\theta}_{T}^{\ast }$
derived from (\ref{CMD-MBB}). Under the conditions of Proposition \ref{Prop
1}, if the proxies $w_{t}$ satisfy the local-to-zero condition (\ref{eq weak
proxxx}), $\Gamma _{T}^{\ast }:=T^{1/2}\hat{V}_{\theta }^{-1/2}(\hat{\theta}%
_{T}^{\ast }-\hat{\theta}_{T})$ converges weakly in distribution to a
non-Gaussian limit.
\end{proposition}

The different asymptotic behaviors of $\Gamma _{T}^{\ast }$ highlighted in
Propositions \ref{prop 3} and \ref{prop 4} and, in particular, the distance
of the cdf of $\Gamma _{T}^{\ast }$ from the Gaussian cdf, are the key
ingredients of our bootstrap test of instrument relevance,\footnote{%
In principle, our approach can also be used to derive alternative estimators
of strength of the proxies $w_{t}$. For example, one can exploit only
subsets of proxy-SVAR moment conditions in (\ref{moment_conditions_crucial}%
). For instance, it is tempting to refer to a MD estimator of the parameters 
$\lambda $ alone, based on the moment conditions $\Omega _{w}=\Lambda
\Lambda ^{\prime }$. Although this is feasible, the estimators obtained
using subsets of moment conditions may fail to incorporate all the pertinent
information required to capture the strength of the proxies. Consequently,
the resulting pre-tests may exhibit relatively low power in finite samples.}
which we consider next.

\subsection{Bootstrap test}

\label{Section_bs_diagnostic_test}Our measure of strength is the cdf,
conditional on the data, of the bootstrap statistic $\hat{\Gamma}_{T}^{\ast
}:=T^{1/2}\hat{V}_{\theta }^{-1/2}(\hat{\theta}_{T}^{\ast }-\hat{\theta}%
_{T}) $. For simplicity and without loss of generality, we consider\textbf{\ 
}one component of the vector $\hat{\Gamma}_{T}^{\ast }$, say its first
element, $\hat{\Gamma}_{1,T}^{\ast }$; its cdf, conditional on the data, is
denoted by $\digamma _{T}^{\ast }(\cdot )$.

By Proposition \ref{prop 3}, if the proxies satisfy condition (\ref{eq
strong proxxx}), $\hat{\Gamma}_{1,T}^{\ast }$ converges to a standard normal
random variable; hence, $\digamma _{T}^{\ast }(x)-\digamma _{\mathcal{G}%
}\left( x\right) \rightarrow _{p}0$ uniformly in $x\in \mathbb{R}$ as $%
T\rightarrow \infty $, where $\digamma _{\mathcal{G}}\left( \cdot \right) $
denotes the $N(0,1)$ cdf. Our approach simply consists in evaluating, for
large $T$, how `close or distant' $\digamma _{T}^{\ast }(x)$ is from $%
\digamma _{\mathcal{G}}\left( x\right) $. To do so, consider a set of $N$
i.i.d. (conditionally on the original data)\ bootstrap replications, say $%
\hat{\Gamma}_{1,T:1}^{\ast },\ldots ,\hat{\Gamma}_{1,T:N}^{\ast }$, and the
corresponding estimator of $\digamma _{T}^{\ast }(x)$, given by%
\begin{equation}
\digamma _{T,N}^{\ast }\left( x\right) :=\frac{1}{N}\sum\nolimits_{b=1}^{N}%
\mathbb{I(}\hat{\Gamma}_{1,T:b}^{\ast }\leq x\mathbb{)}\text{, }x\in \mathbb{%
R}.  \label{eq bootstrap EDF}
\end{equation}%
For any $x$, deviation of $\digamma _{T,N}^{\ast }\left( x\right) $ from the
standard normal distribution can be evaluated by considering the distance $%
|\digamma _{T,N}^{\ast }(x)-\digamma _{\mathcal{G}}\left( x\right) |$. By
standard arguments, and regardless of the strength of the proxies, as $%
N\rightarrow \infty $ (keeping $T$ fixed)%
\begin{equation}
N^{1/2}(\digamma _{T,N}^{\ast }(x)-\digamma _{T}^{\ast }(x))\overset{d^{\ast
}}{\rightarrow }_{p}N\left( 0,U_{T}(x)\right)  \label{eq CLT for G*T,Btilde}
\end{equation}%
where $U_{T}(x):=\digamma _{T}^{\ast }(x)(1-\digamma _{T}^{\ast }(x))$. This
suggests that, with $\hat{U}_{T}(x)$ a consistent estimator of $U_{T}(x)$,%
\footnote{%
For instance, one may consider $\hat{U}_{T}(x):=\digamma _{T,N}^{\ast
}\left( x\right) (1-\digamma _{T,N}^{\ast }\left( x\right) )$ for an
arbitrary large value of $N$, or can simply set $\hat{U}_{T}(x)$ to its
theoretical value under normality; i.e., $\hat{U}_{T}(x):=U_{\mathcal{G}%
}(x)=\digamma _{\mathcal{G}}(x)(1-\digamma _{\mathcal{G}}(x)).$} we may
consider the normalized statistic:%
\begin{equation}
\tau _{T,N}^{\ast }(x):=N^{1/2}\hat{U}_{T}(x)^{-1/2}(\digamma _{T,N}^{\ast
}(x)-\digamma _{\mathcal{G}}\left( x\right) ).  \label{eq def SNTB}
\end{equation}%
The next two propositions establish the limit behavior of $\tau _{T,N}^{\ast
}(x)$ in the two scenarios of interest: under the conditions of Proposition %
\ref{prop 3}, where the proxy-SVAR is identified and strong proxy
asymptotics holds, and under the conditions of Proposition \ref{prop 4},
where weak proxy asymptotics \`{a} la Staiger and Stock (1997) holds.

\begin{proposition}
\label{prop 5}Assume that 
\begin{equation}
T,N\rightarrow \infty \text{ jointly and }NT^{-1}=o\left( 1\right) .
\label{rate_B_T}
\end{equation}%
Under the conditions of Proposition \ref{prop 3}, if $\digamma _{T}^{\ast
}(x)$ admits the standard Edgeworth expansion\footnote{%
The Edgeworth expansion here assumed is also maintained in e.g. Bose (1988)\
and Kilian (1988). It is typical in the presence of asymptotically normal
statistics, see e.g. Horowitz (2001, p. 3171) and Hall (1992).} $\digamma
_{T}^{\ast }(x)-\digamma _{\mathcal{G}}\left( x\right) $ $=O_{p}(T^{-1/2})$;
conditional on the data, then $\tau _{T,N}^{\ast }(x)\overset{d^{\ast }}{%
\rightarrow }_{p}N(0,1)$.
\end{proposition}

\begin{proposition}
\label{prop 6}Assume that (\ref{rate_B_T}) holds. Under the conditions of
Proposition \ref{prop 4}, $\tau _{T,N}^{\ast }(x)$ diverges at the rate $%
N^{1/2}$.
\end{proposition}

Together, Propositions 5 and 6 form the basis of our approach to testing
instrument relevance: precisely, a straightforward test can be conducted by
directly comparing $\tau _{T,N}^{\ast }(x)$ with critical values derived
from the standard normal distribution, regardless of the number of shocks
being instrumented. The rejection of the null hypothesis indicates the
presence of weak proxies. A few remarks about the test are as follows.

\medskip

\noindent (i) The condition \eqref{rate_B_T} is a specificity of the
suggested approach: $N$ should be large for power consideration but, at the
same time, $N$ should not be too large relatively to $T$, otherwise the
noise generated by the $N$ random draws from the bootstrap distribution will
cancel the signal about the form of such distribution, which depends on $T$;
see below and the proof of Proposition \ref{prop 5}. As a practical rule, we
suggest using $N=[T^{1/2}]$; see the next section.

\medskip

\noindent (ii) Consistency of the test is preserved despite the asymptotic
randomness of $\digamma _{T}^{\ast }(\cdot )$, which makes the power of the
test random. The asymptotic randomness of $\digamma _{T}^{\ast }(\cdot )$
introduces complexity in analyzing the local power of the test, which
exceeds the scope of this paper.

\medskip

\noindent (iii) The scalar test statistic $\tau _{T,N}^{\ast }(x)$ defined
in (\ref{eq def SNTB}) can be built by considering the cdf of any single
components of the vector $\hat{\Gamma}_{T}^{\ast }$; moreover, the results
in Propositions \ref{prop 5} and \ref{prop 6} can be extended to
multivariate counterparts of $\tau _{T,N}^{\ast }(x)$, constructed on whole
vector $\hat{\Gamma}_{T}^{\ast }$. That is, one can check relevance of the
proxies by using both multivariate and univariate normality tests.\footnote{%
In principle, a sup-type test based on $\tau _{T,N}^{\ast }(x)$ could be
constructed by considering the classical Kolmogorov-Smirnov-type statistic $%
N^{1/2}||\digamma _{T,N}^{\ast }-\digamma _{\mathcal{G}}||_{\infty
}=N^{1/2}\sup_{x\in \mathbb{R}}|\digamma _{T,N}^{\ast }(x)-\digamma _{%
\mathcal{G}}(x)|$. A $CvM$-type measure of discrepancy delivers $N||\digamma
_{T,N}^{\ast }-\digamma _{\mathcal{G}}||_{2}^{2}=N\int_{\mathbb{R}}(\digamma
_{T,N}^{\ast }(x)-\digamma _{\mathcal{G}}(x))^{2}dx$, while $N\int_{\mathbb{R%
}}\frac{(\digamma _{T,N}^{\ast }(x)-\digamma _{\mathcal{G}}(x))^{2}}{\hat{U}%
_{T}(x)}dx=N\int_{\mathbb{R}}\tau _{T,N}^{\ast }(x)^{2}dx$ leads to an
Anderson-Darling-type statistic. In all cases, the test rejects for large
values of the test statistic. Further tests of normality are considered in
sections \ref{Section_MC_results} and \ref{Section_empirical_illustrations}.}

\medskip

\noindent (iv) The test can be further simplified, \textit{ceteris paribus},
by considering the estimator $\hat{\theta}_{T}^{\ast }$ in place of its
normalized version $\hat{\Gamma}_{T}^{\ast }$. Henceforth, we use $\hat{%
\vartheta}_{T}^{\ast }$ to denote any of the following statistics that can
be alternatively used to test relevance by a normality test: (a) $\hat{%
\vartheta}_{T}^{\ast }\equiv \hat{\theta}_{T}^{\ast }$; (b) $\hat{\vartheta}%
_{T}^{\ast }\equiv \hat{\Gamma}_{T}^{\ast }$; (c) any sub-vector of $\hat{%
\theta}_{T}^{\ast }$ (e.g., $\hat{\vartheta}_{T}^{\ast }\equiv \hat{\beta}%
_{2,T}^{\ast }$, $\hat{\vartheta}_{T}^{\ast }\equiv \hat{\lambda}_{T}^{\ast
} $, or $\hat{\vartheta}_{T}^{\ast }\equiv \hat{\theta}_{i,T}^{\ast }$, $%
\hat{\theta}_{i,T}^{\ast }$\ being the $i$-th element of $\hat{\theta}%
_{T}^{\ast } $); (d) any sub-vector of $\hat{\Gamma}_{T}^{\ast }$.

\medskip

\noindent (v) The testing principle developed in this section can in fact be
applied to \emph{any }bootstrap statistic built from the proxy-SVAR,
provided it is (asymptotically) Gaussian under the strong proxy condition (%
\ref{eq strong proxxx}), and (asymptotically) non-Gaussian under the weak
proxy condition (\ref{eq weak proxxx}). For instance, when one proxy is used
for one structural shock our approach can also be applied to the bootstrap
(normalized) IRFs in Jentsch and Lunsford (2022), which satisfy these two
conditions; see their Corollary 4.1 and Theorem 4.3(i)(a).

\medskip

\noindent (vi) As a concluding remark, it is worth noting that our suggested
pre-test can, in principle, be applied to the original proxies $z_{t}$ for
the target shocks, similar to how it is applied to the proxies $w_{t}$ for
the non-target shocks. Proposition \ref{prop 7} in Section \ref%
{Sub_section_screening} below guarantees that there are no pre-testing
issues in the subsequent inference.

\subsection{Monte Carlo results}

\label{Section_MC_results}In this section, we investigate by Monte Carlo
simulations the finite sample properties of the bootstrap test of relevance
discussed in the previous section.\footnote{Simulations have been performed with Matlab 2021b. Codes, including the ones that replicate the empirical illustrations, are available upon request from the authors.}

The DGP belongs to a SVAR system with $n=3$ variables, featuring a single
target shock $\varepsilon _{1,t}$ ($k=1$) and two non-target shocks ($n-k=2$%
). The dynamic causal effects produced by the target shock $\varepsilon
_{1,t}$ are recovered by the indirect-MD approach developed in Section \ref%
{Section_indirect_approach_MD}, i.e., by estimating the structural equation $%
A_{1\bullet }u_{t}=\alpha _{1,1}u_{1,t}+\alpha _{1,2}u_{2,t}+\alpha
_{1,3}u_{3,t}=\varepsilon _{1,t}$ using a proxy $w_{t}$ for one of the two
non-target shocks, along with the maintained hypothesis (valid in the DGP)
that $\alpha _{1,2}=0$; hence, $k=1$ and $s=1<n-k=2$. The proxy $w_{t}$ is
uncorrelated with the target shock $\varepsilon _{1,t}$ as well as with the
other non-instrumented, non-target shock of the system; see Supplement,
Section \ref{Section_Supplement_DGP} for details. The strength of the proxy $%
w_{t}$ is tested on samples of length $T=250$ and $T=1,000$, with $\eta
_{t}:=(u_{t}^{\prime },w_{t})^{\prime }$ being either i.i.d. or a GARCH-type
process. All elements of the DGP are described in detail in the Supplement,
Section \ref{Section_Supplement_DGP}.

Table 1 summarizes the empirical rejection frequencies of the bootstrap
diagnostic test computed on 20,000 simulations in three different scenarios,
see below. All normality tests are carried out at the $5\%$ nominal
significance level, considering bootstrap replications of elements of the
MBB-CMD estimator $\hat{\theta}_{T}^{\ast }:=(\hat{\beta}_{2,T}^{\ast \prime
},\hat{\lambda}_{T}^{\ast \prime })^{\prime }$.\footnote{%
As already observed, in the MBB\ algorithm we fix the parameter $\ell $ (see
Supplement, Section \ref{Section_Supplementary_MBB_algorithm})\ to the
largest integer smaller than the value $5.03T^{1/4}$; see Jentsch and
Lunsford (2019) and Mertens and Ravn (2019). In their simulation
experiments, Jentsch and Lunsford (2022) use $\ell =4$ in samples of $T=200$
observations; we checked that the results of our simulation experiments
based on $T=250$ observations do not change substantially with $\ell =4$.}
We apply Doornik and Hansen's (2008) multivariate test of normality (DH in
the table) to the sequence of bootstrap replications $\{\hat{\vartheta}%
_{T:1}^{\ast },$ $\hat{\vartheta}_{T:2}^{\ast },\ldots ,\hat{\vartheta}%
_{T:N}^{\ast }\}$, where $\hat{\vartheta}_{T}^{\ast }$ is selected as $\hat{%
\vartheta}_{T}^{\ast }\equiv \hat{\beta}_{2,T}^{\ast }$ (see (iii) in
Section \ref{Section_bs_diagnostic_test}); further, we apply\ Lilliefors'
(1967) version of univariate Kolmogorov-Smirnov (KS in the table) tests of
normality to the sequence $\{\hat{\vartheta}_{T:1}^{\ast },$ $\hat{\vartheta}%
_{T:2}^{\ast },\ldots ,\hat{\vartheta}_{T:N}^{\ast }\}$, with $\hat{\vartheta%
}_{T}^{\ast }$ selected as $\hat{\vartheta}_{T}^{\ast }\equiv \hat{\theta}%
_{i,T}^{\ast }$, for $i=1,\ldots ,q_{\theta }$, $\hat{\theta}_{i,T}^{\ast }$
being the $i$-th scalar component of $\hat{\theta}_{T}^{\ast }$ (again, see
(iii) in Section \ref{Section_bs_diagnostic_test}). In Table 1, rejection
frequencies not in parentheses refer to the case in which $\eta
_{t}:=(u_{t}^{\prime },w_{t})^{\prime }$ is generated as an i.i.d. process;
rejection frequencies in parentheses refer to the case in which each
component in the vector $\eta _{t}:=(u_{t}^{\prime },w_{t})^{\prime }$ is
generated from univariate GARCH(1,1) processes, independent across
equations. The tuning parameter $N$ is set to $N=[T^{1/2}]$.\footnote{%
Building upon the findings in Angelini \emph{et al.} (2022), we investigate
the selection of $N$ out of $T$ through several additional simulation
experiments, which are not presented here to save space. Results suggest
that the choice $N=[T^{1/2}]$ strikes a satisfactory balance between
controlling the size and maximizing power in samples of lengths commonly
encountered in practical settings.}

Results in the upper panel of Table 1 refer to a `strong proxy' scenario. In
this scenario, the correlation between the `indirect' proxy $w_{t}$ and the
instrumented non-target shocks $\tilde{\varepsilon}_{2,t}$ is set to 59\%
and, in line with the strong proxy condition (\ref{eq strong proxxx}), does
not change with the sample size. Overall, it is evident that the test
effectively controls nominal size reasonably well.

The middle panel of Table 1 presents the rejection frequencies computed
under a `moderately weak proxy' scenario. In this framework, the covariance
between $w_{t}$ and $\tilde{\varepsilon}_{2,t}$ is of the form $\lambda
_{T}=cT^{-1/2}$, see (\ref{eq weak proxxx}), with $c$ chosen such that the
correlation between $w_{t}$ and $\tilde{\varepsilon}_{2,t}$ is 25\% with $%
T=250$, and collapses, \textit{ceteris paribus}, to 13\% with $T=1,000$. Our
test behaves reasonably well: when $T=250$, the test based on $\hat{\vartheta%
}_{T}^{\ast }\equiv \hat{\beta}_{2,T}^{\ast }$ detects the weak proxy with
rejection frequencies fluctuating in the range 20\%--22\%; importantly, the
empirical rejection frequencies increase to 63\%--80\% as $T$ increases.

Finally, the results in the lower panel of Table 1 refer to a `weak proxy'
scenario, where $c$ is such that the correlation between $w_{t}$ and $\tilde{%
\varepsilon}_{2,t}$ is $5\%$ for $T=250$ and reduces, \textit{ceteris paribus%
}, to $2\%$ for $T=1,000$. The table shows that the test detects weak
proxies with high accuracy, regardless of whether the disturbances $\eta
_{t} $ are i.i.d. or follow GARCH(1,1)-type processes. The power of the test
approaches one as the sample size increases, indicating its effectiveness in
detecting weak proxies.

\subsection{Post-test inference on the IRFs}

\label{Sub_section_screening}As is known from the literature on IV
regressions, caution is needed when choosing among instruments on the basis
of their first-stage significance, as screening worsens small sample bias;
see, e.g., Zivot \emph{et al.} (1998), Hausman \emph{et al.} (2005) and
Andrews \emph{et al.} (2019). Hence, one important way to assess the overall
performance of our novel bootstrap pre-test is to examine, in addition to
the rejection frequencies in Table 1, the reliability of post-test
inferences. In this section, we focus, in particular, on the post-test
coverage of confidence intervals for IRFs obtained by the indirect-MD
approach.

In the following, $\rho _{T}$ denotes any statistic based on the proxy-SVAR
estimates from the original sample. For instance, $\rho _{T}$ can be a
Wald-type statistic used for testing restrictions on the proxy-SVAR\
parameters; for a given time horizon $h$ and estimated IRF $\hat{\gamma}%
_{i,j}(h)$ in (\ref{IRF_j}), $\rho _{T}$ might be given by $\rho
_{T}:=T^{1/2}(\hat{\gamma}_{i,j}(h)-\gamma _{i,j,0}(h))/\hat{V}_{\gamma
_{i,j}}^{1/2}$, with $\gamma _{i,j,0}(h)$ being the postulated true null
value and $\hat{V}_{\gamma _{i,j}}$ an estimator of the asymptotic variance.
With $\tau _{T,N}^{\ast }:=\tau (\hat{\theta}_{T:1}^{\ast },\ldots $, $\hat{%
\theta}_{T:N}^{\ast })$, $\tau (\cdot )$ being a continuous function, we
denote any statistic computed from a sequence of $N$ bootstrap replications
of the MBB-CMD estimator, $\hat{\theta}_{T}^{\ast }$. For ease of reference,
in the following we assume that $\tau _{T,N}^{\ast }$ coincides with the
statistic $\tau _{T,N}^{\ast }(x)$ defined in (\ref{eq def SNTB}). Note that 
$\tau _{T,N}^{\ast }$ depends on the original data through its (conditional)
distribution function $\digamma _{T}\left( \cdot \right) $ only.

The following proposition establishes that the statistics $\rho _{T}$ and $%
\tau _{T,N}^{\ast }$ are asymptotically independent (as $T,N\rightarrow
\infty $). We implicitly assume that the data and the auxiliary variables
used to generate the bootstrap data are defined jointly on an extended
probability space.

\begin{proposition}[Asymptotic independence]
\label{prop 7}Let $\rho _{T}$ and $\tau _{T,N}^{\ast }$ be as defined above.
For any $x_{1},x_{2}\in \mathbb{R}$ and $T,N\rightarrow \infty $, it holds
that%
\begin{equation}
P(\left\{ \rho _{T}\leq x_{1}\right\} \cap \left\{ \tau _{T,N}^{\ast }\leq
x_{2}\right\} )-P(\rho _{T}\leq x_{1})P(\tau _{T,N}^{\ast }\leq x_{2})%
\overset{}{\longrightarrow }0,  \label{asymptotic_independence}
\end{equation}%
provided that the conditions of Proposition \ref{prop 5} or Proposition \ref%
{prop 6} hold.
\end{proposition}

The main implication of Proposition \ref{prop 7} is that, under strong
proxies or under weak proxies as in (\ref{eq weak proxxx}), large-sample
inference in the proxy-SVAR based on the statistic $\rho _{T}$ is not
affected by the outcomes of the bootstrap-based statistic $\tau _{T,N}^{\ast
}$. Thus, if the pre-test does not reject the null of relevance, post-test
inference on the proxy-SVAR parameters can be conducted by standard
asymptotic methods without relying on Bonferroni-type adjustments. Moreover,
if the bootstrap pre-test rejects the null of relevance, the investigator
can still apply weak-instrument robust methods, no matter whether they
instrument the target shocks\ $z_{t}$ or the non-target shocks $w_{t}$. In
any case, post-test inference will not be affected asymptotically by the
outcome of the test. Note that here we do not consider sequences of
parameters converging to zero at a rate different from $T^{-1/2}$; see, for
instance, Andrews and Cheng (2012). Accordingly, we do not claim here that
the asymptotic result in Proposition \ref{prop 7} holds uniformly.

To illustrate this important implication of Proposition \ref{prop 7},
consider the DGP discussed in Section \ref{Section_bs_diagnostic_test}.
Figure 1 plots, for samples of $T=250$ observations and for $h=0,1,\ldots
,12 $, the empirical coverage probabilities of $90\%$ confidence intervals
constructed for the response of $Y_{3,t+h}$ to the target shock $\varepsilon
_{1,t}$. Empirical coverage probabilities are estimated using 20,000 Monte
Carlo draws.

The black line (labeled as `Strong, indirect-MD') in the graph, which is
mostly overlapped by the pale blue line (see below), depicts the empirical
coverage probabilities obtained through our indirect-MD approach,
implemented as discussed in the Monte Carlo Section \ref%
{Section_bs_diagnostic_test}. Thus, given the estimated structural
parameters $\hat{A}_{1\bullet }:=(\hat{\alpha}_{1,1},0,\hat{\alpha}%
_{1,3})^{\prime }$ (recall that $\alpha _{1,2}=0$ is imposed)\ and the
implied IRFs $\hat{\gamma}_{3,1}(h)$, $h=0,1,\ldots ,12,$ $\hat{\gamma}%
_{3,1}(h)$ being the third element of $\hat{\gamma}_{\bullet
1}(h):=(S_{n}^{\prime }(\widehat{\mathcal{C}_{y}})^{h}S_{n})\hat{\Sigma}%
_{u,T}\hat{A}_{1\bullet }^{\prime }$, we build $90\%$ confidence intervals
for the true response $\gamma _{3,1,0}(h)$, using the statistic $\rho _{T}$
described above. The setup corresponds to the `strong proxy' scenario
analyzed in the upper panel of Table 1.

Figure 1 shows that, unconditionally, the finite sample coverage of IRFs is
satisfactory. The pale blue line refers to conditional probabilities
(labelled as `Strong, indirect-MD$\mid $DH${\small \leq }$cv'); i.e.,
empirical coverage probabilities conditionally on the bootstrap pre-test,
based on $\tau _{T,N}^{\ast }\equiv DH$ and $N=[T^{1/2}]$, failing to reject
the null that $w_{t}$ is relevant for the instrumented non-target shock. The
graphs in Figure 1 support the result in Proposition \ref{prop 7}:
unconditional and conditional empirical coverage probabilities tend to
coincide.

To further appreciate the asymptotic independence result in Proposition \ref%
{prop 7}, we now consider the coverage of weak-instrument robust methods
when our pre-test \emph{rejects} the relevance condition. As already
observed, when the strong proxy condition for $w_{t}$ is rejected,
researchers can proceed by relying on weak-instrument robust methods as in
Montiel Olea \emph{et al.} (2021). To do so, they can use either the (weak)
proxies $z_{t}$ available for the target shocks, or the (weak) proxies $%
w_{t} $ available for the non-target shocks.

We focus on the case in which the strong proxy condition for $w_{t}$ is
rejected, and the responses of $Y_{3,t+h}$ to $\varepsilon _{1,t}$ are
estimated by the direct approach; i.e., by directly instrumenting the target
shocks $\varepsilon _{1,t}$ with the weak\ proxy $z_{t}$. We specify a DGP
for $z_{t}$ which mimics the `weak proxy' scenario already considered for $%
w_{t}$. In particular, we set $Cov(z_{t},\varepsilon _{1,t})=\phi
_{T}=cT^{-1/2}$, and fix the magnitude of the location parameter $c$ such
that the correlation between $z_{t}$ and $\varepsilon _{1,t}$ is 4.5\% in
samples with $T=250$. Several key findings can be derived from this analysis.

First, when constructing `plug-in' confidence intervals under the maintained
that $z_{t}$ serves as a relevant instrument for $\varepsilon _{1,t}$, the
resulting coverage, represented by the red line in Figure 1 (labelled as
`Weak'), is unsatisfactory.

Second, if one pre-tests the weakness of $z_{t}$ by the first-stage F-test
approach\ and compute confidence intervals for the target responses only
when the first-stage F-test rejects the null of weak proxy, the coverage
probabilities, corresponding to the green line in Figure 1 (labelled as `Weak%
$\mid $F${\small >}$cv'), are unsatisfactory. That is, screening on the
first-stage F-test worsens coverage.

Third, in this scenario, weak-instrument robust (Anderson-Rubin) confidence
intervals based on Montiel Olea \emph{et al.} (2021)'s approach using $z_{t}$
as an instrument have empirical coverage probabilities, summarized by the
blue line in Figure 1 (labeled as "Weak, A\&R"), that closely match the
nominal level.

Fourth, if weak-instrument robust confidence intervals are computed only
when our bootstrap pre-test rejects the relevance of $w_{t}$, \emph{%
conditional }empirical coverage probabilities, given by the orange line in
Figure 1 (labelled as `Weak, A\&R$\mid $DH${\small >}$cv'), are close to the
unconditional ones (blue line). This result aligns with the asymptotic
independence result in Proposition \ref{prop 7}. Similar results obtain if
the bootstrap pre-test is applied to $z_{t}$ rather than $w_{t}$.

\section{Empirical illustrations}

\label{Section_empirical_illustrations}We demonstrate the relevance of our
identification and estimation strategy for proxy-SVARs by reexamining some
empirical illustrations previously discussed in the literature through the
lens of our indirect-MD approach. In Section \ref%
{Section_empirical_illustration_oil} we concentrate on Kilian's (2009) model
for global crude oil production. Section \ref%
{Section_empirical_illustration_uncertainty} examines the joint
identification of financial and macroeconomic uncertainty shocks using
Ludvigson, Ma and Ng's (2021) data and reduced form VAR. A third empirical
illustration, which pertains to a fiscal proxy-SVAR, is deferred to the
Supplement.

\subsection{Oil supply shock}

\label{Section_empirical_illustration_oil}Kilian (2009) considers a
three-equation ($n=3$) SVAR for $Y_{t}:=(prod_{t},rea_{t}$, $%
rpo_{t})^{\prime }$, where $prod_{t}$ is the percentage change in global
crude oil production, $rea_{t}$ is a global real economic activity index of
dry goods shipments and $rpo_{t}$ is the real oil price. Using monthly data
for the period 1973:M1-2007:M12 and a Choleski decomposition based on the
above ordering of the variables, he identifies three structural shocks: an
oil supply shock, $\varepsilon _{t}^{S}$, an aggregate demand shock, $%
\varepsilon _{t}^{AD}$, and an oil-specific demand shock, $\varepsilon
_{t}^{OSD}$, respectively. Montiel Olea {\emph{et al.}} (2021) focus on the
identification of the oil supply shock $\varepsilon _{t}^{S}$ alone, using
Kilian's (2009) reduced form VAR and Kilian's (2008) measure of `exogenous
oil supply shock', $z_{t}$, as external instrument for the shock of
interest, $\varepsilon _{t}^{S}$.

In our notation, $\varepsilon _{1,t}=\varepsilon _{t}^{S}$ ($k=1$) is the
target structural shock, $z_{t}$ is Kilian's (2008)\ proxy directly used for 
$\varepsilon _{1,t}$, and $\varepsilon _{2,t}=(\varepsilon
_{t}^{AD},\varepsilon _{t}^{OSD})^{\prime }$ ($n-k=2$) collects the
non-target shocks of the system. The counterpart of the representation (\ref%
{partition_B}) of the proxy-SVAR is given by the system%
\begin{equation*}
u_{t}:=\left( 
\begin{array}{c}
u_{t}^{prod} \\ 
u_{t}^{rea} \\ 
u_{t}^{rpo}%
\end{array}%
\right) =\left( 
\begin{array}{c}
\beta _{1,1} \\ 
\beta _{2,1} \\ 
\beta _{3,1}%
\end{array}%
\right) \varepsilon _{t}^{S}+B_{\bullet 2}\varepsilon _{2,t}
\end{equation*}%
where $u_{t}$ is the vector of VAR\ disturbances, and $B_{\bullet 1}\equiv
(\beta _{1,1},\beta _{2,1},\beta _{3,1})^{\prime }$ captures the
instantaneous impact of the oil supply shock on the variables. The
counterpart of the linear measurement equation (\ref{equation_link}) is
given by $z_{t}=\phi \varepsilon _{t}^{S}+\omega _{z,t}$, where $\phi $ is
the relevance parameter and $\omega _{z,t}$ is a measurement error,
uncorrelated with all other structural shocks of the system. Since $k=1$, no
additional restriction on the proxy-SVAR parameters is needed to build
weak-instrument robust confidence intervals.

For comparison purposes, we start from the direct approach, which is based
on instrumenting the oil supply shock with the proxy $z_{t}$. Since $z_{t}$
is available on the period 1973:M1-2004:M9, following Montiel Olea {\emph{et
al.}} (2021), we use the common sample period 1973:M1-2004:M9 ($T=381$
monthly observations) for estimation. Montiel Olea {\emph{et al.}} (2021)
report a robust first-stage F statistic for the proxy $z_{t}$ equal to 9.4.
We complement their analysis with our bootstrap pre-test for instrument
relevance. More precisely, we apply Doornik and Hansen's (2008) multivariate
test of normality ($\tau _{T,N}^{\ast }\equiv DH$) on the sequence of MBB
replications $\{\hat{\vartheta}_{T:1}^{\ast },$ $\hat{\vartheta}_{T:2}^{\ast
},\ldots ,\hat{\vartheta}_{T:N}^{\ast }\}$, fixing the tuning parameter at $%
N=[T^{1/2}]=19$. The bootstrap estimator $\hat{\vartheta}_{T}^{\ast }$ is
obtained as follows.\ First, we consider $\hat{\vartheta}_{T}^{\ast }\equiv 
\hat{\theta}_{T}^{\ast }$, where $\hat{\theta}_{T}^{\ast }=(\hat{\beta}%
_{1,T}^{\ast \prime },\hat{\phi}_{T}^{\ast })^{\prime }$ is the MBB-CMD
estimator discussed in Section \ref{Section_indirect_approach_MD}.\footnote{%
Since in this case we are testing the strength of a proxy which directly
instruments the target shock, the test is based on the MBB-CMD estimator in (%
\ref{CMD-MBB}) computed from the moment conditions $\Sigma _{z,u}=\phi
B_{1}^{\prime }$, $\Omega _{z}=\phi B_{1}^{\prime }(BB^{\prime
})^{-1}B_{1}^{\prime }\phi =\phi ^{2}$, which capture the strength of the
proxy $z_{t}$ for the oil supply shock.} The multivariate normality test
yields a p-value of 0.04. Subsequently, considering the choice $\hat{%
\vartheta}_{T}^{\ast }\equiv \hat{\beta}_{1,T}^{\ast }$, the multivariate
normality test returns a p-value of 0.004 (univariate normality tests
corroborate this result). Overall, the bootstrap pre-test provides evidence
countering the hypothesis that Kilian's (2008) proxy $z_{t}$ serves as a
relevant instrument for the oil supply shock. This result lends support to
the employment of the weak-instrument robust approach developed in Montiel
Olea \emph{et al.} (2021).

The\ blue lines plotted in Figure 2 are the estimated dynamic responses to
the oil supply shock identified by Kilian's (2008) proxy $z_{t}$. More
precisely, the graph quantifies the responses of the variables in $%
Y_{t}:=(prod_{t},rpo_{t},rea_{t})^{\prime }$ to an oil supply shock that
increases oil production of 1\% on-impact (the responses plotted for $%
prod_{t}$ are cumulative percent changes). The blue shaded areas depict the
corresponding 68\% (in panel A) and 95\% (in panel B) Anderson-Rubin
weak-instrument robust confidence intervals. They closely resemble the IRFs
plotted in panels A and B of Figure 1 in Montiel Olea \emph{et al.} (2021).
The orange dotted lines represent Jentsch and Lunsford's (2021) 68\% (in
panel A) and 95\% (in panel B) `grid MBB AR' confidence intervals. It is
evident that the use of the MBB enhances the precision of weak-instrument
robust inference on the dynamic causal effects induced by the oil supply
shock.

We now move to our indirect-MD approach, which requires instrumenting the
non-target shocks $\varepsilon _{2,t}=(\varepsilon _{t}^{AD},\varepsilon
_{t}^{OSD})^{\prime }$. The counterpart of system (\ref{structural_sub_A_bis}%
) is given by the equation:%
\begin{equation}
A_{1\bullet }u_{t}=\alpha _{1,1}u_{t}^{prod}+(\alpha _{1,2},\alpha
_{1,3})\left( 
\begin{array}{c}
u_{t}^{rea} \\ 
u_{t}^{rpo}%
\end{array}%
\right) =\varepsilon _{t}^{S}  \label{first_equation_Kilian}
\end{equation}%
where $A_{1\bullet }=(\alpha _{1,1},$ $\alpha _{1,2},$ $\alpha _{1,3})$.
Equation (\ref{first_equation_Kilian})\ provides the moment condition $%
A_{1\bullet }\Sigma _{u}A_{1\bullet }^{\prime }=1$, see (\ref%
{moments_A1_variance2}). If, as in Assumption 4, there exist at least $%
s=n-k=2$ proxies $w_{t}$ for the two non-target shocks $\varepsilon
_{2,t}=(\varepsilon _{t}^{AD},\varepsilon _{t}^{OSD})^{\prime }\equiv \tilde{%
\varepsilon}_{2,t}$, there are two additional moment conditions of the form (%
\ref{moments_A1}) that can be exploited for inference, i.e. $A_{1\bullet
}\Sigma _{u,w}=0_{1\times 2}$, where $\Sigma _{u,w}:=${${E}$}$%
(u_{t}w_{t}^{\prime })$. Overall, there are three moment conditions ($m=%
\frac{1}{2}k(k+1)+ks=3$) that can be used to estimate the three structural
parameters in $A_{1\bullet }$ ($a=3$)\ by the method discussed in Section %
\ref{Section_indirect_approach_MD}.

Following the arguments in Kilian (2009) and Montiel Olea {\emph{et al.}}
(2021), our Assumption 1 is considered valid. Assumption 2 is investigated
by a set of diagnostic tests on the VAR residuals (the VAR is estimated with 
$l=24$ lags), which suggest that the residuals are conditionally
heteroskedastic but serially uncorrelated. Assumption 3 is maintained. The
validity of the proxies in the sense of Assumption 4 is discussed below.\ 

The proxies selected for the two non-target shocks are $%
w_{t}:=(w_{t}^{RV},w_{t}^{Br})^{\prime }$, where $w_{t}^{RV}$ represents the
logarithmic difference of the World Steel Index (WSI) introduced by
Ravazzolo and Vespignani (2020), and $w_{t}^{Br}$ represents the logarithmic
difference of the Brent Oil Futures. The proxy $w_{t}^{RV}$ serves as an
instrument for the aggregate demand shock, $\varepsilon _{t}^{AD}$, and the
proxy $w_{t}^{Br}$ is used as an instrument for the oil-specific demand
shock, $\varepsilon _{t}^{OSD}$. Since $w_{t}^{RV}$ is available on the
shorter sample, 1990:M2-2004:M9, we employ the entire sample period
1973:M1-2004:M9 to estimate $\Sigma _{u}$ and the shorter sample period,
1990:M2-2004:M9 ($T=176$ monthly observations), to estimate $\Sigma _{u,w}$.
Then, the MD estimates of the structural parameters in equation (\ref%
{first_equation_Kilian}) follow from (\ref{distance_g})-(\ref{MD_A_model}).

We pre-test the strength of the proxies $w_{t}$ by our bootstrap test. In
this case, to estimate the parameters that capture the strength of the
proxies, $\hat{\theta}_{T}^{\ast }=(\hat{\beta}_{2,T}^{\ast \prime },\hat{%
\lambda}_{T}^{\ast })^{\prime }$, we consider the sample common to both
instruments in $w_{t}$, 1990:M2-2004:M9. We apply the multivariate normality
test $\tau _{T,N}^{\ast }\equiv DH$ to the sequence of bootstrap
replications $\{\hat{\vartheta}_{T:1}^{\ast },$ $\hat{\vartheta}_{T:2}^{\ast
},\ldots ,\hat{\vartheta}_{T:N}^{\ast }\}$, where $N=[T^{1/2}]=13$ and $\hat{%
\vartheta}_{T}^{\ast }\equiv \hat{\theta}_{T}^{\ast }$, with $\hat{\theta}%
_{T}^{\ast }$ $=(\hat{\beta}_{2,T}^{\ast \prime },\hat{\lambda}_{T}^{\ast
})^{\prime }$ being the MBB-CMD estimator discussed in Section \ref%
{Section_indirect_approach_MD}.\footnote{\label{footnote_B2}Since $s=2$, at
least one restriction must be imposed on the parameters of $\tilde{B}%
_{\bullet 2}$ and/or $\Lambda $ to obtain the CMD estimators $\hat{\theta}%
_{T}$ and $\hat{\theta}_{T}^{\ast }$, respectively; see Supplement, proof of
Lemma \ref{Lemma S.4}, equation (\ref{restrictions_theta}). We specify the
matrix $\Lambda $ upper triangular (hence imposing one zero restriction).
This implies that the proxy $w_{t}^{RV}$ is allowed to instrument the
aggregate demand shock $\varepsilon _{t}^{AD}$ alone, while the proxy $%
w_{t}^{Br}$ can instrument both the oil-specific demand shock, $\varepsilon
_{t}^{OSD}$, and the aggregate demand shock, $\varepsilon _{t}^{AD}$. Note
that in the MD estimation problem (\ref{MD_A_model}) we need a consistent
estimator of the matrix $\Sigma _{u,w}$, say $\hat{\Sigma}_{u,w}$:=$\frac{1}{%
T}\sum_{t=1}^{T}\hat{u}_{t}w_{t}^{\prime }$, and can ignore the possible
restrictions that characterize the matrices $\Lambda $ and $\tilde{B}%
_{\bullet 2}$, see footnote 10.} The corresponding p-value is 0.67 which
does not reject the null hypothesis. As robustness check, we repeat the test
using $\hat{\vartheta}_{T}^{\ast }\equiv \hat{\beta}_{2,T}^{\ast \prime }$,
obtaining a p-value equal to 0.73. We conclude that the null hypothesis that
the proxies $w_{t}:=(w_{t}^{RV},w_{t}^{Br})^{\prime }$ are relevant for the
shocks $\tilde{\varepsilon}_{2,t}=(\varepsilon _{t}^{AD},\varepsilon
_{t}^{OSD})^{\prime }$ in the sense of condition (\ref{eq strong proxxx}) is
not rejected by the data. An indirect check of the exogeneity condition is
discussed at the end of this section.

The IRFs estimated by the indirect-MD approach correspond to the red lines
plotted in Figure 2. They are surrounded by the red shaded areas
representing the 68\%-MBB (panel A) and 95\%-MBB\ (panel B) pointwise
confidence intervals, computed by using Hall's percentile method.
Proposition \ref{prop 7} ensures that no Bonferroni-type adjustment is
needed; see Section \ref{Sub_section_screening}.

From Figure 2, we derive two important observations. First, the MBB
confidence intervals obtained by the indirect-MD approach using the strong
proxies $w_{t}$ for the non-target shocks -- estimated on a shorter sample
-- are `more informative' than both the Anderson-Rubin weak-instrument
robust confidence intervals and the grid MBB AR confidence intervals
obtained by instrumenting the oil supply shock directly with Kilian's (2008)
proxy $z_{t}$. Differences become marked when considering $95\%$ confidence
intervals, see panel B. Second, our empirical results line up with Kilian's
(2009) main results. In Kilian's (2009) Choleski-SVAR, both real economic
activity and the real price of oil exhibit limited, temporary, and
statistically insignificant responses to the oil supply shock. This finding
is also evident from our estimated IRFs. Kilian's (2009) recursive SVAR
implies the testable restrictions $A_{1,2}\equiv (\alpha _{1,2}$, $\alpha
_{1,3})=(0,0)$ in the structural equation (\ref{first_equation_Kilian}).
These restrictions imply a vertical short run oil supply curve. Under the
conditions outlined in Proposition \ref{prop 2} and with the support of our
pre-test that does not reject the relevance of the instruments, a standard
Wald-type test conducted on these restrictions produces a bootstrap p-value
of 0.68. This evidence aligns with Kilian's (2009) recursive SVAR.
Importantly, according to Proposition \ref{prop 7}, the outcome of the Wald
test remains unaffected by the failure of the bootstrap pre-test to reject
the null hypothesis. As a result, there is no need for Bonferroni
adjustments.

To assess the exogeneity (orthogonality) of the proxies $w_{t}$ with respect
to the oil supply shock $\varepsilon _{t}^{S}$, we adopt a commonly employed
approach in the empirical proxy-SVAR literature. Examples include, e.g.,
Caldara and Kamps (2017) and Piffer and Podstawki (2018). This involves
approximating the shocks of interest by proxies or shocks derived from other
studies, or identification methods. In our framework, a natural solution is
to calculate the correlations between the proxies $w_{t}$ and Kilian's
(2008) instrument $z_{t}$. We obtain the correlations $\widehat{Corr}%
(w_{t},z_{t})=(0.0047,$ $-0.09)^{\prime }$ on the common sample
1990:M2-2004:M9, which are not statistically significant at any conventional
significance level. An alternative method to assess the exogeneity condition
is as follows. The empirical results discussed in this section support
Kilian's (2009) original triangular SVAR specification on the sample
1990:M2-2004:M9, featuring a vertical short run oil supply curve. Other
studies suggest, using different identification schemes, that a
Choleski-SVAR for $Y_{t}:=(prod_{t},rea_{t},rpo_{t})^{\prime }$ represents a
good approximation of the data also on periods longer than the estimation
sample 1990:M2-2004:M9; see, e.g., Kilian and Murphy (2012). This suggests
that we can interpret the time series $\hat{\varepsilon}_{t}^{S,Chol}$, $%
t=1,\ldots ,T$, recovered from the first equation of Kilian's (2009)
Choleski-SVAR, as a reasonable approximation of an oil supply shock. Also in
this case, the correlations computed on the common period 1990:M2-2004:M9,
equal to $\widehat{Corr}(w_{t},\hat{\varepsilon}_{t}^{S,Chol})=(-0.059,$ $%
0.038)^{\prime }$, are not statistically significant at any conventional
significance level.

\subsection{Financial and macroeconomic uncertainty shocks}

\label{Section_empirical_illustration_uncertainty}In this second empirical
illustration, we emphasize the merit of the indirect-MD approach in
situations where finding valid multiple instruments for multiple target
shocks can be problematic.

Our objective is to track the dynamic causal effects produced by financial
and macroeconomic uncertainty shocks ($k=2$) on a measure of the real
economic activity. As in Ludvigson {\emph{et al.}} (2021), we consider a
small-scale VAR\ model with $n=3$ variables: $%
Y_{t}:=(U_{F,t},U_{M,t},a_{t})^{\prime }$, where $U_{F,t}$ is an index of
(1-month ahead) financial uncertainty, $U_{M,t}$ is the index of (1-month
ahead) macroeconomic uncertainty, and $a_{t}$ is a measure of real economic
activity, proxied by the growth rate of industrial production. The two
uncertainty indexes are analyzed and discussed in Ludvigson \emph{et al.}
(2021), where the authors contend that unraveling the relative impacts of
these two distinct sources of uncertainty is crucial for understanding how
they are transmitted to the business cycle.

We focus on the `Great Recession + Slow Recovery' period 2008:M1-2015:M4 ($%
T=88$ monthly observations). The dataset is the same as in Ludvigson {\emph{%
et al.}} (2021) and Angelini {\emph{et al.}} (2019). The decision to focus
on the period following the Global Financial Crisis is based on the
empirical findings presented in Angelini \emph{et al.} (2019), where it was
discovered that the VAR model for $Y_{t}:=(U_{F,t},U_{M,t},a_{t})^{\prime }$
exhibits two significant breaks in unconditional volatility over the
extended period from 1960 to 2015, resulting in three distinct volatility
regimes.

The reduced form VAR model for $Y_{t}$ includes a constant and $l=4$ lags.
The VAR residuals display neither serial correlation, nor conditionally
heteroskedasticity on the sample 2008:M1-2015:M4.

The target structural shocks are collected in the vector $\varepsilon
_{1,t}:=(\varepsilon _{F,t},\varepsilon _{M,t})^{\prime }$, where $%
\varepsilon _{F,t}$ denotes the financial uncertainty shock and $\varepsilon
_{M,t}$ the macroeconomic uncertainty shock. The non-target shock of the
system is the `non-uncertainty' shock $\varepsilon _{a,t}\equiv \tilde{%
\varepsilon}_{2,t}$ $(n-k=1)$, which can be interpreted as a shock
reflecting forces related to real economic activity. In this model, the
counterpart of (\ref{partition_B}) is as follows:%
\begin{equation}
\underset{u_{t}}{\underbrace{\left( 
\begin{array}{c}
u_{F,t} \\ 
u_{M,t} \\ 
u_{a,t}%
\end{array}%
\right) }}=\underset{B_{\bullet 1}}{\underbrace{\left( 
\begin{array}{cc}
\beta _{F,F} & \beta _{F,M} \\ 
\beta _{M,F} & \beta _{M,M} \\ 
\beta _{a,F} & \beta _{a,B}%
\end{array}%
\right) }}\underset{\varepsilon _{1,t}}{\underbrace{\left( 
\begin{array}{c}
\varepsilon _{F,t} \\ 
\varepsilon _{M,t}%
\end{array}%
\right) }}+\underset{B_{\bullet 2}}{\underbrace{\left( 
\begin{array}{c}
b_{F,a} \\ 
b_{M,a} \\ 
b_{a,a}%
\end{array}%
\right) }}\underset{\varepsilon _{2,t}}{\underbrace{\left( \varepsilon
_{a,t}\right) }}  \label{uncertainty_B}
\end{equation}%
where $u_{t}:=(u_{F,t},u_{M,t},u_{a,t})^{\prime }$ is the vector of VAR
reduced form disturbances. The implementation of the direct identification
approach presents a challenge in identifying two reliable external
instruments for the two uncertainty shocks $\varepsilon _{1,t}:=(\varepsilon
_{F,t},\varepsilon _{M,t})^{\prime }$. Ludvigson {\emph{et al.}} (2021, p.
6) acknowledge that in this application `\textit{Instrumental variable
analysis is challenging, since instruments that are credibly exogenous are
difficult if not impossible to find...}'.\footnote{%
Driven by this idea, Ludvigson {\emph{et al.}} (2021) develop a novel
identification strategy which combines `external variable constraints' with
inequality constraints. In their approach, proxies are not required to be
`strong' as defined in (\ref{eq strong proxxx}), nor do they need to be
uncorrelated with the non-instrumented structural shocks.}

We show that the indirect-MD approach simplifies the process of inferring
the effects of macroeconomic and financial uncertainty shocks on real
economic activity. Indeed, the indirect approach enables us to shift the
issue of identifying (at least) two valid proxies for the two uncertainty
shocks to the task of finding (at least) \emph{one} valid instrument for the
shock in real economic activity. This requires considering the equations%
\begin{equation}
A_{1\bullet }u_{t}\equiv \underset{A_{1,1}}{\underbrace{\left( 
\begin{array}{cc}
\alpha _{F,F} & \alpha _{F,M} \\ 
\alpha _{M,F} & \alpha _{M,M}%
\end{array}%
\right) }}\underset{u_{1,t}}{\underbrace{\left( 
\begin{array}{c}
u_{F,t} \\ 
u_{M,t}%
\end{array}%
\right) }}+\underset{A_{1,2}}{\underbrace{\left( 
\begin{array}{c}
\alpha _{F,a} \\ 
\alpha _{M,a}%
\end{array}%
\right) }}\underset{u_{2,t}}{\underbrace{\left( u_{a,t}\right) }}\text{ }=%
\text{ }\varepsilon _{1,t}\equiv \left( 
\begin{array}{c}
\varepsilon _{F,t} \\ 
\varepsilon _{M,t}%
\end{array}%
\right)  \label{A_form_uncertainty}
\end{equation}%
which represents the counterpart of system (\ref{structural_sub_A_bis}).
Since $k=2$, point-identifica\-tion of the target uncertainty shocks
requires at least $\frac{1}{2}k(k-1)=1$ extra\textbf{\ }restriction on the
elements of the matrix $A_{1\bullet }$. Equation (\ref{A_form_uncertainty})
provides $\frac{1}{2}k(k+1)=3$ moment conditions implied by the expression $%
A_{1\bullet }^{\prime }\Sigma _{u}A_{1\bullet }=I_{2}$. As $n-k=1$, we need
at least one external instrument for the non-target shock; i.e., a variable $%
w_{t}$ ($s=n-k=1$) that satisfies the linear measurement equation%
\begin{equation}
w_{t}=\lambda \varepsilon _{a,t}+\omega _{w,t}
\label{strong_proxies_uncertainty}
\end{equation}%
where $\tilde{\varepsilon}_{2,t}=\varepsilon _{a,t}$, $\lambda $ is the
relevance parameter and $\omega _{w,t}$ is a measurement error term,
uncorrelated with structural shocks. Equation (\ref%
{strong_proxies_uncertainty}) is the counterpart of (\ref{strong_proxies})
in Assumption \ref{Assn 4} and provides two additional moment restrictions, $%
A_{1\bullet }^{\prime }\Sigma _{u,w}=0_{2\times 1}$, where $\Sigma _{u,w}:=${%
${E}$}$(u_{t}w_{t})$. By jointly considering the restrictions $A_{1\bullet
}^{\prime }\Sigma _{u}A_{1\bullet }=I_{2}$ and $A_{1\bullet }^{\prime
}\Sigma _{u,w}=0_{2\times 1}$, we obtain a total of $m=3+2=5$ distinct and
independent moment conditions which can be used to estimate $a=5$ structural
parameters in $A_{1\bullet }$. To impose the necessary identification
constraint on $A_{1\bullet }$, we borrow the restriction$\ \beta _{F,M}=0$
(on $B_{\bullet 1}$) from Angelini {\emph{et al.}} (2019). Using a
methodology based on changes in volatility regimes and considering the
extended period 1960-2015, Angelini {\emph{et al.}} (2019) explore the idea
that instantaneous causality between uncertainty shocks solely runs from
financial to macroeconomic uncertainty. They test the hypothesis that
financial uncertainty does not respond instantaneously to macroeconomic
uncertainty shocks ($\beta _{F,M}=0$) and do not reject this hypothesis for
the sample period 2008:M1-2015:M4. By using the relationship (\ref%
{crucial_relationship_partial}), the restriction $\beta
_{F,M}=e_{3,1}^{\prime }(B_{\bullet 1})e_{2,2}=0$ (recall that, e.g., $%
e_{3,1}$ is the $3\times 1$ vector containing `1' in the position $1$ and
zero elsewhere$)$\ can be mapped to the elements of $A_{1\bullet }^{\prime }$
via $e_{3,1}^{\prime }(\Sigma _{u}A_{1\bullet }^{\prime })e_{2,2}=0$, and
properly expressed in the form (\ref{restrictions_A1}) once $\Sigma _{u}$ is
replaced by its consistent estimator $\hat{\Sigma}_{u}:=T^{-1}\sum_{t=1}^{T}%
\hat{u}_{t}\hat{u}_{t}^{\prime }$. This allows to estimate $a=5$ free
structural parameters in the matrix $A_{1\bullet }$ by or MD\ approach. On
the other hand, the constraint $\beta _{F,M}=0$ can be directly incorporated
in the estimation of the proxy-SVAR by relying on the alternative
indirect-MD estimation method discussed in the Supplement, Section \ref%
{Section_supplementary_restrictions_on_B1}.

To find a valid proxy $w_{t}$ for the real economic activity shock $%
\varepsilon _{a,t}$, we follow Angelini and Fanelli (2019). Let $house_{t}$
be the log of new privately owned housing units started on the estimation
period 2008:M1-2015:M4 (source: Fred). We take the `raw' growth rate of new
privately owned housing units started, $\Delta house_{t}$, and estimate an
auxiliary dynamic linear regression model of the form $\Delta house_{t}=${${E%
}$}$(\Delta house_{t}\mid \mathcal{F}_{t-1})+err_{t}$, where $\mathcal{F}%
_{t-1}$ denotes the information set available to the econometrician at time $%
t-1$, {${E}$}$(\Delta house_{t}\mid \mathcal{F}_{t-1})$ denotes the linear
projection of $\Delta house_{t}$ on the past information set, and $err_{t}$
can be interpreted as the `innovation component' of the dynamic auxiliary
model for the external instrument. The residuals, denoted as $w_{t}:=$ $%
\widehat{err}_{t}$, $t=1,\ldots ,T$, resulting from regressing $\Delta
house_{t}$ on past information, serve as our approximation for the shock in
real economic activity.

We pre-test the strength of the proxy $w_{t}$ by computing our bootstrap
test of instrument relevance. We apply the DH multivariate normality test to
the bootstrap replications $\{\hat{\vartheta}_{T:1}^{\ast },\hat{\vartheta}%
_{T:2}^{\ast },\ldots ,\hat{\vartheta}_{T:N}^{\ast }\}$, where $\hat{%
\vartheta}_{T:b}^{\ast }\equiv \hat{\beta}_{2,T:b}^{\ast }$, $b=1,\ldots ,N$%
, $N=[T^{1/2}]=9$, and $\hat{\theta}_{T}^{\ast }=(\hat{\beta}_{2,T}^{\ast
\prime },\hat{\lambda}_{T}^{\ast })^{\prime }\ $is the MBB-CMD estimator
discussed in Section \ref{Section_indirect_approach_MD}.\ The DH
multivariate normality test yields a p-value of 0.38, indicating no
rejection of the null hypothesis of relevant proxy.

To indirectly assess the exogeneity condition, we examine the correlation
between our proxy variable $w_{t}$ and time series data of macroeconomic and
financial uncertainty shocks, as determined by Angelini \emph{et al. }(2019)
using their approach based on changes in unconditional volatility.
Specifically, we consider their estimated time series $\hat{\varepsilon}%
_{F,t}$ and $\hat{\varepsilon}_{M,t}$, $t=1,\ldots ,T$. The resulting
correlations, computed over the sample period 2008:M1-2015:M4, are $\widehat{%
Corr}(w_{t},(\hat{\varepsilon}_{F,t},\hat{\varepsilon}_{M,t})^{\prime
})=(-0.092,$ $-0.096)^{\prime }$ and are not statistically significant at
any conventional level.

After estimating the model using the indirect-MD approach, we generate IRFs
for a 40-month period. In Figure 3, the red lines (labelled as `indirect-MD
approach') represent the dynamic responses of the growth rate of industrial
production to identified financial (upper panel) and macroeconomic (lower
panel) uncertainty shocks. These responses are based on one-standard
deviation uncertainty shocks and are surrounded by 90\% MBB confidence
intervals (depicted as red shaded areas), calculated using Hall's percentile
method. According to Proposition \ref{prop 7}, the asymptotic coverage of
these confidence intervals remains unaffected by pre-testing bias. To allow
for easy comparison with a benchmark, Figure 3 also incorporates the
responses obtained by Angelini {\emph{et al.}} (2019), shown in blue and
identified as `Angelini, Bacchiocchi, Caggiano, and Fanelli (2019)' (refer
to their Figure 5). These responses are also based on one-standard deviation
uncertainty shocks. The blue shaded region in Figure 3 represents the 90\%
bootstrap confidence intervals computed by Angelini {\emph{et al.}} (2019)
over the period 2008:M1-2015:M4, using the i.i.d. bootstrap method.

Figure 3 unveils two important findings. First, both the indirect-MD
approach and Angelini \emph{et al.}'s (2019) method reveal a significant
effect of macroeconomic and financial uncertainty shocks in restraining
economic activity during the post-Great Recession period. Secondly,
substantial disparities emerge in the estimated impact of the macroeconomic
uncertainty shock on industrial production growth. Using the indirect-MD
approach, the estimated peak response of industrial production growth to the
macroeconomic uncertainty shock is both significant and instantaneous, equal
to -0.32\%. Conversely, the method based on changes in volatility indicates
that the peak response, also statistically significant, occurs five months
post-shock, with a magnitude of -0.15\%. In both the indirect-MD approach
and the volatility-based approach, the peak response of industrial
production growth to the financial uncertainty shock is significant, equal
to -0.17\%. Upon examination of the 90\% bootstrap confidence intervals, it
becomes evident that the dynamic causal effects resulting from macroeconomic
and financial uncertainty shocks are more precisely estimated through the
indirect-MD approach. \ \ \ \ \ \ \ \ \ \ \ \ \ \ \ \ \ \ \ \ \ \ \ \ \ \ \
\ \ \ \ \ \ \ \ \ \ \ \ \ \ \ \ \ \ \ \ \ \ \ \ \ \ \ \ \ \ \ \ \ \ \ \ \ \
\ \ \ \ \ \ \ \ \ \ \ \ \ \ \ \ \ \ \ \ \ \ \ \ \ \ \ \ \ \ \ \ \ \ \ \ \ \
\ \ \ \ \ \ \ \ \ \ \ \ \ \ \ \ \ \ \ \ \ \ \ \ \ \ \ \ \ \ \ \ \ \ \ \ \ \
\ \ \ \ \ \ \ \ \ \ \ \ \ \ \ \ \ \ \ \ \ \ \ \ \ \ \ \ \ \ \ \ \ \ \ \ \ \
\ 

\section{Conclusions}

\label{Sez_Conclusions}We have designed a MD estimation strategy for
proxy-SVARs in which strong proxies for the non-target shocks are used to
identify the target shocks. This approach proves particularly effective when
the instruments available for the target shocks are weak. It becomes
especially advantageous when, faced with multiple target shocks, the
application of weak-instrument robust methods necessitates imposing a large
number of restrictions which might lack economic motivation and/or could
pose challenges in terms of testing their validity. Furthermore, we have
enriched this proposed strategy with a novel, computationally
straightforward diagnostic pre-test for instrument relevance which relies on
bootstrap resampling and does not introduce any pre-testing bias.

It could be argued that in models of the dimensions typically encountered in
practice, obtaining valid proxies for the non-target shocks and establishing
additional credible identifying restrictions that are sufficient to uniquely
point-identify the target structural shocks can be challenging. However, the
empirical illustrations revisited in this paper demonstrate the potential
benefits and effectiveness of the suggested approach in cases of interest.
One question that arises is whether it is appropriate to solely instrument
the non-target shocks without considering any information from available
weak proxies for the target shocks, as this approach may overlook
potentially valuable identifying information. In principle, one may use both
proxies for the non-target shocks and proxies for the target shocks jointly.
Intuitively, in such situations, the strong proxies for the non-target
shocks act as a form of `insurance' against potential identification issues
that could arise if the proxies for the target shocks were weak, allowing
for more reliable inference. Exploring this intriguing issue further will be
the focus of our future research.

\subsection*{Acknowledgements}

We thank Luca Gambetti, Iliyan Georgiev,
Alexander Kriwoluzky, Lutz Kilian, Daniel Lewis, Helmut L\"{u}tkepohl,
Sophocles Mavroeidis, Mikkel Plagborg-M\o ller, Ben Schumann, Lorenzo
Trapani, as well as seminar participants at the SIdE Webinar Series (March
2021), the Granger Centre for Time Series Econometrics at the University of
Nottingham (March 2022), and participants to the Workshop `Advances in
Structural Shocks Identification: Information, Fundamentalness and
Recoverability' (Barcelona GSE Summer Forum,\ June 2021), the IAAE 2021
Annual Meeting (Rotterdam, June 2021), the EEA-ESEM 2021 (Virtual, August
2021), the DIW\ Berlin `Macroeconometric Workshop' (Berlin, May 2022). We
gratefully acknowledge financial support from MIUR (PRIN 2017, Grant
2017TA7TYC) and the University of Bologna (RFO grants).

\section*{References}

\begin{description}
\item Anderson, T.W. and Rubin, H. (1949), Estimation of the parameters of a
single equation in a complete system of stochastic equations, \emph{Annals
of Mathematical Statistics} 20, 46-63.

\item Andrews, D.W. K. and Cheng, X. (2012) Estimation and inference with
weak, semi-strong, and strong identification, \emph{Econometrica} 80,
2153-2211.

\item Andrews, I., Stock, J.H. and Sun, L. (2019), Weak instruments in
instrumental variables regression: Theory and practice, \emph{Annual Review
of Economics} 11, 727-753.

\item Angelini, G. and Fanelli, L. (2019), Exogenous uncertainty and the
identification of Structural Vector Autoregressions with external
instruments, \emph{Journal of Applied Econometrics} 34, 951-971.

\item Angelini, G., Bacchiocchi, E., Caggiano, G., and Fanelli, L. (2019),
Uncertainty across volatility regimes, \emph{Journal of Applied Econometrics 
}34, 437-455.

\item Angelini, G., Cavaliere, G. and Fanelli, L. (2022), Bootstrap
inference and diagnostics in state space models: with applications to
dynamic macro models, \emph{Journal of Applied Econometrics} 37, 3-22.

\item Arias, J.E., Rubio-Ramirez, J.F. and Waggoner, D.F. (2021), Inference
in Bayesian Proxy-SVARs, \emph{Journal of Econometrics} 225, 88-106.

\item Bacchiocchi, E. and Kitagawa, T. (2022), Locally- but not
globally-identified SVARs, \emph{Quaderni - Working Paper DSE} n. 1171, 2022,%
{\small \ available at SSRN: https://ssrn.com/abstract=4124228 or
http://dx.doi.org/10.2139/ssrn.4124228.}

\item Bose, A. (1988), Edgeworth correction by bootstrap in Autoregressions, 
\emph{Annals of Statistics} 16, 1709-1722.

\item Boubacar Mainnasara, Y. and Francq, C. (2011), Estimating structural
VARMA models with uncorrelated but non-independent error terms, \emph{%
Journal of Multivariate Analysis} 102, 496-505.

\item Br\"{u}ggemann, R., Jentsch, C. and Trenkler, C. (2016), Inference in
VARs with conditional volatility of unknown form, \emph{Journal of
Econometrics} 191, 69-85.

\item Caldara, D. and Herbst, E. (2019), Monetary policy, real activity, and
credit spreads: Evidence from Bayesian Proxy SVARs, \emph{American Economic
Journal: Macroeconomics} 11, 157-92.

\item Caldara, D. and Kamps, C. (2017), The analytics of SVARs: A unified
framework to measure fiscal multipliers, \emph{Review of Economic Studies}
84, 1015-1040.

\item Cavaliere, G.\ and Georgiev, I. (2020), Inference under random limit
bootstrap measures, \emph{Econometrica} 88, 2547-2974.

\item Francq, C. and Ra\"{\i}ssi, H. (2006), Multivariate portmanteau test
for autoregressive models with uncorrelated but nonindependent errors, \emph{%
Journal of Time Series Analysis} 28, 454-470.

\item Giacomini, R., Kitagawa, T. and Read, M. (2022), Robust Bayesian
inference in Proxy SVARs, \emph{Journal of Econometrics} 228, 107-126.

\item Hahn, J. and Hausman, J. (2002), A new specification test for the
validity of instrumental variables, \emph{Econometrica} 70, 163-189.

\item Hall, P. (1992), \emph{The bootstrap and Edgeworth expansion},
Springer-Verlag, Berlin.

\item Hausman, J., Stock, J.H. and Yogo, M. (2005), Asymptotic properties of
the Hahn-Hausman test for weak instruments, \emph{Economic Letters} 89,
333-342.

\item Horowitz, J.L. (2001), The bootstrap, \emph{Handbook of Econometrics}
Vol. 5, Heckman, J.J. and E. Leamer (eds.), Ch. 52.

\item Jentsch, C. and Lunsford, K.C. (2019), The dynamic effects of personal
and corporate income tax changes in the United States: Comment, \emph{%
American Economic Review} 109, 2655-2678.

\item Jentsch, C. and Lunsford, K.C. (2022), Asymptotic valid bootstrap
inference for Proxy SVARs, \emph{Journal of Business and Economic Statistics}
40, 1876--1891.\emph{\ }

\item Kilian, L. (1998), Small-sample confidence intervals for impulse
response functions, \emph{Review of Economics and Statistics} 80, 218-230.

\item Kilian, L. (2008), Exogenous oil supply shocks: How big are they and
how much do they matter for the U.S. economy? \emph{Review of Economics and
Statistics} 90, 216-240.

\item Kilian, L. (2009), Not all oil shocks are alike: Disentangling demand
and supply shocks in the crude oil market, \emph{American Economic Review}
99, 1053-1069.

\item Kilian, L. and Murphy, D. P. (2012), Why agnostic sign restrictions
are not enough: Understanding the dynamics of oil market VAR\ models, \emph{%
Journal of the European Economic Association} 10, 1166-1188.

\item Leeper, E.M, Sims, C.A. and Zha, T. (1996), What does monetary policy
do? \emph{Brooking Papers of Economic Activity} 27, 1-78.

\item Lilliefors, H. (1967), On the Kolmogorov-Smirnov test for normality
with mean and variance unknown, \emph{Journal of the American Statistical
Association} 62, 399-402.

\item Ludvigson, S.C., Ma, S. and Ng, S. (2021), Uncertainty and business
cycles: exogenous impulse or endogenous response? \emph{American Economic
Journal:\ Macroeconomics} 13, 369-410.

\item Lunsford, K. G. (2016), Identifying Structural VARs with a proxy
variable and a test for a weak proxy, \emph{Federal Reserve Bank of
Cleveland Working Paper} no. 15-28 (version dated October 2016).

\item Mertens, K. and Ravn, M. (2013), The dynamic effects of personal and
corporate income tax changes in the United States, \emph{American Economic
Review} 103, 1212-1247.

\item Mertens, K. and Ravn, M. (2019), The dynamic effects of personal and
corporate income tax \ changes in the United States: Reply to Jentsch and
Lunsford, \emph{American Economic Review} 109, 2679-2691.

\item Montiel Olea, J.L. and Pflueger, C. (2013), A robust test for weak
instruments, \emph{Journal of Business and Economic Statistics} 31, 358-369.

\item Montiel Olea, J.L., Stock, J.H. and Watson, M.W. (2021), Inference in
SVARs identified with an external instrument, \emph{Journal of Econometrics}
225, 74-87.

\item Newey, W.K. and McFadden, D. (1994), Large sample estimation and
hypothesis testing, \emph{Handbook of Econometrics}, R.F. Engle and D.L.
McFadden (eds.), Vol. IV, Chap. 36.

\item Piffer, M. and Podstawki, M. (2018), Identifying uncertainty shocks
using the price of gold, \emph{The Economic Journal}, 128, 3266-3284.

\item Ravazzolo, F., Vespignani, J. (2020), World steel production: A new
monthly indicator of global real economic activity, \emph{Canadian Journal
of Economics} 53, 743-766.

\item Staiger, D. and Stock, J. H. (1997), Instrumental variables
regressions with weak instruments, \emph{Econometrica} 65, 557-586.

\item Stock, J.H. (2008), What's New in Econometrics-Time Series, \emph{%
Lecture 7. Structural VARs}, Cambridge, MA., NBER.

\item Stock, J.H. and Watson, M.W. (2012), Disentangling the channels of the
2007--2009 recession, in \emph{Brookings Panel of Economic Activity}, Spring
2012, 81-135.

\item Stock J.H. and Yogo, M. (2005), Testing for weak instruments in linear
IV regression. In \emph{Identification and Inference for Econometric Models:
Essays in Honor of Thomas Rothenberg}, ed. D.W.K. Andrews, J.H. Stock,
80--108. Cambridge, UK: Cambridge University Press.

\item Stock J.H. and Wright, J. (2000), GMM with weak identification, \emph{%
Econometrica} 68, 1055--96.

\item Zivot, E., Startz, R. and Nelson, C.R. (1998), Valid confidence
regions and inference in the presence of weak instruments, \emph{%
International Economic Review} 39, 1119-1146.
\end{description}

\clearpage

\newpage

\begin{table}[h!]
\centering
\begin{tabular}{ccccc}
\multicolumn{5}{c}{\textbf{Rejection frequencies}} \\ \hline\hline
\multicolumn{5}{c}{\textbf{Strong proxy}} \\ 
& \multicolumn{2}{c}{$T=250$} & \multicolumn{2}{c}{$T=1000$} \\ 
& \multicolumn{2}{c}{$corr=0.59$} & \multicolumn{2}{c}{$corr=0.59$} \\ 
${\theta}$ & ${DH}$ & ${KS}$ & ${DH}$ & ${KS}$ \\ 
$\beta_{2,1}$ & \multirow{3}{*}{0.05(0.06)} & 0.05(0.06) & %
\multirow{3}{*}{0.05(0.05)} & 0.05(0.05) \\ 
$\beta_{2,2}$ &  & 0.05(0.06) &  & 0.05(0.05) \\ 
$\beta_{2,3}$ &  & 0.05(0.05) &  & 0.05(0.05) \\ 
$\lambda$ &  & 0.05(0.05) &  & 0.05(0.05) \\ \hline
\multicolumn{5}{c}{\textbf{Moderately weak proxy}} \\ 
& \multicolumn{2}{c}{$T=250$} & \multicolumn{2}{c}{$T=1000$} \\ 
& \multicolumn{2}{c}{$corr=0.25$} & \multicolumn{2}{c}{$corr=0.13$} \\ 
${\theta}$ & ${DH}$ & ${KS}$ & ${DH}$ & ${KS}$ \\ 
$\beta_{2,1}$ & \multirow{3}{*}{0.22(0.22)} & 0.21(0.23) & %
\multirow{3}{*}{0.80(0.63)} & 0.36(0.35) \\ 
$\beta_{2,2}$ &  & 0.27(0.29) &  & 0.38(0.39) \\ 
$\beta_{2,3}$ &  & 0.20(0.24) &  & 0.30(0.33) \\ 
$\lambda$ &  & 0.09(0.09) &  & 0.10(0.12) \\ \hline
\multicolumn{5}{c}{\textbf{Weak proxy}} \\ 
& \multicolumn{2}{c}{$T=250$} & \multicolumn{2}{c}{$T=1000$} \\ 
& \multicolumn{2}{c}{$corr=0.05$} & \multicolumn{2}{c}{$corr=0.02$} \\ 
${\theta}$ & ${DH}$ & ${KS}$ & ${DH}$ & ${KS}$ \\ 
$\beta_{2,1}$ & \multirow{3}{*}{0.72(0.75)} & 0.80(0.78) & %
\multirow{3}{*}{0.98(0.98)} & 0.93(0.93) \\ 
$\beta_{2,2}$ &  & 0.85(0.83) &  & 0.95(0.95) \\ 
$\beta_{2,3}$ &  & 0.82(0.83) &  & 0.95(0.95) \\ 
$\lambda$ &  & 0.24(0.26) &  & 0.50(0.51) \\ \hline\hline
\end{tabular}%
\caption{Empirical rejection frequencies of the bootstrap pre-test of
instrument relevance. \newline
\newline
{\protect\small {Notes: Results are based on $20,000$ simulations and tuning
parameter $N:=[T^{1/2}]$. $corr=corr(w_t, \protect\varepsilon_{2,t} )$ is
the correlation between the instrument $w_t$ and the non-target structural
shock $\protect\varepsilon_{2,t}$. KS is Lilliefors' (1967) version of
Kolgomorov-Smirnov univariate normality test; $DH$ is Doornik and Hansen's
(2008) multivariate normality test. Results (not) in parenthesis refer to
(iid) GARCH-type VAR disturbances and proxies. The block size in the MBB
algorithm is $l=5.03T^{1/4}$, see footnote 18. All tests are computed at the
5\% nominal significance level.}}}
\label{teble:MonteCarlo_RejFreq}
\end{table}

\newpage \clearpage 

\begin{figure}[h]
\centering
\begin{tabular}{c}
\hline\hline
\\ 
\includegraphics[width=12cm,keepaspectratio]{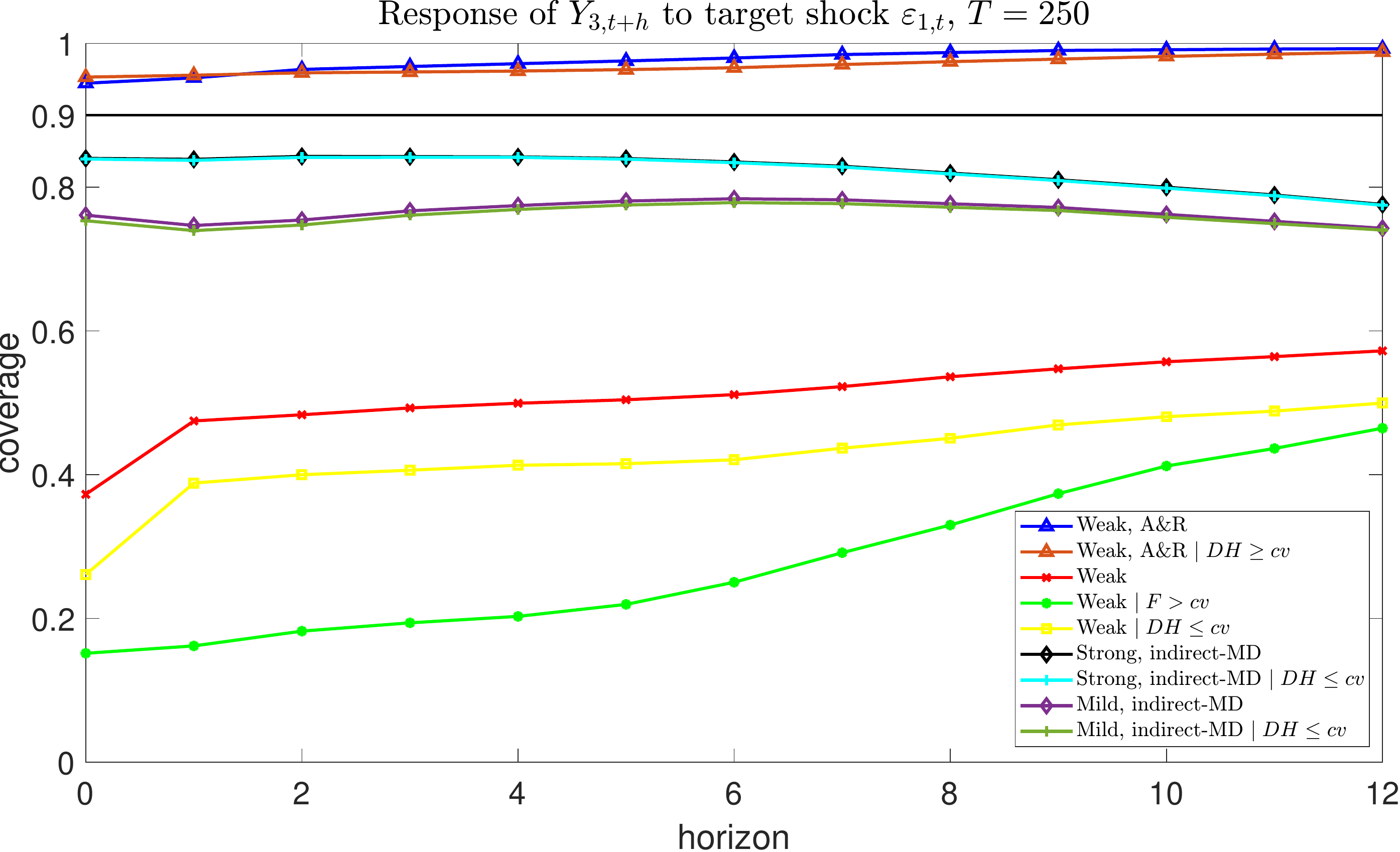} \\ 
\\ \hline\hline
\end{tabular}%
\caption{Empirical coverage probabilities of IRFs calculated on $20,000$
simulations (90\% nominal). IRFs refer to the response of the variable $%
Y_{3,t+h}$ to the target shock $\protect\varepsilon_{1,t}$, $h=0,1,...,12$.}
\label{figure:MonteCarlo_Coverages}
\end{figure}

\newpage \clearpage 

\begin{landscape}

\begin{figure}[h]
\centering
\begin{tabular}{c c c}
\hline
\hline
\\
\textbf{A. 68\% confidence intervals} && \textbf{B. 95\% confidence intervals} \\
\\
\includegraphics[width=9cm,keepaspectratio]{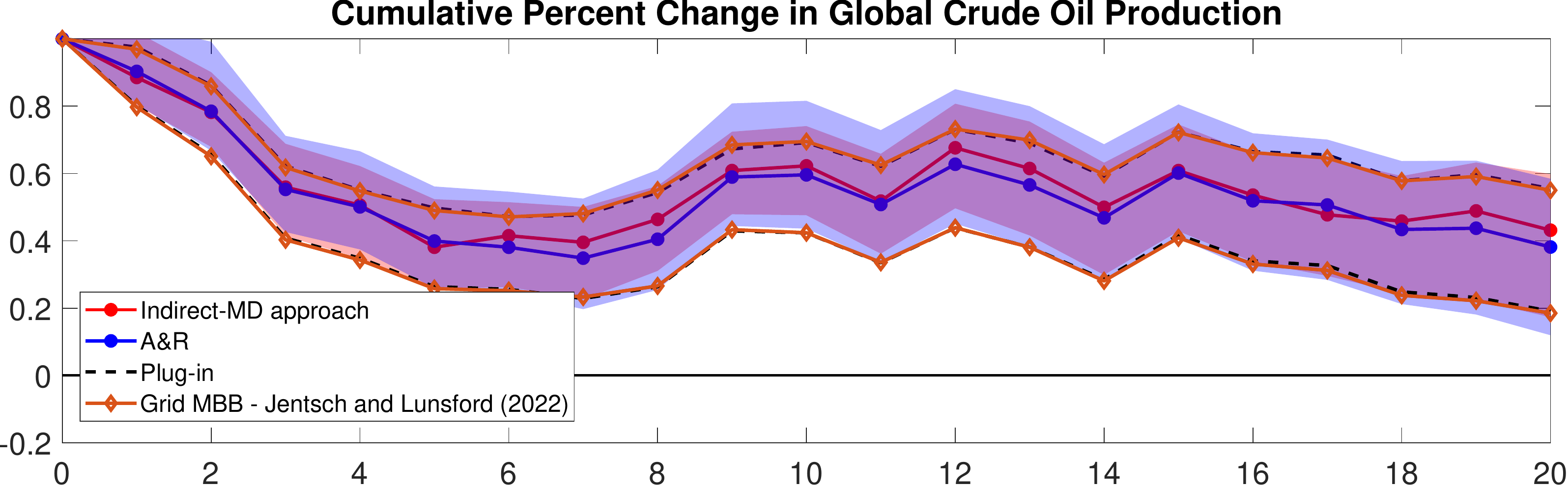} && \includegraphics[width=9cm,keepaspectratio]{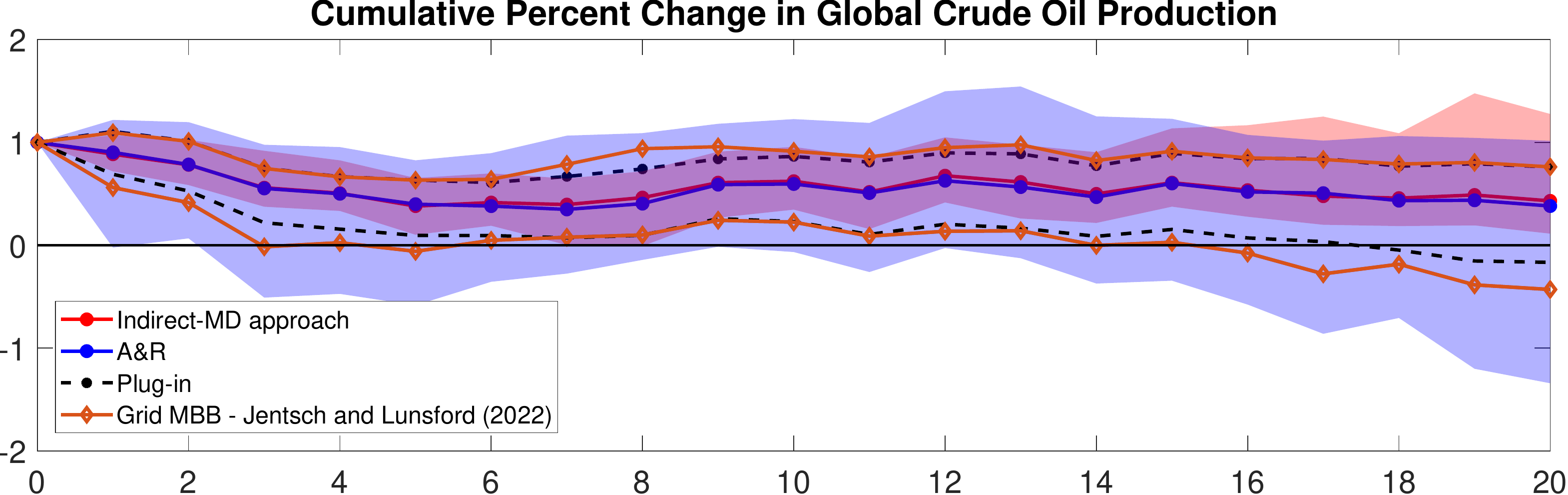}     \\
\\
\includegraphics[width=9cm,keepaspectratio]{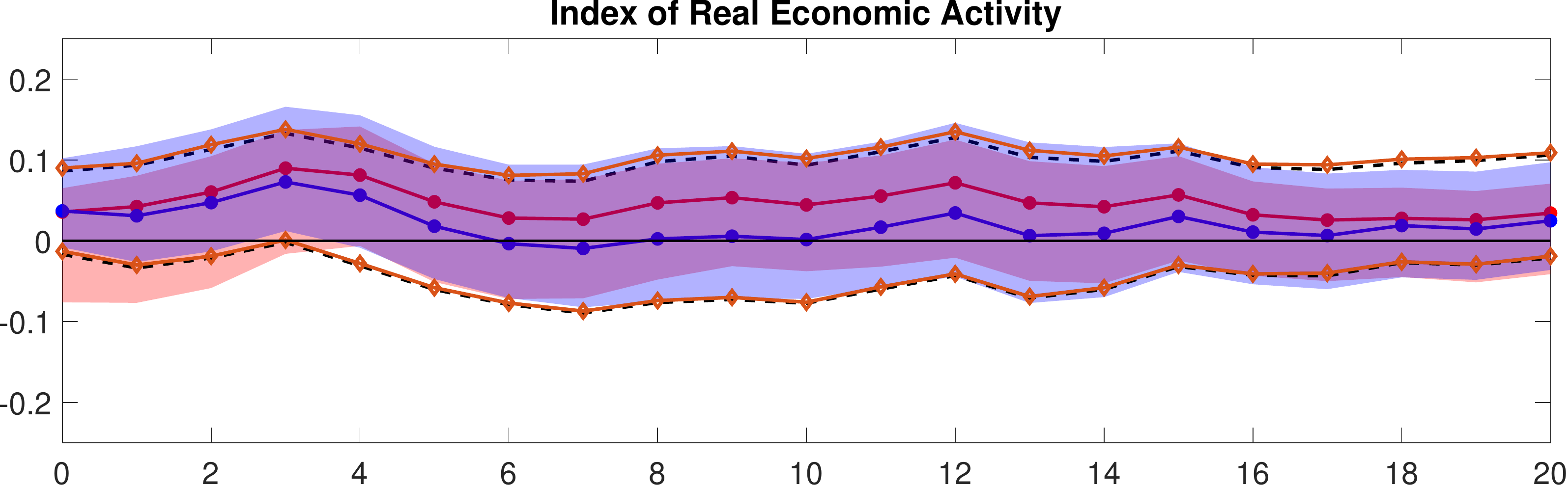} && \includegraphics[width=9cm,keepaspectratio]{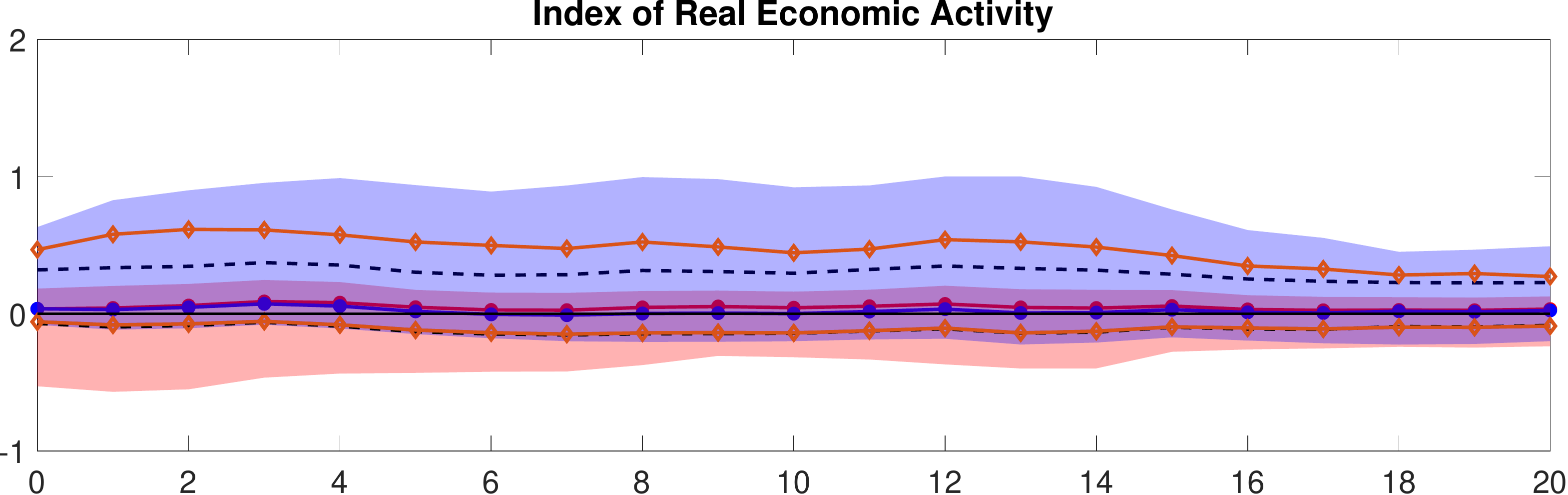}     \\
\\
\includegraphics[width=9cm,keepaspectratio]{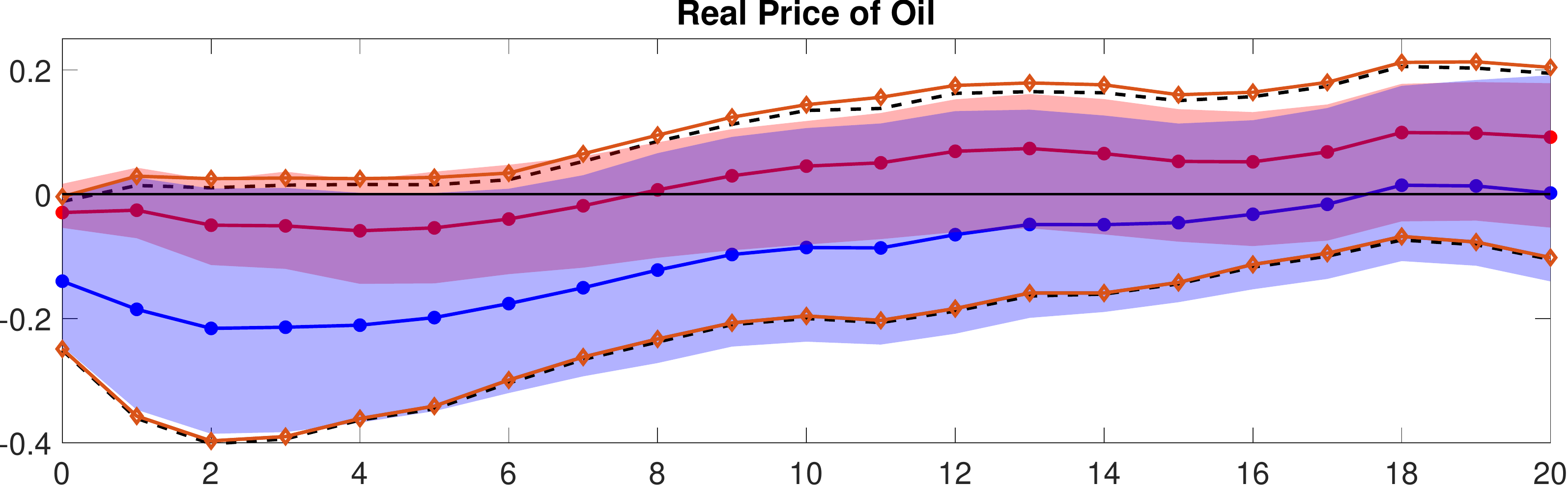} && \includegraphics[width=9cm,keepaspectratio]{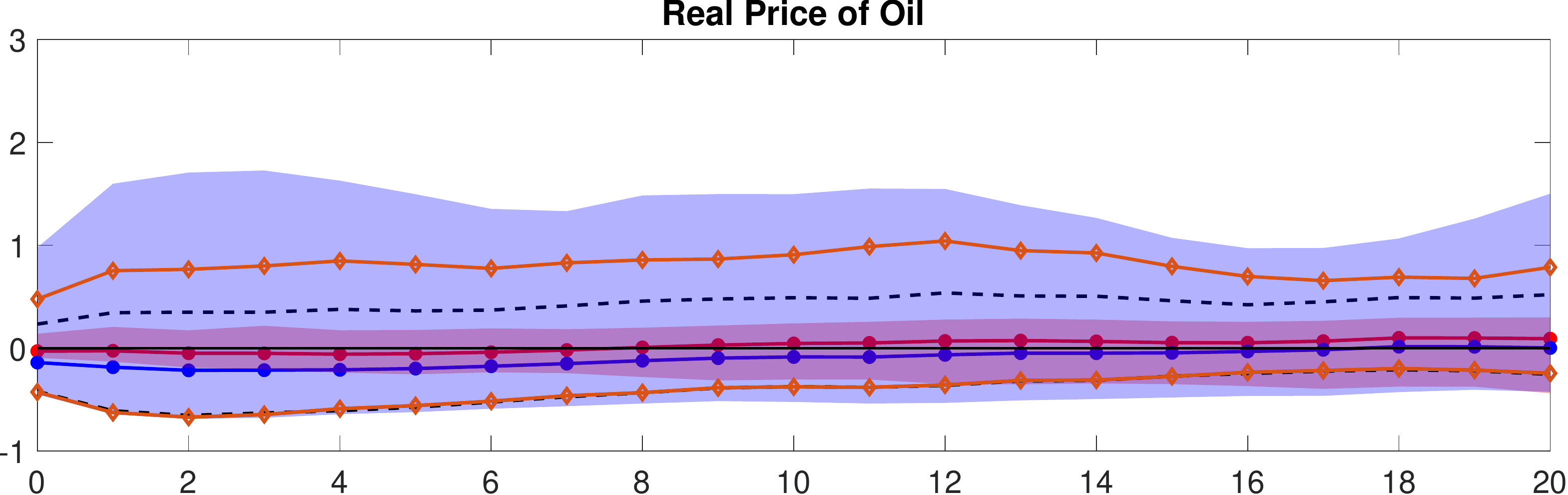}     \\
\\
\hline
\hline
\end{tabular}
\caption{Impulse responses to an oil-supply shock. Red dotted lines correspond to the IRFs estimated with our indirect-MD approach; red shaded areas are the corresponding 68\% and 95\% MBB confidence intervals; blue dotted lines correspond to the Plug-in IRFs obtained pretending that Kilian's (2008) proxy is a strong instrument for the oil supply shock; black dashed lines are the 68\% and 95\% Plug-in confidence intervals; blue shaded areas are the corresponding 68\% and 95\% weak instruments robust confidence intervals; orange dotted lines correspond to the 68\% and 95\% ``Grid MBB'' weak instruments robust confidence intervals.}
\label{figure:IRF_OilShock}
\end{figure}

\end{landscape}

\newpage \clearpage 

\begin{figure}[h]
\centering
\begin{tabular}{c}
\hline\hline
\\ 
\includegraphics[width=12cm,keepaspectratio]{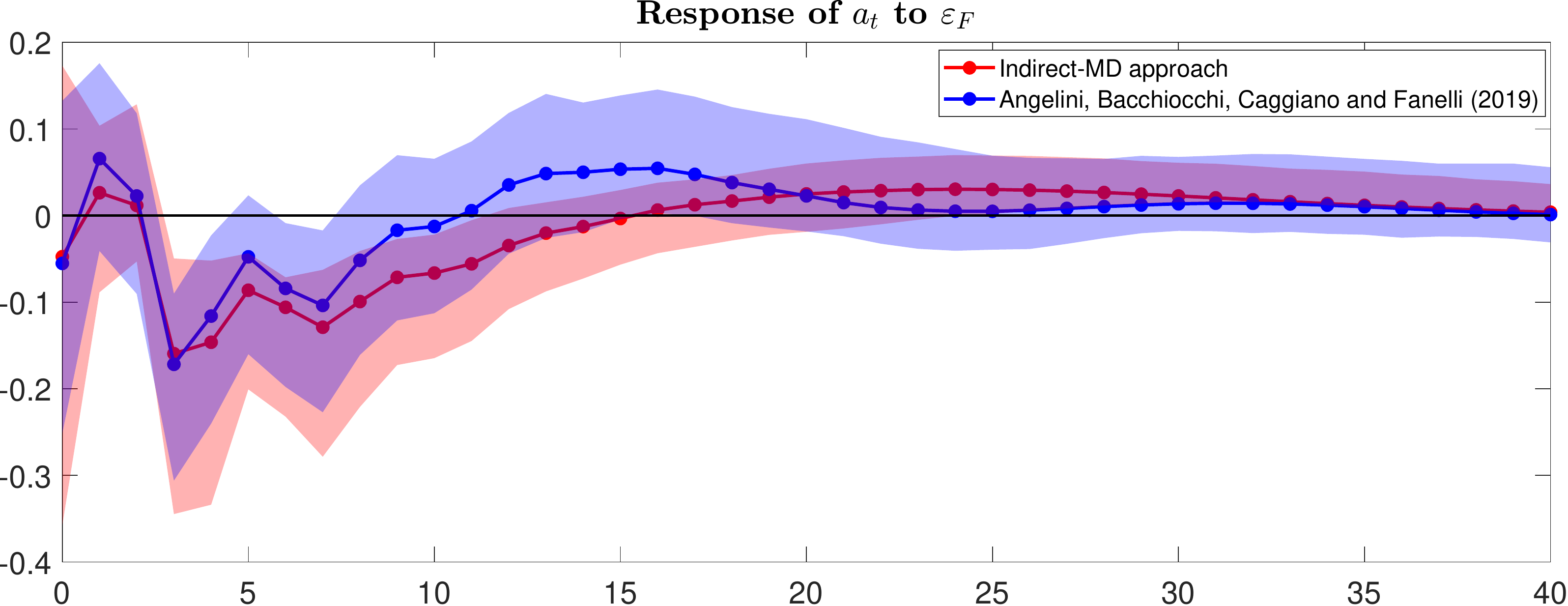} \\ 
\\ 
\includegraphics[width=12cm,keepaspectratio]{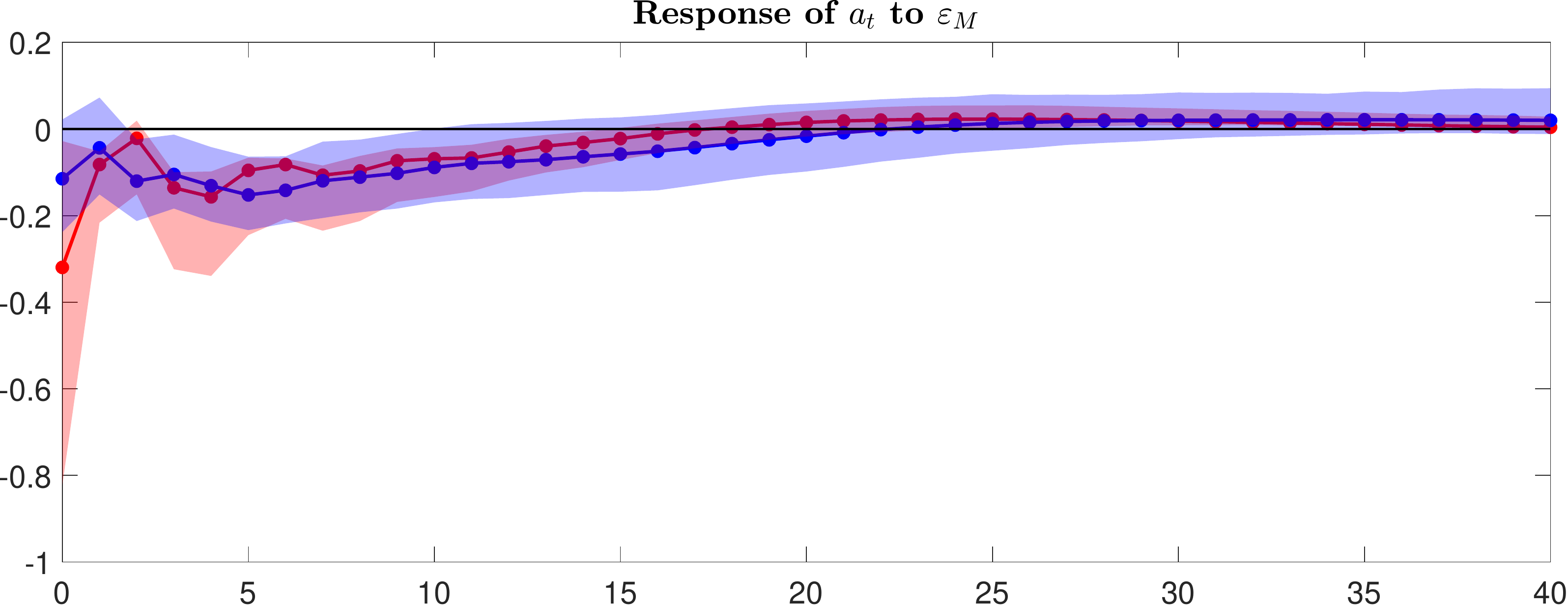} \\ 
\hline\hline
\end{tabular}
\caption{Impulse responses of industrial production growth ($a_t$) to a one
standard deviation financial ($\protect\varepsilon_F$) and a macro ($\protect%
\varepsilon_M$) uncertainty shocks. Red dotted lines correspond to the IRFs
estimated with our indirect-MD approach; red shaded areas are the
corresponding 90\% MBB confidence intervals; blue dotted lines correspond to
the IRFs obtained by Angelini et at. (2019); blue shaded areas correspond to
their 90\% (iid, bootstrap) confidence intervals.}
\label{figure:IRF_UncertaintyShock}
\end{figure}

\clearpage
\newpage

\begin{center}
\textsc{SUPPLEMENT\ TO: }

\medskip \textsc{`AN IDENTIFICATION AND TESTING STRATEGY FOR PROXY-SVARS
WITH WEAK PROXIES'}

\renewcommand{\thefootnote}{} \addtocounter{footnote}{-1} %
\renewcommand{\thefootnote}{\arabic{footnote}}

\medskip

{\large \textsc{By Giovanni Angelini, Giuseppe Cavaliere, Luca Fanelli}}%
{\normalsize \vspace{0.45cm}}

First draft: September{\small \ }2021{\small .\ }First revision: September
2022{\small .\ } \\[0pt]
Second revision: July 2023{\small .\ } This version: October 2023
\end{center}

\appendix
\setcounter{page}{1} 
\setcounter{section}{0} 
\setcounter{page}{1}
\setcounter{equation}{0} 
\setcounter{theorem}{0} 
\setcounter{proposition}{0}
\setcounter{lemma}{0} 
\setcounter{figure}{0} 
\setcounter{footnote}{0}
\renewcommand{\thesection}{S.\arabic{section}}
\renewcommand{\thetheorem}{S.\arabic{theorem}}
\renewcommand{\theproposition}{S.\arabic{proposition}}
\renewcommand{\theequation}{S.\arabic{equation}}
\renewcommand{\thelemma}{S.\arabic{lemma}}
\renewcommand{\thefigure}{S.\arabic{figure}}%

\section{Introduction}

This supplement complements the results of the paper along several
dimensions. Section \ref{Section_bs_notation} summarizes the notation used
for the bootstrap as well as some additional matrix notation. Section \ref%
{Section_supplementary_Lemmas} presents the auxiliary lemmas used to prove
the main propositions in the paper, and Section \ref%
{Section_Supplementary_Proofs} contains the proofs of lemmas and
propositions.

Section \ref{Section_supplementary_restrictions_on_B1} revisits the
indirect-MD approach discussed in Section \ref{Section_indirect_approach_MD}%
, using a different parameterization of the proxy-SVAR. Section \ref%
{Section_supplementary_IV_comparison} compares the MD estimation method with
the IV approach. Section \ref{Section_Supplementary_MBB_algorithm} sketches
the MBB algorithm mentioned in the paper and used to build our test of
instrument relevance. Section \ref{Section_Supplement_DGP} provides details
on the DGPs used in the Monte Carlo experiments in Section \ref%
{Section_MC_results} of the paper. Finally, Section \ref%
{Section_supplementary_fiscal_proxy-SVAR} provides an additional empirical
illustration, where a fiscal proxy-SVAR is used to infer US\ fiscal
multipliers on quarterly data.

Unless differently specified, all references -- except those starting with
`S.' -- pertain to sections, assumptions, equations, and results in the main
paper.

\section{Notation}

\label{Section_bs_notation}

\noindent \textsc{Bootstrap. }We use $P$ to denote the probability measure
for the data, and ${E}(\cdot )$ and ${Var}(\cdot )$ to denote expectations
and variance computed under $P$, respectively. We use $P^{\ast }$ to denote
the probability measure induced by the bootstrap; i.e., conditional on the
original sample. Expectation and variance computed under $P^{\ast }$ are
denoted by ${E}^{\ast }(\cdot )$ and ${Var}^{\ast }(\cdot )$, respectively.

Let, for any $\varsigma >0$, $p_{T}^{\ast }(\varsigma ):=P^{\ast }(||\hat{%
\theta}_{T}^{\ast }-\hat{\theta}_{T}||>\varsigma )$, where $\hat{\theta}%
_{T}^{\ast }$ is the bootstrap analog of the estimator $\hat{\theta}_{T}$,
and let $\left\Vert \cdot \right\Vert $ denote the Euclidean norm. With the
notation `$\hat{\theta}_{T}^{\ast }-\hat{\theta}_{T}\overset{p^{\ast }}{%
\rightarrow }_{p}0$', which reads `$\hat{\theta}_{T}^{\ast }-\hat{\theta}%
_{T} $ converges in $P^{\ast }$ to $0$, in probability', we mean that the
(stochastic)\ sequence $\left\{ p_{T}^{\ast }(\varsigma )\right\} $
converges in probability to zero ($p_{T}^{\ast }(\varsigma )\overset{p}{%
\rightarrow }0$).

Consider a scalar a random variable $X$, with associated cdf $\digamma
_{X}(x):=P(X\leq x)$, and a bootstrap sequence $\{X_{T}^{\ast }\}$, where $%
X_{T}^{\ast }$ has associated cdf (conditional on the data) $\digamma
_{X_{T}^{\ast }}^{\ast }(x):=P^{\ast }(X_{T}^{\ast }\leq x)$. We say that $%
X_{T}^{\ast }$ `converges in conditional distribution to $X$, in
probability', denoted `$X_{T}^{\ast }\overset{d^{\ast }}{\rightarrow }_{p}X$%
', if $\digamma _{X_{T}^{\ast }}^{\ast }(x)\overset{p}{\rightarrow }\digamma
_{T}(x)$ for each $x$ at which $\digamma _{X}(x)$ is continuos. Notice that
if $\digamma _{X}(\cdot )$ is continuous, then the latter convergence also
implies that $\sup_{x\in \mathbb{R}}|\digamma _{X_{T}^{\ast }}^{\ast
}(x)-\digamma _{X}(x)|\overset{p}{\rightarrow }0$ by P\'{o}lya's theorem.
These definitions can be extended to the multivariate framework in the
conventional way.

\smallskip

\noindent \textsc{Matrices.} In the results and proofs that follow we refer
the following matrices (Magnus and Neudecker, 1999): $D_{n}$ is the $n$%
-dimensional duplication matrix ($D_{n}{vech}(M)={vec}(M)$, $M$ being an $%
n\times n$ matrix) and $D_{n}^{+}:=(D_{n}^{\prime }D_{n})^{-1}D_{n}$ is the
Moore-Penrose generalized inverse of $D_{n}$; $K_{ns}$ is the $ns$%
-dimensional commutation matrix ($K_{ns}{vec}(M)={vec}(M^{\prime })$, $M$
being $n\times s$).

\section{\protect\smallskip Auxiliary lemmas}

\label{Section_supplementary_Lemmas}This section summarizes the lemmas
useful for the propositions considered in the paper. We initially represent
the proxy-SVAR in a form that facilitates the derivation of the reduced form
parameter estimator.

\smallskip

\noindent \textsc{Estimator of the reduced form parameters. }By coupling the
VAR for $Y_{t}$ in equation (\ref{VAR-RF2}) with the proxies available for
the non-target shocks $w_{t}$ in equation (\ref{strong_proxies})\ (see
Assumption \ref{Assn 4}), the proxy-SVAR can be represented as the `large',
parametrically constrained, VAR model:%
\begin{equation}
\left( 
\begin{array}{cc}
I_{n}-\Pi (L) & 0 \\ 
0 & I_{s}%
\end{array}%
\right) \underset{W_{t}}{\underbrace{\left( 
\begin{array}{c}
Y_{t} \\ 
w_{t}%
\end{array}%
\right) }}=\underset{\eta _{t}}{\underbrace{\left( 
\begin{array}{c}
u_{t} \\ 
w_{t}%
\end{array}%
\right) }}\text{, \ }\Sigma _{\eta }:=\left( 
\begin{array}{cc}
\Sigma _{u} & \Sigma _{u,w} \\ 
\Sigma _{w,u} & \Sigma _{w}%
\end{array}%
\right)  \label{large_SVAR_general}
\end{equation}%
where $\Pi (L):=\Pi _{1}L+$ $\ldots $ $+\Pi _{l}L^{l}$. In (\ref%
{large_SVAR_general}), the proxies in $w_{t}$ are expressed in innovation
form; i.e., they are serially uncorrelated. In applications, however, it may
happen that the `raw' observed proxy $w_{t}$ is serially autocorrelated and
generated by a dynamic model of the form: $w_{t}={E}_{t-1}w_{t}+\rho _{w,t}$%
, where ${E}_{t-1}w_{t}$ may depend on variables in the information set a
time $t-1$, and $\rho _{w,t}$ is the associated `unsystematic component'
innovation; in this case, $\rho _{w,t}$ is assumed to satisfy the same $%
\alpha $-mixing conditions postulated for the VAR\ innovations $u_{t}$ in
Assumption 2. System (\ref{large_SVAR_general}) can be\ generalized to the
representation%
\begin{equation}
\left( 
\begin{array}{cc}
I_{n}-\Pi (L) & 0 \\ 
\Xi _{w,y}(L) & I_{s}-\Xi _{w,w}(L)%
\end{array}%
\right) \underset{W_{t}}{\underbrace{\left( 
\begin{array}{c}
Y_{t} \\ 
w_{t}%
\end{array}%
\right) }}=\underset{\eta _{t}}{\underbrace{\left( 
\begin{array}{c}
u_{t} \\ 
\rho _{w,t}%
\end{array}%
\right) }}\text{, \ }\Sigma _{\eta }:=\left( 
\begin{array}{cc}
\Sigma _{u} & \Sigma _{u,w} \\ 
\Sigma _{w,u} & \Sigma _{w}%
\end{array}%
\right)  \label{large_SVAR_general_further}
\end{equation}%
where $\Xi _{w,y}(L)$ and $\Xi _{w,w}(L)$ are matrix polynomials in the lag
operator assumed, without loss of generality, of order not larger than $l$
and such that the roots of the characteristic equation $\det (I_{s}-\Xi
_{w,w}(x))=0$ satisfy the condition $\left\vert x\right\vert >1$. Under
Assumption 1, the stability condition on $I_{s}-\Xi _{w,w}(L)$ ensures that
system (\ref{large_SVAR_general}) remains asymptotically stable. Regardless
of whether we consider system (\ref{large_SVAR_general}) or (\ref%
{large_SVAR_general_further}), the proxy-SVAR innovations $\eta
_{t}:=(u_{t}^{\prime },w_{t}^{\prime })^{\prime }$ or $\eta
_{t}:=(u_{t}^{\prime },\rho _{w,t}^{\prime })^{\prime }$ satisfy the $\alpha 
$-mixing properties in Assumption 2.

Given $W_{t}:=(Y_{t}^{\prime },w_{t}^{\prime })^{\prime }$ of dimension $%
(n+s)\times 1$, we compact the proxy-SVAR (either system (\ref%
{large_SVAR_general}) or (\ref{large_SVAR_general_further}))\ as%
\begin{equation}
W_{t}=\Psi _{1}W_{t-1}+\Psi _{2}W_{t-2}+\ldots +\Psi _{l}W_{t-l}+\eta _{t}%
\text{ }  \label{large_B_SVAR}
\end{equation}%
where each $\Psi _{i}$, $i=1,\ldots ,l$, has a triangular structure.
Henceforth, we denote with $\delta _{\psi }$ the vector that collects the
non-zero autoregressive parameters in the matrices $\Psi _{i}$, $i=1,\ldots
,l$, and with $\delta _{\eta }$ the vector that collects the non-repeated
elements in the covariance matrix $\Sigma _{\eta }$. Jointly, the reduced
form parameters of the proxy-SVAR are in the vector $\delta :=(\delta _{\psi
}^{\prime },\delta _{\eta }^{\prime })^{\prime }$ of dimension $q\times 1,$
with $q=q_{\psi }+q_{\eta }$; $q_{\psi }$ is the dimension of $\delta _{\psi
}$ and $q_{\eta }$ the dimension of $\delta _{\eta }$. $\delta _{0}:=(\delta
_{\psi ,0}^{\prime },\delta _{\eta ,0}^{\prime })^{\prime }$ is the true
value of $\delta $ and $\hat{\delta}_{T}:=(\hat{\delta}_{\psi ,T}^{\prime },%
\hat{\delta}_{\eta ,T}^{\prime })^{\prime }$ the quasi-maximum likelihood
[QML] estimator.\footnote{%
The QML estimator of $\delta $ is computed by maximizing the Gaussian
quasi-likelihood function associated with model (\ref{large_SVAR_general})
along the lines described, e.g., in Section 3 in Boubacar Mainassara and
Francq (2011). Observe, indeed, that the reduced form model in (\ref%
{large_B_SVAR}) reads as a special case of Boubacar Mainnasara and Francq's
(2011) structural VARMA models.} Further, we consider a MBB analog of the
QML estimator of $\delta :=(\delta _{\psi }^{\prime },\delta _{\eta
}^{\prime })^{\prime }$, denoted $\hat{\delta}_{T}^{\ast }:=(\hat{\delta}%
_{\psi ,T}^{\ast \prime },\hat{\delta}_{\eta ,T}^{\ast \prime })^{\prime }$.
A sequence of $N$ bootstrap replications of this estimator, $\{\hat{\delta}%
_{T:1}^{\ast },\ldots \hat{\delta}_{T:N}^{\ast }\}$, can be obtained with
the MBB algorithm sketched in Section \ref%
{Section_Supplementary_MBB_algorithm}.

\smallskip

\textsc{Lemmas.} Lemma \ref{Lemma S.1} deals with the asymptotic properties
of the non-bootstrap and bootstrap estimators of the parameters $\delta
:=(\delta _{\psi }^{\prime },\delta _{\eta }^{\prime })^{\prime }$. Below, $%
\ell $ denotes the parameter that governs the block length in MBB
resampling, see Jentsch and Lunsford (2019, 2022) and Section \ref%
{Section_Supplementary_MBB_algorithm}.

\begin{lemma}
\label{Lemma S.1}Consider the proxy-SVAR model (\ref{large_B_SVAR}). Let $%
\hat{\delta}_{T}:=(\hat{\delta}_{\psi ,T}^{\prime },\hat{\delta}_{\eta
,T}^{\prime })^{\prime }$ and $\hat{\delta}_{T}^{\ast }:=(\hat{\delta}_{\psi
,T}^{\ast \prime },\hat{\delta}_{\eta ,T}^{\ast \prime })^{\prime }$ be as
defined above. Under Assumptions 1, 2 and 4, for sequences of models in
which ${E}(w_{t}\tilde{\varepsilon}_{2,t}^{\prime })=\Lambda _{T}\rightarrow
\Lambda $: \newline
(i)%
\begin{equation}
\hat{\delta}_{T}-\delta _{0}\overset{p}{\rightarrow }0_{q\times 1};
\label{consistency_reduced_form}
\end{equation}%
\begin{equation}
T^{1/2}\left( 
\begin{array}{c}
\hat{\delta}_{\psi ,T}-\delta _{\psi ,0} \\ 
\hat{\delta}_{\eta ,T}-\delta _{\eta ,0}%
\end{array}%
\right) \overset{d}{\rightarrow }N(0_{q\times 1},V_{\delta })\ ,\ V_{\delta
}:=\left( 
\begin{array}{cc}
V_{\psi } & V_{\psi ,\eta } \\ 
V_{\psi ,\eta }^{\prime } & V_{\eta }%
\end{array}%
\right) ;  \label{asym_normal_reduced_form2}
\end{equation}%
\newline
(ii)\ under the additional condition $\ell ^{3}/T\rightarrow 0$: 
\begin{equation}
\hat{\delta}_{T}^{\ast }-\hat{\delta}_{T}\overset{p^{\ast }}{\rightarrow }%
_{p}0_{q\times 1}  \label{consistency_boot}
\end{equation}%
\begin{equation}
T^{1/2}V_{\delta }^{-1/2}\left( 
\begin{array}{c}
\hat{\delta}_{\psi ,T}^{\ast }-\hat{\delta}_{\psi ,T} \\ 
\hat{\delta}_{\eta ,T}^{\ast }-\hat{\delta}_{\eta ,T}%
\end{array}%
\right) \overset{d^{\ast }}{\rightarrow }_{p}N(0_{q\times 1},I_{q}).
\label{asym_norm_boot}
\end{equation}
\end{lemma}

The results in Lemma \ref{Lemma S.1} are robust to the strength of the
proxies; i.e., they hold regardless of whether the proxies $w_{t}$ satisfy
the condition (\ref{eq strong proxxx}) or (\ref{eq weak proxxx}) in Section %
\ref{Section_Assumptions}. The structure of the asymptotic covariance matrix 
$V_{\delta }$ in (\ref{asym_normal_reduced_form2}) is specified in detail in
Br\"{u}ggemann, Jentsch and Trenkler (2016).\ It can be proved it has a
`sandwich' form $V_{\delta }:=\mathcal{A}_{0}^{-1}\mathcal{B}_{0}\mathcal{A}%
_{0}^{-1\prime }$, where $\mathcal{A}_{0}:=\lim_{T\rightarrow \infty }\big(%
\frac{\partial ^{2}}{\partial \delta \partial \delta ^{\prime }}\log
L_{T}(\delta _{0})\big)$, $\mathcal{B}_{0}:=\lim_{T\rightarrow \infty }{Var}%
\big(\frac{\partial }{\partial \delta }\log L_{T}(\delta _{0})\big)$, and $%
\log L_{T}(\delta _{0})$ is the Gaussian log-likelihood associated with the
reduced form model in (\ref{large_SVAR_general}), see Theorem 1 in Boubacar
Mainassara and Francq (2011). A consistent estimator of $V_{\delta }$ has
HAC-type form: $\hat{V}_{\delta }^{HAC}:=\mathcal{\hat{A}}^{-1}\mathcal{\hat{%
B}}^{HAC}\mathcal{\hat{A}}^{-1\prime }$; Boubacar Mainassara and Francq
(2011) discuss the computation of $\mathcal{\hat{A}}$ and $\mathcal{\hat{B}}%
^{HAC}$, see in particular their Theorem 3.\footnote{%
When Assumption 2 can be replaced with the stronger i.i.d. condition for $%
\eta _{t}$, or when $\eta _{t}$ is a MDS (${E}(\eta _{t}\mid \mathcal{F}%
_{t-1})=0_{q\times 1}$) and is also conditionally homoskedastic (${E}(\eta
_{t}\eta _{t}^{\prime }\mid \mathcal{F}_{t-1})=\Sigma _{\eta }$), one has $%
V_{\psi ,\eta }=0_{q_{\psi }\times q_{\eta }}$ in (\ref%
{asym_normal_reduced_form2}), which implies easily manageable expressions
for the asymptotic covariance matrices $V_{\psi }$ and $V_{\eta }$. For
instance, $V_{\eta }:=2D_{q_{\eta }}^{+}(\Sigma _{\eta }\otimes \Sigma
_{\eta })D_{q_{\eta }}^{+\prime }$ when $\eta _{t}$ is a conditionally
homoskedastic MDS; see Section \ref{Section_bs_notation} for $D_{q_{\eta
}}^{+}$. The simulation studies in Br\"{u}ggemann, Jentsch and Trenkler
(2016) show that the MBB is `robust' in the sense that it performs
satisfactorily well in finite samples also when the true data generating
process for $\eta _{t}$ is i.i.d. and therefore it would be `natural'
applying the residual-based i.i.d. bootstrap. In this respect, the MBB is
`robust' to $\alpha $-mixing and i.i.d. conditions and, as such, it
represents an ideal method of inference in proxy-SVARs.} Under fairly
general conditions, a consistent estimator of $V_{\delta }$ can also be
obtained from MBB replications of the estimator $\hat{\delta}_{T}^{\ast }$,
see Jentsch and Lunsford (2019, 2022). In the following, we denote with $%
\hat{V}_{\delta }$ a consistent estimator of the covariance matrix $%
V_{\delta }$.

Lemma \ref{Lemma S.1}(i) allows us to derive the asymptotic distribution of
the estimator of the parameters in the vector $\sigma ^{+}:=({vech}(\Sigma
_{u})^{\prime },{vec}(\Sigma _{w,u})^{\prime })^{\prime }$, which plays an
important role in the MD estimation problem discussed in Section \ref%
{Section_indirect_approach_MD}. Note that $\sigma ^{+}:=M_{\sigma
^{+}}\delta _{\eta }$, $M_{\sigma ^{+}}$ being a full row rank selection
matrix. Hence, by a simple delta-method argument:%
\begin{equation}
T^{1/2}(\hat{\sigma}_{T}^{+}-\sigma _{0}^{+})\overset{d}{\rightarrow }%
N(0_{a\times 1},V_{\sigma ^{+}})\text{ \ , \ }V_{\sigma ^{+}}=M_{\sigma
^{+}}V_{\eta }M_{\sigma ^{+}}^{\prime }  \label{sigma_reduced_form_estimator}
\end{equation}%
where the positive definite asymptotic covariance matrix $V_{\sigma ^{+}}$
can be estimated consistently by $\hat{V}_{\sigma ^{+}}=M_{\sigma ^{+}}\hat{V%
}_{\eta }M_{\sigma ^{+}}^{\prime }$, $\hat{V}_{\eta }$ denoting the (2,2)
block of $\hat{V}_{\delta }$, see (\ref{asym_normal_reduced_form2}).

The next two lemmas derive the asymptotic distribution of the estimator of
the reduced form proxy-SVAR parameters in the vector $\mu :=({vech}(\Omega
_{w})^{\prime },{vec}(\Sigma _{w,u})^{\prime })^{\prime }$, where $\Omega
_{w}:=\Sigma _{w,u}\Sigma _{u}^{-1}\Sigma _{u,w}$, when the proxy-SVAR\ is
identified according to Proposition \ref{Prop 1}, and when the instruments
satisfy the weak proxies condition in equation (\ref{eq weak proxxx}),
respectively. These lemmas are important because, recall, $\mu $ is a
nonlinear function of the covariance parameters in $\sigma ^{+}:=M_{\sigma
^{+}}\delta _{\eta }$ and, as shown in Section \ref{Section_bs_strength},
the estimator of $\mu $ plays a crucial role in the derivation of the CMD
estimator used to build our bootstrap pre-test of instrument relevance, see
below. In what follows, we exploit the functional dependence of $\mu $ on
the $m\times 1$ vector $\sigma ^{+}:=({vech}(\Sigma _{u})^{\prime },{vec}%
(\Sigma _{w,u})^{\prime })^{\prime }$, which in turn depends on $\delta
_{\eta }$, $\sigma ^{+}:=M_{\sigma ^{+}}\delta _{\eta }$. Furthermore, we
decompose $\mu $ as $\mu :=(\omega ^{\prime },\varpi ^{\prime })^{\prime }$,
where $\omega ={vech}(\Omega _{w})$ is $o_{1}\times 1$, $o_{1}=\frac{1}{2}%
s(s+1)$, and $\varpi :={vec}(\Sigma _{w,u})$ is $o_{2}\times 1$, $o_{2}=ns$.
Thus, $\mu $ is an $o\times 1$ vector, $o=o_{1}+o_{2}$. $\mu _{0}=\mu
(\sigma _{0}^{+})\equiv (\omega _{0}^{\prime },\varpi _{0}^{\prime
})^{\prime }$ denotes the true value of $\mu $ and $\sigma _{0}^{+}$ is the
true value of $\sigma ^{+}.$ The QML\ estimator of $\mu $, $\hat{\mu}_{T}:=(%
\hat{\omega}_{T}^{\prime },\hat{\varpi}_{T}^{\prime })^{\prime }$, obtains
from $\hat{\delta}_{\eta ,T}$ and has the same asymptotic properties as the
estimator $\hat{\delta}_{\eta ,T}$ stated in Lemma \ref{Lemma S.1}(i) by a
delta-method argument. Given sequences of models in which ${E}(w_{t}\tilde{%
\varepsilon}_{2,t}^{\prime })=\Lambda _{T}\rightarrow \Lambda $, we denote
with $\mathcal{N}_{\Lambda }$ a neighborhood of the parameters in the limit
matrix $\Lambda $.

\begin{lemma}
\label{Lemma S.2}Under the conditions of Lemma \ref{Lemma S.1}: \newline
(i) $\hat{\mu}_{T}-\mu _{0}\overset{p}{\rightarrow }0$ (regardless of the
strength of the proxies); \newline
(ii) if the proxy-SVAR is identified according to Proposition \ref{Prop 1},%
\begin{equation*}
T^{1/2}(\hat{\mu}_{T}-\mu _{0})\overset{d}{\rightarrow }J_{\sigma ^{+}}%
\mathbb{G}_{\sigma ^{+}}
\end{equation*}%
where $\mathbb{G}_{\sigma ^{+}}\sim N(0,V_{\sigma ^{+}})$, $V_{\sigma
^{+}}:=(M_{\sigma ^{+}}V_{\eta }M_{\sigma ^{+}}^{\prime })$ with $V_{\eta }$
is defined in (\ref{asym_normal_reduced_form2}), and 
\begin{equation*}
J_{\sigma ^{+}}:=\frac{\partial \mu }{\partial \sigma ^{+\prime }}=\left( 
\begin{array}{cc}
-D_{s}^{+}\left( \Sigma _{w,u}\Sigma _{u}^{-1}\otimes \Sigma _{w,u}\Sigma
_{u}^{-1}\right) D_{n} & 2D_{s}^{+}(\Sigma _{w,u}\Sigma _{u}^{-1}\otimes
I_{s}) \\ 
0 & I_{ns}%
\end{array}%
\right)
\end{equation*}%
is an $o\times m$ Jacobian matrix of full row rank, $rank[J_{\sigma ^{+}}]=o$%
.
\end{lemma}

\begin{lemma}
\label{Lemma S.3}Under the conditions of Lemma \ref{Lemma S.1}, if the
proxies $w_{t}$ satisfy the local-to-zero condition (\ref{eq weak proxxx}),
the component $\hat{\omega}_{T}-\omega _{0}$ of the vector $\hat{\mu}%
_{T}-\mu _{0}$ is distributed as follows:%
\begin{equation*}
T(\hat{\omega}_{T}-\omega _{0})\overset{d}{\rightarrow }J^{(1)}\mathbb{G}%
_{\sigma ^{+}}+\tfrac{1}{2}(I_{o_{1}}\otimes \mathbb{G}_{\sigma
^{+}}^{\prime })H_{\sigma ^{+}}^{(1)}\mathbb{G}_{\sigma ^{+}}\text{,}
\end{equation*}%
where $J^{(1)}$ is a Jacobian matrix that satisfies the condition $%
T^{1/2}J_{\sigma ^{+}}^{(1)}\rightarrow J^{(1)}$, $J_{\sigma ^{+}}^{(1)}$ is
the $o_{1}m\times m$ upper block of the Jacobian matrix $J_{\sigma ^{+}}$
reported in Lemma \ref{Lemma S.2}, $H_{\sigma ^{+}}^{(1)}$ is the $%
o_{1}m\times m$ upper block of the $om\times m$ Hessian matrix $H_{\sigma
^{+}}:=\frac{\partial }{\partial \sigma ^{+\prime }}{vec}\{(\frac{\partial
\mu }{\partial \sigma ^{+\prime }})^{\prime }\}$, and is different from zero.
\end{lemma}

Lemma \ref{Lemma S.2} ensures that when the proxy-SVAR is (locally)\
identified, the estimator $\hat{\mu}_{T}$ in (\ref{CMD_problem})\ (see also (%
\ref{CMD_problem_bis})\ below) satisfies `standard' regularity conditions.
Conversely, Lemma \ref{Lemma S.3} shows that this is not the case when the
proxies are local-to-zero. Indeed, Lemma \ref{Lemma S.3} ensures that under
the weak proxies condition, the asymptotic distribution of $T(\hat{\omega}%
_{T}-\omega _{0})$ is a mixture of Gaussian and $\chi ^{2}$-type random
variables and, because of convergence at the $T$ rate, $T^{1/2}(\hat{\omega}%
_{T}-\omega _{0})\overset{p}{\rightarrow }0_{o_{1}\times 1}$. This in turn
implies that the vector $T(\hat{\mu}_{T}-\mu _{0})\equiv (T(\hat{\omega}%
_{T}-\omega _{0})^{\prime },T(\hat{\varpi}_{T}-\varpi _{0})^{\prime
})^{\prime }$ is asymptotically non-Gaussian. Our proof of Lemma \ref{Lemma
S.3} in Section \ref{Section_Proof_LemmaS3} is presented for the case in
which all the $s$ proxies in the vector $w_{t}$ satisfy the local-to-zero
embedding in (\ref{eq weak proxxx}); when only a subset of the $s$ proxies
satisfies that condition, the asymptotic distribution of $T(\hat{\mu}%
_{T}-\mu _{0})$ is still not Gaussian.\footnote{%
Results are available upon request.}

In the two lemmas that follow we present the asymptotic distribution of the
random vector $\Gamma _{T}:=T^{1/2}V_{\theta }^{-1/2}(\hat{\theta}%
_{T}-\theta _{0})$, where $\hat{\theta}_{T}$ is the CMD estimator resulting
from the problem (\ref{CMD_problem}), here reported for convenience:%
\begin{equation}
\hat{\theta}_{T}:=\arg \min_{\theta \in \mathcal{P}_{\theta }}\hat{Q}%
_{T}(\theta )\text{, \ \ }\hat{Q}_{T}(\theta ):=(\hat{\mu}_{T}-f(\theta
))^{\prime }\hat{V}_{\mu }^{-1}(\hat{\mu}_{T}-f(\theta ))
\label{CMD_problem_bis}
\end{equation}%
and where the vector $\theta :=(\beta _{2}^{\prime },\lambda ^{\prime
})^{\prime }$ contains the (free)\ parameters in the matrix $(\tilde{B}%
_{2}^{\prime }$ $\vdots $ $\Lambda ^{\prime })^{\prime }$. The asymptotic
distribution of $T^{1/2}V_{\theta }^{-1/2}(\hat{\theta}_{T}-\theta _{0})$ is
derived considering instruments that satisfy the strong proxies condition in
(\ref{eq strong proxxx}) and Staiger and Stock's (1997) embedding in (\ref%
{eq weak proxxx}), respectively. Below $\mathcal{N}_{\theta _{0}}$
represents a neighborhood of $\theta _{0}$.

\begin{lemma}
\label{Lemma S.4}Under the conditions of Lemma \ref{Lemma S.1} and
Proposition \ref{Prop 1}: \newline
(i) $\hat{\theta}_{T}-\theta _{0}\overset{p}{\rightarrow }0$;\newline
(ii) $T^{1/2}(\hat{\theta}_{T}-\theta _{0})\overset{d}{\rightarrow }%
N(0,V_{\theta })$, where $V_{\theta }:=\left( J_{\theta }^{\prime }V_{\mu
}^{-1}J_{\theta }\right) ^{-1}$ and $J_{\theta }$ is a Jacobian matrix of
full column rank in $\mathcal{N}_{\theta _{0}}$.
\end{lemma}

\begin{lemma}
\label{Lemma S.5}Under the conditions of Lemma \ref{Lemma S.1}, if the
proxies $w_{t}$ satisfy the local-to-zero condition (\ref{eq weak proxxx}), $%
T^{1/2}(\hat{\theta}_{T}-\theta _{0})$ is not asymptotically Gaussian.
\end{lemma}

\section{Proofs of lemmas, corollaries and propositions}

\label{Section_Supplementary_Proofs}

\subsection{Proof of Lemma \protect\ref{Lemma S.1}}

(i) The result follow from Theorem 1 in Boubacar Mainassara and Francq
(2011) by setting the matrices $B_{01},\ldots ,B_{0q}$ in the VARMA model of
their equation (3) to zero, and the matrices $A_{00}$ and $B_{00}$ to the
identity matrix; see also Theorem 2.1 in Br\"{u}ggemann {\emph{et al.}}
(2016). (ii) The result follows from Theorem 4.1 in Br\"{u}ggemann {\emph{et
al.}} (2016). $\blacksquare $

\subsection{Proof of Lemma \protect\ref{Lemma S.2}}

\label{Section_Proof_LemmaS2}(i) $\mu =\mu (\sigma ^{+})$ is a smooth
function of $\sigma ^{+}$ and therefore of $\delta _{\eta }$ (recall that $%
\sigma ^{+}=M_{\sigma ^{+}}\delta _{\eta }$, $M_{\sigma ^{+}}$ being a
selection matrix of full row rank). The result follows from Lemma \ref{Lemma
S.1}(i) and Slutsky's Theorem.

\smallskip

(ii)\ Since $\sigma ^{+}=M_{\sigma ^{+}}\delta _{\eta }$, Lemma \ref{Lemma
S.1}(i) implies (\ref{sigma_reduced_form_estimator}). Consider the following
quadratic expansion of $\hat{\mu}_{T}=\mu (\hat{\sigma}_{T}^{+})$ around $%
\sigma _{0}^{+}$:%
\begin{equation}
T^{1/2}\left( \hat{\mu}_{T}-\mu _{0}\right) =J_{\sigma _{0}^{+}}(\sigma
_{0}^{+})T^{1/2}(\hat{\sigma}_{T}^{+}-\sigma _{0}^{+})+\tfrac{1}{2}%
T^{1/2}R_{T}(\ddot{\sigma}_{T}^{+})  \label{mu_expansion}
\end{equation}%
where $J_{\sigma _{0}^{+}}(\sigma _{0}^{+})$ is the $o\times m$ Jacobian
matrix $J_{\sigma _{0}^{+}}:=\frac{\partial \mu }{\partial \sigma ^{+\prime }%
}$\ evaluated at $\sigma _{0}^{+}$; the remainder term $R_{T}(\ddot{\sigma}%
_{T}^{+})$ has representation:%
\begin{eqnarray*}
&&R_{T}(\ddot{\sigma}_{T}^{+})\overset{}{:=}\left( I_{o}\otimes (\hat{\sigma}%
_{T}^{+}-\sigma _{0}^{+})^{\prime }\right) H_{\sigma ^{+}}(\ddot{\sigma}%
_{T}^{+})(\hat{\sigma}_{T}^{+}-\sigma _{0}^{+})\text{,} \\
&&H_{\sigma ^{+}}(\ddot{\sigma}_{T}^{+})\overset{}{:=}\frac{\partial }{%
\partial \sigma ^{+\prime }}{vec}\{\left. (\tfrac{\partial \mu }{\partial
\sigma ^{+\prime }})^{\prime }\right\vert _{\sigma ^{+}=\ddot{\sigma}%
_{T}^{+}}\}
\end{eqnarray*}%
where $H_{\sigma ^{+}}(\ddot{\sigma}_{T}^{+})$ is the $om\times m$ Hessian
matrix evaluated at $\ddot{\sigma}_{T}^{+}$, an intermediate vector value
between $\hat{\sigma}_{T}^{+}$ and $\sigma _{0}^{+}$. By construction, the
last $o_{2}$ components of the vector $T^{1/2}\left( \hat{\mu}_{T}-\mu
_{0}\right) $ coincide with the last elements in the vector $T^{1/2}(\hat{%
\sigma}_{T}^{+}-\sigma _{0}^{+})$ (i.e. $T^{1/2}(\hat{\varpi}_{T}-\varpi
_{0})$), hence the the Jacobian $J_{\sigma _{0}^{+}}(\sigma _{0}^{+})$ and
the remainder term $R_{T}(\ddot{\sigma}_{T}^{+})$ have representations: 
\begin{equation}
J_{\sigma _{0}^{+}}(\sigma _{0}^{+}):=\left( 
\begin{array}{c}
J_{\sigma _{0}^{+}}^{(1)} \\ 
J_{\sigma _{0}^{+}}^{(2)}%
\end{array}%
\right) \equiv \left( 
\begin{array}{cc}
J_{\sigma _{0}^{+}}^{(1,1)} & J_{\sigma _{0}^{+}}^{(1,2)} \\ 
0 & I_{ns}%
\end{array}%
\right)  \label{partition1}
\end{equation}%
and 
\begin{equation}
R_{T}(\ddot{\sigma}_{T}^{+})\equiv \left( 
\begin{array}{c}
R_{1,T}(\ddot{\sigma}_{T}^{+}) \\ 
0%
\end{array}%
\right) \text{ }%
\begin{array}{c}
o_{1}\times 1 \\ 
o_{2}\times 1%
\end{array}
\label{partition2}
\end{equation}%
where%
\begin{equation*}
\text{ }R_{1,T}(\ddot{\sigma}_{T}^{+}):=\left( I_{o_{1}}\otimes (\hat{\sigma}%
_{T}^{+}-\sigma _{0}^{+})^{\prime }\right) H_{\sigma ^{+}}^{(1)}(\ddot{\sigma%
}_{T}^{+})(\hat{\sigma}_{T}^{+}-\sigma _{0}^{+})\text{,}
\end{equation*}%
and $H_{\sigma ^{+}}^{(1)}(\ddot{\sigma}_{T}^{+}):=\frac{\partial }{\partial
\sigma ^{+\prime }}{vec}[J_{\ddot{\sigma}_{T}^{+}}^{(1)\prime }]$ is the $%
o_{1}m\times m$ upper block of the Hessian $H_{\sigma ^{+}}(\ddot{\sigma}%
_{T}^{+})$ defined above.

To prove the result, we show that the Jacobian $J_{\sigma _{0}^{+}}(\sigma
_{0}^{+})$ in (\ref{mu_expansion}) is constant and has full row rank, while
the remainder term $\tfrac{1}{2}T^{1/2}R_{T}(\ddot{\sigma}_{T}^{+})$ is $%
o_{p}(1)$ as $\hat{\sigma}_{T}^{+}$ (and hence $\ddot{\sigma}_{T}^{+}$)
converges in probability to $\sigma _{0}^{+}.$

By using standard matrix derivative rules (Magnus and Neudecker, 1999), it
is seen that the blocks $J_{\sigma _{0}^{+}}^{(1,1)}$ and $J_{\sigma
_{0}^{+}}^{(1,2)}$ of $J_{\sigma _{0}^{+}}(\sigma _{0}^{+})$ in (\ref%
{partition1}) are given by%
\begin{equation}
J_{\sigma _{0}^{+}}^{(1,1)}=-D_{s}^{+}\left( \Sigma _{w,u}\Sigma
_{u}^{-1}\otimes \Sigma _{w,u}\Sigma _{u}^{-1}\right) D_{n}\text{ ; \ }%
J_{\sigma _{0}^{+}}^{(1,2)}=2D_{s}^{+}(\Sigma _{w,u}\Sigma _{u}^{-1}\otimes
I_{s}).  \label{Jacob_form1}
\end{equation}%
Without loss of generality (ordering is not crucial for the arguments that
follow), partition the matrix $B$ as $B=(\tilde{B}_{\bullet 1}$ , $\tilde{B}%
_{\bullet 2})$, where $\tilde{B}_{\bullet 1}$ collects the columns of $B$
associated with the $n-s$ non-instrumented structural shocks (note that, in
general, $\tilde{B}_{\bullet 1}$ will include some of the columns of the
matrix $B_{\bullet 1}$). Likewise, partition the matrix $A=B^{-1}$ as $A=(%
\tilde{A}_{1\bullet }^{\prime }$ $,$ $\tilde{A}_{2\bullet }^{\prime
})^{\prime }$, where $\tilde{A}_{1\bullet }$ is the block associated with
the $n-s$ non-instrumented structural shocks (notice that $\tilde{A}%
_{1\bullet }$ is different from the matrix $A_{1\bullet }$ that plays a key
role in the paper) and $\tilde{A}_{2\bullet }$ is the block associated with
the $s$ instrumented structural shocks; $rank[\tilde{A}_{2\bullet }]=s$
under Assumption 3. Under sequences of models for which ${E}(w_{t}\tilde{%
\varepsilon}_{2,t}^{\prime })=\Lambda _{T}\rightarrow \Lambda $, by imposing
the proxy-SVAR restrictions $\Sigma _{w,u}=\Lambda \tilde{B}_{\bullet
2}^{\prime }$, $\Sigma _{u}=BB^{\prime }$ and using the above partitions, it
turns out that $\Sigma _{w,u}\Sigma _{u}^{-1}=\Lambda \tilde{B}_{\bullet
2}^{\prime }(BB^{\prime })^{-1}=\Lambda (0$ , $I_{s})A=\Lambda \tilde{A}%
_{2\bullet }$. Hence, the Jacobian is equal to%
\begin{equation}
J_{\sigma ^{+}}(\sigma _{0}^{+}):=\left( 
\begin{array}{cc}
-D_{s}^{+}\left( \Lambda \tilde{A}_{2\bullet }\otimes \Lambda \tilde{A}%
_{2\bullet }\right) D_{n} & 2D_{s}^{+}(\Lambda \tilde{A}_{2\bullet }\otimes
I_{s}) \\ 
0 & I_{ns}%
\end{array}%
\right)  \label{J_at_true}
\end{equation}%
and it is therefore constant and of full column rank ($rank[\Lambda ]=s$ in $%
\mathcal{N}_{\Lambda }$) if the identification conditions in Proposition \ref%
{Prop 1} hold, which implies strong proxies as in (\ref{eq strong proxxx}).

To prove that the remainder term $\tfrac{1}{2}T^{1/2}R_{T}(\ddot{\sigma}%
_{T}^{+})$ is $o_{p}(1)$ as $\hat{\sigma}_{T}^{+}$ (and hence $\ddot{\sigma}%
_{T}^{+}$) converges in probability to $\sigma _{0}^{+}$, we prove that the
block $H_{\sigma ^{+}}^{(1)}(\ddot{\sigma}_{T}^{+}):=\frac{\partial }{%
\partial \sigma ^{+\prime }}{vec}[J_{\ddot{\sigma}_{T}^{+}}^{(1)\prime }]$
of the Hessian in (\ref{partition2}) does not depend on $T$. It is useful to
note that%
\begin{equation}
H_{\sigma ^{+}}^{(1)}(\ddot{\sigma}_{T}^{+})^{\prime }:=\frac{\partial }{%
\partial \sigma ^{+\prime }}{vec}[J_{\ddot{\sigma}_{T}^{+}}^{(1)}]\equiv
\left( 
\begin{array}{c}
\frac{\partial }{\partial \sigma ^{+\prime }}{vec}[J_{\ddot{\sigma}%
_{T}^{+}}^{(1,1)}] \\ 
\frac{\partial }{\partial \sigma ^{+\prime }}{vec}[J_{\ddot{\sigma}%
_{T}^{+}}^{(1,2)}]%
\end{array}%
\right) \equiv \left( 
\begin{array}{cc}
H_{11}^{(1)} & H_{12}^{(1)} \\ 
H_{21}^{(1)} & H_{22}^{(1)}%
\end{array}%
\right)  \label{Hessian_partial_block}
\end{equation}%
and that, by applying standard matrix derivative rules:%
\begin{equation*}
H_{11}^{(1)}:=\frac{1}{\partial {vech}(\Sigma _{u})^{\prime }}\partial {vec}%
[J_{\ddot{\sigma}_{T}^{+}}^{(1,1)}],\text{ }H_{12}^{(1)}:=\frac{1}{\partial {%
vec}(\Sigma _{w,u})^{\prime }}\partial {vec}[J_{\ddot{\sigma}%
_{T}^{+}}^{(1,1)}],
\end{equation*}%
\begin{equation*}
H_{21}^{(1)}:=\frac{1}{\partial {vech}(\Sigma _{u})^{\prime }}\partial {vec}%
[J_{\ddot{\sigma}_{T}^{+}}^{(1,2)}],\text{ }H_{22}^{(1)}:=\frac{1}{\partial {%
vec}(\Sigma _{w,u})^{\prime }}\partial {vec}[J_{\ddot{\sigma}%
_{T}^{+}}^{(1,2)}],
\end{equation*}%
one can notice that $H_{11}^{(1)}$, $H_{12}^{(1)}$, $H_{21}^{(1)}$ and $%
H_{22}^{(1)}$ depend only on $\Sigma _{u}$ and $\Sigma _{w,u}$, not on $T$
under the strong proxies condition.

Summing up, asymptotic normality follows from (\ref{mu_expansion}), the
result 
\begin{equation*}
J_{\sigma _{0}^{+}}(\sigma _{0}^{+})T^{1/2}(\hat{\sigma}_{0,T}^{+}-\sigma
_{0}^{+})\overset{d}{\rightarrow }J_{\sigma ^{+}}\mathbb{G}_{\sigma ^{+}}
\end{equation*}%
and the fact that the term $\tfrac{1}{2}T^{1/2}R_{T}(\ddot{\sigma}_{T}^{+})$%
\ in the expansion (\ref{mu_expansion})\ is $o_{p}(1)$. $\blacksquare $

\subsection{Proof of Lemma \protect\ref{Lemma S.3}}

\label{Section_Proof_LemmaS3}From the expansion (\ref{mu_expansion}), we
isolate the block associated with the component $T^{1/2}\left( \hat{\omega}%
_{T}-\omega _{0}\right) $: 
\begin{equation}
T^{1/2}\left( \hat{\omega}_{T}-\omega _{0}\right) =(J_{\sigma
_{0}^{+}}^{(1,1)}\text{ },\text{ }J_{\sigma _{0}^{+}}^{(1,2)})T^{1/2}(\hat{%
\sigma}_{0,T}^{+}-\sigma _{0}^{+})+\tfrac{1}{2}T^{1/2}R_{1,T}(\ddot{\sigma}%
_{T}^{+})  \label{mu_expansion2}
\end{equation}%
and show that, if the instruments $w_{t}$ are weak for $\tilde{\varepsilon}%
_{2,t}$ in the sense of equation (\ref{eq weak proxxx}), then for $%
T\rightarrow \infty $ :%
\begin{equation*}
T\left( \hat{\omega}_{T}-\omega _{0}\right) =\underset{=J^{(1)}+o(1)}{%
\underbrace{T^{1/2}(J_{\sigma _{0}^{+}}^{(1,1)}\text{ },\text{ }J_{\sigma
_{0}^{+}}^{(1,2)})}}\underset{O_{p}(1)}{\underbrace{T^{1/2}(\hat{\sigma}%
_{0,T}^{+}-\sigma _{0}^{+})}}
\end{equation*}%
\begin{equation}
+\tfrac{1}{2}(I_{o_{1}}\otimes \underset{O_{p}(1)}{\underbrace{T^{1/2}(\hat{%
\sigma}_{0,T}^{+}-\sigma _{0}^{+})^{\prime }}})H_{\sigma ^{+}}^{(1)}(\ddot{%
\sigma}_{T}^{+})\underset{O_{p}(1)}{\underbrace{T^{1/2}(\hat{\sigma}%
_{0,T}^{+}-\sigma _{0}^{+})}}  \label{expansion_weak2}
\end{equation}%
where $J^{(1)}:=T^{1/2}J_{\sigma ^{+}}^{(1)}\equiv T^{1/2}(J_{\sigma
_{0}^{+}}^{(1,1)}$ $,$ $J_{\sigma _{0}^{+}}^{(1,2)})$ and $H_{\sigma
^{+}}^{(1)}(\ddot{\sigma}_{T}^{+})\neq 0$ and does not depend on $T.$

We start by proving that in (\ref{expansion_weak2}), $T^{1/2}(J_{\sigma
_{0}^{+}}^{(1,1)},J_{\sigma _{0}^{+}}^{(1,2)})\rightarrow J^{(1)}$, where $%
J^{(1)}$ is independent of $T$. From (\ref{Jacob_form1}) and (\ref{J_at_true}%
), we have 
\begin{eqnarray*}
&&T^{1/2}(J_{\sigma _{0}^{+}}^{(1,1)},J_{\sigma _{0}^{+}}^{(1,2)}) \\
&&\hspace{1cm}\overset{}{=}T^{1/2}(-D_{s}^{+}(\Lambda _{T}\tilde{A}%
_{2\bullet }\otimes \Lambda _{T}\tilde{A}_{2\bullet })D_{n},\text{ }%
2D_{s}^{+}(\Lambda _{T}\tilde{A}_{2\bullet }\otimes I_{s})) \\
&&\hspace{1cm}\overset{}{=}T^{1/2}D_{s}^{+}(\Lambda _{T}\tilde{A}_{2\bullet
}\otimes I_{s})(-(I_{s}\otimes \Lambda _{T}\tilde{A}_{2\bullet })D_{n},\text{
}2(I_{s}\otimes I_{s}))
\end{eqnarray*}%
Hence, for $\Lambda _{T}:=CT^{-1/2}$, $C$ being an $s\times s$ matrix with
finite norm, $\left\Vert C\right\Vert <\infty $: 
\begin{eqnarray*}
&&T^{1/2}(J_{\sigma _{0}^{+}}^{(1,1)},J_{\sigma _{0}^{+}}^{(1,2)})\overset{}{%
:=}T^{1/2}D_{s}^{+}(T^{-1/2}C\tilde{A}_{2\bullet }\otimes I_{s}) \\
&&\hspace{2.3cm}\overset{}{\times }[-(I_{s}\otimes T^{-1/2}C\tilde{A}%
_{2\bullet })D_{n}\text{ },\text{ }2(I_{s}\otimes I_{s})]
\end{eqnarray*}%
and, as $T\rightarrow \infty $,%
\begin{equation*}
T^{1/2}(J_{\sigma _{0}^{+}}^{(1,1)}\text{ },\text{ }J_{\sigma
_{0}^{+}}^{(1,2)})\rightarrow J^{(1)}:=D_{s}^{+}(C\tilde{A}_{2\bullet
}\otimes I_{s})\left[ 0\text{ },\text{ }2I_{s^{2}}\right]
\end{equation*}%
which does not depend on $T$.

Next, we show that in the expansion (\ref{expansion_weak2}), $H_{\sigma
^{+}}^{(1)}(\ddot{\sigma}_{T}^{+})\neq 0$ and does not depend on $T$. From
the inspection of the Hessian matrix in (\ref{Hessian_partial_block}), it
follows that while $H_{11}^{(1)}$, $H_{22}^{(1)}$ and $H_{21}^{(1)}$ depend
on $\Sigma _{w,u}=T^{-1/2}C\tilde{B}_{\bullet 2}^{\prime }$ and converge to
zero as $T\rightarrow \infty $, $H_{22}^{(1)}$ is given by the expression:%
\begin{eqnarray*}
&&H_{22}^{(1)}\overset{}{:=}\frac{1}{\partial {vec}(\Sigma _{w,u})^{\prime }}%
\partial {vec}[J_{\ddot{\sigma}_{T}^{+}}^{(1,2)}]\overset{}{=}\frac{1}{%
\partial {vec}(\Sigma _{w,u})^{\prime }}\partial {vec}[2D_{s}^{+}(\Sigma
_{w,u}\Sigma _{u}^{-1}\otimes I_{s})] \\
&&\hspace{1.3cm}\overset{}{=}(I_{ns}\otimes 2D_{s}^{+})\frac{1}{\partial {vec%
}(\Sigma _{w,u})^{\prime }}\partial {vec}[(\Sigma _{w,u}\Sigma
_{u}^{-1}\otimes I_{s})]
\end{eqnarray*}%
which shows that $H_{22}^{(1)}$ does not depend on the covariance matrix $%
\Sigma _{w,u}$ because of the derivative; hence $H_{22}^{(1)}\neq 0$ for any 
$T.$

Finally, note that if $C=0_{s\times s}$ (i.e.\thinspace , the instruments $%
w_{t}$ are totally irrelevant for $\tilde{\varepsilon}_{2,t})$, then $\hat{%
\omega}_{T}\overset{p}{\rightarrow }0$; the first term in the expansion (\ref%
{expansion_weak2}) is zero, therefore $T\hat{\omega}_{T}=O_{p}(1)$ and $%
T^{1/2}\hat{\omega}_{T}\overset{p}{\rightarrow }0$. $\blacksquare $

\subsection{Proof of Lemma \protect\ref{Lemma S.4}}

\label{Section_Proof_LemmaS4}The proof of this lemma requires some
preliminary steps. First, given the distance function $d(\mu ,\theta )=\mu
-f(\theta )$ minimized in (\ref{CMD_problem_bis}) (see also equation (\ref%
{CMD_problem})), when $s>1$ (multiple instrumented shocks) it is necessary
to consider the following set of identification restrictions on the
parameters in the matrix $(\tilde{B}_{\bullet 2}^{\prime }$ $,$ $\Lambda
^{\prime })^{\prime }$ (see footnote 10 in the paper): 
\begin{equation}
\left( 
\begin{array}{c}
{vec}(\Lambda ) \\ 
{vec}(\tilde{B}_{\bullet 2}^{\prime })%
\end{array}%
\right) =\left( 
\begin{array}{cc}
S_{\Lambda } & 0 \\ 
0 & S_{\tilde{B}_{2}}%
\end{array}%
\right) \theta +\left( 
\begin{array}{c}
s_{\Lambda } \\ 
s_{\tilde{B}_{2}}%
\end{array}%
\right)  \label{restrictions_theta}
\end{equation}%
where $S_{\Lambda }$ and $S_{\tilde{B}_{2}}$ and are known selection
matrices of full column rank, and $s_{\Lambda }$ and $s_{\tilde{B}_{2}}$ are
possibly non-zero vectors containing known elements; see Angelini and
Fanelli (2019) for details. Second, by standard matrix derivative rules, the
Jacobian matrix $J_{\theta }:=\frac{\partial f(\theta )}{\partial \theta
^{\prime }}$ has structure 
\begin{equation}
J_{\theta }:=\left( 
\begin{array}{cc}
2D_{s}^{+}(\Lambda \otimes I_{s}) & 0 \\ 
(\tilde{B}_{\bullet 2}\otimes I_{s}) & (I_{n}\otimes \Lambda )K_{ns}%
\end{array}%
\right) \left( 
\begin{array}{cc}
S_{\Lambda } & 0 \\ 
0 & S_{\tilde{B}_{2}}%
\end{array}%
\right) .  \label{Jacob_theta}
\end{equation}%
Equation (\ref{Jacob_theta}) shows that the Jacobian matrix $J_{\theta }$
has full column rank in $\mathcal{N}_{\theta _{0}}$ under the strong proxies
condition (\ref{eq strong proxxx}) and has reduced rank in $\mathcal{N}%
_{\theta _{0}}$ under the weak proxies condition (\ref{eq weak proxxx}).

\smallskip (i) Given the CMD problem in (\ref{CMD_problem_bis}), under the
strong instrument condition (\ref{eq strong proxxx}), the consistency result
follows from the same arguments used in the proof of Proposition \ref{prop 2}
to establish the consistency of the MD estimator $\hat{\alpha}_{T}$.

\smallskip (ii) The first-order conditions associated with the problem (\ref%
{CMD_problem_bis}) are given by 
\begin{equation*}
J_{\hat{\theta}_{T}}^{\prime }\hat{V}_{\mu }^{-1}(\hat{\mu}_{T}-f(\hat{\theta%
}_{T}))=0
\end{equation*}%
where $J_{\hat{\theta}_{T}}$ is the Jacobian (\ref{Jacob_theta}) evaluated
at the CMD estimator $\hat{\theta}_{T}$. By using a mean-value expansion of $%
f(\hat{\theta}_{T})$ around $\theta _{0}$, the first-order conditions are%
\begin{equation*}
J_{\hat{\theta}_{T}}^{\prime }\hat{V}_{\mu }^{-1}(\hat{\mu}_{T}-\mu _{0}-J_{%
\ddot{\theta}}(\hat{\theta}_{T}-\theta _{0}))=0
\end{equation*}%
where $\ddot{\theta}$ is an intermediate vector between $\hat{\theta}_{T}$
and $\theta _{0}$, and $\mu _{0}=f(\theta _{0})$. By re-arranging the
expression above, we obtain the equation%
\begin{equation}
\{J_{\hat{\theta}_{T}}^{\prime }\hat{V}_{\mu }^{-1}J_{\ddot{\theta}%
}\}T^{1/2}(\hat{\theta}_{T}-\theta _{0})=J_{\hat{\theta}_{T}}^{\prime }\hat{V%
}_{\mu }^{-1}T^{1/2}(\hat{\mu}_{T}-\mu _{0})  \label{expression}
\end{equation}%
which shows that the asymptotic distribution of $T^{1/2}(\hat{\theta}%
_{T}-\theta _{0})$ depends on two main components: the asymptotic
distribution of $T^{1/2}(\hat{\mu}_{T}-\mu _{0})$ derived in Lemma \ref%
{Lemma S.2} and Lemma \ref{Lemma S.3}, and the property of the matrix $J_{%
\hat{\theta}_{T}}^{\prime }\hat{V}_{\mu }^{-1}J_{\ddot{\theta}}$ for $%
T\rightarrow \infty $.

Under the strong instrument condition (\ref{eq strong proxxx}), the
consistency result implies that $J_{\hat{\theta}_{T}}\overset{p}{\rightarrow 
}J_{\theta _{0}}$ and $J_{\ddot{\theta}}\overset{p}{\rightarrow }J_{\theta
_{0}}$; asymptotic normality follows from Lemma \ref{Lemma S.2}(i) which
ensures that $\hat{V}_{\mu }\overset{p}{\rightarrow }V_{\mu }$, and Lemma %
\ref{Lemma S.2}(ii). $\blacksquare $

\subsection{Proof of Lemma \protect\ref{Lemma S.5}}

\label{Section_Proof_LemmaS5}To prove that $T^{1/2}(\hat{\theta}_{T}-\theta
_{0})$ is not asymptotically Gaussian under the weak instrument condition in
equation (\ref{eq weak proxxx}), it suffices to consider the expression in (%
\ref{expression}), the partition $T^{1/2}(\hat{\mu}_{T}-\mu _{0})\equiv
(T^{1/2}(\hat{\omega}_{T}-\omega _{0})^{\prime },T^{1/2}(\hat{\varpi}%
_{T}-\varpi _{0})^{\prime })^{\prime }$, and the application of Lemma \ref%
{Lemma S.3} which implies $T^{1/2}(\hat{\omega}_{T}-\omega _{0})\overset{p}{%
\rightarrow }0_{o_{1}\times 1}$. $\blacksquare $

\subsection{Proof of Proposition \protect\ref{Prop 1}}

(i)\ Under Assumptions 1-2 and 4 and sequences of models for which ${E}(w_{t}%
\tilde{\varepsilon}_{2,t}^{\prime })=\Lambda _{T}\rightarrow \Lambda $, $%
\hat{\sigma}_{T}^{+}\overset{p}{\rightarrow }\sigma _{0}^{+}$ by Lemma \ref%
{Lemma S.1}(i), hence, by Slutsky's Theorem, $g_{T}(\hat{\sigma}%
_{T}^{+},\alpha )\overset{p}{\rightarrow }$ $g(\sigma _{0}^{+},\alpha )$.
For $\hat{V}_{\sigma ^{+}}$ consistent estimator of $V_{\sigma ^{+}}$, and $%
\alpha ,$ $\bar{\alpha}\in \mathcal{P}_{\alpha }$, $\hat{Q}_{T}(\alpha
):=g_{T}(\hat{\sigma}_{T}^{+},\alpha )^{\prime }\hat{V}_{gg}(\bar{\alpha}%
)^{-1}g_{T}(\hat{\sigma}_{T}^{+},\alpha )\overset{p}{\rightarrow }$ $%
Q_{0}(\alpha ):=g(\sigma _{0}^{+},\alpha )^{\prime }V_{gg,0}^{-1}(\bar{\alpha%
})g(\sigma _{0}^{+},\alpha )$, where $V_{gg,0}(\bar{\alpha})=$ $G_{\sigma
^{+}}(\sigma _{0}^{+},\bar{\alpha})V_{\sigma ^{+}}G_{\sigma ^{+}}(\sigma
_{0}^{+},\bar{\alpha})^{\prime }$ is positive definite because the $m\times
m $ Jacobian matrix $G_{\sigma ^{+}}(\sigma ^{+},\alpha )$ is nonsingular
for any $\sigma ^{+}$. To see that $G_{\sigma ^{+}}(\sigma ^{+},\alpha )$ is
nonsingular, one can apply standard derivative rules (Magnus and Neudecker,
1999) obtaining: 
\begin{equation*}
G_{\sigma ^{+}}(\sigma ^{+},\alpha ):=\underset{m\times m}{\frac{\partial
g(\sigma ^{+},\alpha )}{\partial \sigma ^{+\prime }}}=\left( 
\begin{array}{c}
\frac{\partial {vech}(A_{1\bullet }\Sigma _{u}A_{1\bullet }^{\prime }-I_{k})%
}{\partial \sigma ^{+\prime }} \\ 
\frac{\partial {vec}(A_{1\bullet }\Sigma _{u,w})}{\partial \sigma ^{+\prime }%
}%
\end{array}%
\right) =\left( 
\begin{array}{c}
D_{k}^{+}\frac{\partial {vec}(A_{1\bullet }\Sigma _{u}A_{1\bullet }^{\prime
}-I_{k})}{\partial \sigma ^{+\prime }} \\ 
\frac{\partial {vec}(A_{1\bullet }\Sigma _{u,w})}{\partial \sigma ^{+\prime }%
}%
\end{array}%
\right)
\end{equation*}%
\begin{equation*}
=\left( 
\begin{array}{cc}
D_{k}^{+}\frac{\partial {vec}(A_{1\bullet }\Sigma _{u}A_{1\bullet }^{\prime
}-I_{k})}{{vech}(\Sigma _{u})^{\prime }} & D_{k}^{+}\frac{\partial {vec}%
(A_{1\bullet }\Sigma _{u}A_{1\bullet }^{\prime }-I_{k})}{{vech}(\Sigma
_{u,w})^{\prime }} \\ 
\frac{\partial {vec}(A_{1\bullet }\Sigma _{u,w})}{{vech}(\Sigma
_{u})^{\prime }} & \frac{\partial {vec}(A_{1\bullet }\Sigma _{u,w})}{{vech}%
(\Sigma _{u,w})^{\prime }}%
\end{array}%
\right)
\end{equation*}%
\begin{equation}
=\left( 
\begin{array}{cc}
D_{k}^{+}(A_{1\bullet }\otimes A_{1\bullet })D_{n} & 0 \\ 
0 & (I_{s}\otimes A_{1\bullet })%
\end{array}%
\right) .  \label{Jacobian_G_tau}
\end{equation}%
Equation (\ref{Jacobian_G_tau}) shows that $G_{\sigma ^{+}}(\sigma
^{+},\alpha )$ does not depend on $\sigma ^{+}$ and is nonsingular because $%
rank[A_{1\bullet }]=k$ (Assumption 3). Since $V_{gg,0}^{-1}(\bar{\alpha})$
is nonsingular, the condition for $Q_{0}(\alpha )$ to have a unique minimum
(of zero) in $\mathcal{N}_{\alpha _{0}}$ is that the first derivative of $%
Q_{0}(\alpha )$, given by $G_{\alpha }(\sigma _{0}^{+},\alpha )^{\prime
}V_{gg,0}^{-1}(\bar{\alpha})g(\sigma _{0}^{+},\alpha )$, satisfies the rank
condition $rank[G_{\alpha }(\sigma ^{+},\alpha )^{\prime }V_{gg,0}^{-1}(\bar{%
\alpha})]=rank[G_{\alpha }(\sigma ^{+},\alpha )]=a$ in $\mathcal{N}_{\alpha
_{0}}$. Again, by standard matrix derivative rules: 
\begin{equation*}
G_{\alpha }(\sigma ^{+},\alpha ):=\frac{\partial g(\sigma ^{+},\alpha )}{%
\partial \alpha ^{\prime }}=\text{ }\frac{\partial g(\sigma ^{+},\alpha )}{%
\partial {vec}(A_{1\bullet })^{\prime }}\times \text{ }S_{A_{1}}
\end{equation*}%
\begin{equation}
=\left( 
\begin{array}{c}
D_{k}^{+}\frac{\partial {vec}(A_{1\bullet }\Sigma _{u}A_{1\bullet }^{\prime
}-I_{k})}{\partial {vec}(A_{1\bullet })^{\prime }} \\ 
\frac{\partial {vec}(A_{1\bullet }\Sigma _{u,w})}{\partial {vec}(A_{1\bullet
})^{\prime }}%
\end{array}%
\right) S_{A_{1}}=\left( 
\begin{array}{c}
2D_{k}^{+}(A_{1\bullet }\Sigma _{u}\otimes I_{k}) \\ 
\Sigma _{w,u}\otimes I_{k}%
\end{array}%
\right) S_{A_{1}}  \label{expression_Jacobian_Ga}
\end{equation}%
which, for $\Sigma _{w,u}=\Lambda \tilde{B}_{2\bullet }$, proves the result.

\smallskip\ \newline
(ii) The restriction $a\leq m$ follows from the rank condition and the fact
that the Jacobian matrix $G_{\alpha }(\sigma ^{+},\alpha )$ is $m\times a$.
We exploit the relationship $f+a=nk$, which establishes that the number of
restrictions placed on the matrix $A_{1\bullet }$, $f$, plus the number of
free (unconstrained) parameters in the matrix $A_{1\bullet }$, $a$, equals
the total number of elements in the matrix $A_{1\bullet }$, $nk$. Since $%
s\leq n-k$, then 
\begin{equation*}
a\leq m=\frac{1}{2}k(k+1)+ks\leq \frac{1}{2}k(k+1)+k(n-k)=nk-\frac{1}{2}%
k(k-1)
\end{equation*}%
so that, for $k>1$:%
\begin{equation*}
f=nk-a\geq nk-\{nk-\frac{1}{2}k(k-1)\}=\tfrac{1}{2}k(k-1).\text{ }%
\blacksquare
\end{equation*}

\subsection{Proof of Corollary 1}

The proof follows from the fact that under sequences of models in which ${E}%
(w_{t}\tilde{\varepsilon}_{2,t}^{\prime })=\Lambda _{T}\rightarrow \Lambda $%
, if the weak proxies condition (\ref{eq weak proxxx}) holds, $\Lambda
=0_{s\times s}$ and the Jacobian $G_{\alpha }(\sigma ^{+},\alpha )\ $in (\ref%
{expression_Jacobian_Ga}) is singular. $\blacksquare $

\subsection{Proof of Proposition \protect\ref{prop 2}}

(i) To prove consistency we observe that: (a) under Assumptions 1-2 and 4,
and if the rank condition in Proposition \ref{Prop 1} holds, $Q_{0}(\alpha
):=g(\sigma _{0}^{+},\alpha )^{\prime }V_{gg,0}^{-1}(\bar{\alpha})g(\sigma
_{0}^{+},\alpha )$ is uniquely maximized at $\alpha _{0}$ in $\mathcal{N}%
_{\alpha _{0}}$; (b) $\mathcal{P}_{\alpha }$\ is compact and $\mathcal{N}%
_{\alpha _{0}}\subseteq \mathcal{P}_{\alpha }$; (c) $Q_{0}(\alpha )$ is
continuous; (d) for any $\bar{\alpha}$, $\hat{Q}_{T}(\alpha ):=g_{T}(\hat{%
\sigma}_{T}^{+},\alpha )^{\prime }\hat{V}_{gg}(\bar{\alpha})^{-1}g_{T}(\hat{%
\sigma}_{T}^{+},\alpha )$ converges uniformly in probability to $%
Q_{0}(\alpha )$. To see that (d) holds, recall that $\hat{\sigma}_{T}^{+}%
\overset{p}{\rightarrow }\sigma _{0}^{+}$ by Lemma \ref{Lemma S.1}(i), hence 
$g_{T}(\hat{\sigma}_{T}^{+},\alpha )\overset{p}{\rightarrow }$ $g(\sigma
_{0}^{+},\alpha )$ and $\hat{V}_{gg}(\bar{\alpha})\overset{p}{\rightarrow }%
V_{gg,0}$ by Slutsky's Theorem. Then, with $\left\Vert \cdot \right\Vert $
denoting the Euclidean norm, by the triangle and Cauchy-Schwartz
inequalities:%
\begin{eqnarray*}
\left\vert \hat{Q}_{T}(\alpha )-Q_{0}(\alpha )\right\vert &\leq &\left\vert
[g_{T}(\hat{\sigma}_{T}^{+},\alpha )-g(\sigma _{0}^{+},\alpha )]^{\prime }%
\hat{V}_{gg}(\bar{\alpha})^{-1}[g_{T}(\hat{\sigma}_{T}^{+},\alpha )-g(\sigma
_{0}^{+},\alpha )]\right\vert \\
&&+\left\vert g(\sigma _{0}^{+},\alpha )^{\prime }[\hat{V}_{gg}(\bar{\alpha}%
)^{-1}+\hat{V}_{gg}(\bar{\alpha})^{\prime -1}][g_{T}(\hat{\sigma}%
_{T}^{+},\alpha )-g(\sigma _{0}^{+},\alpha )]\right\vert \\
&&+\left\vert g(\sigma _{0}^{+},\alpha )^{\prime }[\hat{V}_{gg}(\bar{\alpha}%
)^{-1}-V_{gg,0}^{-1}]g(\sigma _{0}^{+},\alpha )^{\prime }\right\vert \\
&\leq &\left\Vert g_{T}(\hat{\sigma}_{T}^{+},\alpha )-g(\sigma
_{0}^{+},\alpha )\right\Vert ^{2}\left\Vert \hat{V}_{gg}(\bar{\alpha}%
)^{-1}\right\Vert \\
&&+2\left\Vert g(\sigma _{0}^{+},\alpha )\right\Vert \left\Vert g_{T}(\hat{%
\sigma}_{T}^{+},\alpha )-g(\sigma _{0}^{+},\alpha )\right\Vert \left\Vert 
\hat{V}_{gg}(\bar{\alpha})^{-1}\right\Vert \\
&&+\left\Vert g(\sigma _{0}^{+},\alpha )\right\Vert ^{2}\left\Vert \hat{V}%
_{gg}(\bar{\alpha})^{-1}-V_{gg,0}^{-1}\right\Vert
\end{eqnarray*}%
and $\sup_{\alpha \in \mathcal{P}_{\alpha }}|\hat{Q}_{T}(\alpha
)-Q_{0}(\alpha )|\overset{p}{\rightarrow }0$. Given (a), (b), (c), and (d),
the consistency result follows from Theorem 2.1 in Newey and McFadden (1994).

\smallskip

(ii) To prove asymptotic normality, we start from the first-order conditions
implied by the problem (\ref{MD_A_model}):%
\begin{equation}
G_{\alpha }(\hat{\sigma}_{T}^{+},\hat{\alpha}_{T})^{\prime }\hat{V}%
_{gg}^{-1}(\bar{\alpha})g_{T}(\hat{\sigma}_{T}^{+},\hat{\alpha}_{T})=0\text{.%
}  \label{fist-order_condition_MD_A}
\end{equation}%
By expanding $g_{T}(\hat{\sigma}_{T}^{+},\hat{\alpha}_{T})$ around $\alpha
_{0}$ and solving, yields the expression (valid in $\mathcal{N}_{\alpha
_{0}} $):%
\begin{equation*}
\{G_{\alpha }(\hat{\sigma}_{T}^{+},\hat{\alpha}_{T})^{\prime }\hat{V}%
_{gg}^{-1}(\bar{\alpha})G_{\alpha }(\hat{\sigma}_{T}^{+},\breve{\alpha}%
)\}T^{1/2}(\hat{\alpha}_{T}-\alpha _{0})
\end{equation*}%
\begin{equation}
\text{\hspace{1cm}}=-G_{\alpha }(\hat{\sigma}_{T}^{+},\hat{\alpha}%
_{T})^{\prime }\hat{V}_{gg}^{-1}(\bar{\alpha})T^{1/2}g_{T}(\hat{\sigma}%
_{T}^{+},\alpha _{0})  \label{expression_Jac}
\end{equation}%
where $\breve{\alpha}$ is a mean value. From the consistency result in (i),
as $T\rightarrow \infty $, $G_{\alpha }(\hat{\sigma}_{T}^{+},\hat{\alpha}%
_{T})\overset{p}{\rightarrow }G_{\alpha }(\sigma _{0}^{+},\alpha _{0})$ and $%
G_{\alpha }(\hat{\sigma}_{T}^{+},\breve{\alpha})\overset{p}{\rightarrow }%
G_{\alpha }(\sigma _{0}^{+},\alpha _{0})$, respectively. Moreover, the
matrix $G_{\alpha }(\sigma _{0}^{+},\alpha _{0})^{\prime }\hat{V}_{gg}^{-1}(%
\bar{\alpha})G_{\alpha }(\sigma _{0}^{+},\alpha _{0})$ is nonsingular in $%
\mathcal{N}_{\alpha _{0}}$ because of Proposition \ref{Prop 1}. It turns out
that 
\begin{eqnarray*}
&&\{G_{\alpha }(\hat{\sigma}_{T}^{+},\hat{\alpha}_{T})^{\prime }\hat{V}_{gg}(%
\bar{\alpha})^{-1}G_{\alpha }(\hat{\sigma}_{T}^{+},\breve{\alpha}%
)\}^{-1}G_{\alpha }(\hat{\sigma}_{T}^{+},\hat{\alpha}_{T})^{\prime }\hat{V}%
_{gg}^{-1}(\bar{\alpha}) \\
&&\hspace{1cm}\overset{p}{\rightarrow }\{G_{\alpha }(\sigma _{0}^{+},\alpha
_{0})^{\prime }V_{gg}(\bar{\alpha})^{-1}G_{\alpha }(\sigma _{0}^{+},\alpha
_{0})\}^{-1}G_{\alpha }(\sigma _{0}^{+},\alpha _{0})^{\prime }\hat{V}%
_{gg}^{-1}(\bar{\alpha}).
\end{eqnarray*}%
\ Under Assumptions 1, 2 and 4 and Lemma \ref{Lemma S.1}, $T^{1/2}g_{T}(\hat{%
\sigma}_{T}^{+},\alpha _{0})\overset{d}{\rightarrow }N(0_{m\times 1},V_{gg}(%
\bar{\alpha}))$. The result follows by solving (\ref{expression_Jac}) for $%
T^{1/2}(\hat{\alpha}_{T}-\alpha _{0})$\ and applying Slutsky's Theorem. $%
\blacksquare $

\subsection{Proof of Proposition \protect\ref{prop 3}}

$\hat{\mu}_{T}^{\ast }$ is a smooth function of $\hat{\sigma}%
_{T}^{+}{}^{\ast }=M_{\sigma ^{+}}\hat{\delta}_{\eta ,T}^{\ast }$, hence
from Lemma \ref{Lemma S.1}(ii) we have $\hat{\mu}_{T}^{\ast }-\hat{\mu}_{T}%
\overset{p^{\ast }}{\rightarrow }_{p}0_{o\times 1}$. It follows that $\hat{Q}%
_{T}^{\ast }(\theta ):=d(\hat{\mu}_{T}^{\ast },\theta )^{\prime }\hat{V}%
_{\mu }^{-1}d(\hat{\mu}_{T}^{\ast },\theta )=(\hat{\mu}_{T}^{\ast }-f(\theta
))^{\prime }\hat{V}_{\mu }^{-1}(\hat{\mu}_{T}^{\ast }-f(\theta ))$ satisfies 
$\hat{Q}_{T}^{\ast }(\theta )-\hat{Q}_{T}(\theta )\overset{p^{\ast }}{%
\rightarrow }_{p}0$, where $\hat{Q}_{T}(\theta ):=(\hat{\mu}_{T}-f(\theta ))%
\hat{V}_{\mu }^{-1}(\hat{\mu}_{T}-f(\theta ))$ is continuous and, for $%
\theta \in \mathcal{N}_{\theta _{0}}$ and the condition in (\ref{eq strong
proxxx}), uniquely minimized at $\hat{\theta}_{T}$ by Lemma \ref{Lemma S.4}.
Moreover, $\hat{\mu}_{T}^{\ast }-f(\theta )$ is such that ${E}^{\ast }\left[
\sup_{\theta \in \emph{P}_{\theta }}\left\Vert \hat{\mu}_{T}^{\ast
}-f(\theta )\right\Vert \right] <\infty $; then, the result $\hat{\theta}%
_{T}^{\ast }-\hat{\theta}_{T}$ $\overset{p^{\ast }}{\rightarrow }%
_{p}0_{q_{\theta }\times 1}$ follows from Theorem 2.6 in Newey and McFadden
(1994) and Assumption 1.

The first-order conditions associated with the minimization problem in
equation (\ref{CMD-MBB}) are given by%
\begin{equation}
J_{\hat{\theta}_{T}^{\ast }}^{\prime }\hat{V}_{\mu }^{-1}(\hat{\mu}%
_{T}^{\ast }-f(\hat{\theta}_{T}^{\ast }))=0  \label{FOC_boot}
\end{equation}%
where $J_{\hat{\theta}_{T}^{\ast }}^{\prime }$ is the Jacobian in (\ref%
{Jacob_theta}) evaluated at the MBB-CMB estimator $\hat{\theta}_{T}^{\ast }$%
. By a mean-value expansion of $f(\hat{\theta}_{T}^{\ast })$ about $\hat{%
\theta}_{T}$, we obtain%
\begin{equation*}
f(\hat{\theta}_{T}^{\ast })=f(\hat{\theta}_{T})+J_{\dot{\theta}}(\hat{\theta}%
_{T}^{\ast }-\hat{\theta}_{T})
\end{equation*}%
where $\dot{\theta}$ is an intermediate vector value between $\hat{\theta}%
_{T}^{\ast }$ and $\hat{\theta}_{T}$. Using the above expansion in (\ref%
{FOC_boot}) yields%
\begin{equation*}
J_{\hat{\theta}_{T}^{\ast }}^{\prime }\hat{V}_{\mu }^{-1}(\hat{\mu}%
_{T}^{\ast }-f(\hat{\theta}_{T})-J_{\dot{\theta}}(\hat{\theta}_{T}^{\ast }-%
\hat{\theta}_{T}))=0,
\end{equation*}%
hence, for $f(\hat{\theta}_{T})=\hat{\mu}_{T}$, it holds:%
\begin{equation*}
J_{\hat{\theta}_{T}^{\ast }}^{\prime }\hat{V}_{\mu }^{-1}(\hat{\mu}%
_{T}^{\ast }-\hat{\mu}_{T})-J_{\hat{\theta}_{T}^{\ast }}^{\prime }\hat{V}%
_{\mu }^{-1}J_{\dot{\theta}}(\hat{\theta}_{T}^{\ast }-\hat{\theta}_{T})=0
\end{equation*}%
namely%
\begin{equation}
\{J_{\hat{\theta}_{T}^{\ast }}^{\prime }\hat{V}_{\mu }^{-1}J_{\dot{\theta}%
}\}T^{1/2}(\hat{\theta}_{T}^{\ast }-\hat{\theta}_{T})=J_{\hat{\theta}%
_{T}^{\ast }}^{\prime }\hat{V}_{\mu }^{-1}T^{1/2}(\hat{\mu}_{T}^{\ast }-\hat{%
\mu}_{T}).  \label{relationship_boot}
\end{equation}%
Equation (\ref{relationship_boot})\ links the asymptotic distribution of $%
T^{1/2}(\hat{\theta}_{T}^{\ast }-\hat{\theta}_{T})$, conditional on the
data, to the asymptotic distribution of $T^{1/2}(\hat{\mu}_{T}^{\ast }-\hat{%
\mu}_{T})$, conditional on the data, and to the local rank properties of the
Jacobian matrix $J_{\theta }$. If for $\theta \in \mathcal{N}_{\theta _{0}}$
the proxies are strong in the sense of equation (\ref{eq strong proxxx})
then, conditionally on the data, the asymptotic normality of $T^{1/2}(\hat{%
\mu}_{T}^{\ast }-\hat{\mu}_{T})$ in (\ref{relationship_boot}) follows from
the asymptotic normality of $T^{1/2}(\hat{\sigma}_{T}^{+}{}^{\ast }-\hat{%
\sigma}_{T}^{+})$ which is guaranteed by Lemma \ref{Lemma S.1}(ii).
Moreover, as $\hat{\theta}_{T}^{\ast }-\hat{\theta}_{T}=o_{p}^{\ast }(1)$,
in probability, then, in probability, $J_{\hat{\theta}_{T}^{\ast }}-J_{\hat{%
\theta}_{T}}=o_{p}^{\ast }(1)$, $J_{\dot{\theta}}-J_{\hat{\theta}%
_{T}}=o_{p}^{\ast }(1)$ and, accordingly, $J_{\hat{\theta}_{T}^{\ast
}}^{\prime }\hat{V}_{\mu }^{-1}J_{\dot{\theta}}-J_{\hat{\theta}_{T}}^{\prime
}\hat{V}_{\mu }^{-1}J_{\hat{\theta}_{T}}=o_{p}^{\ast }(1)$, in probability,
where the $q_{\theta }\times q_{\theta }$ matrix $J_{\hat{\theta}%
_{T}}^{\prime }\hat{V}_{\mu }^{-1}J_{\hat{\theta}_{T}}$ is positive definite
asymptotically. This proves the result. $\blacksquare $

\subsection{Proof of Proposition \protect\ref{prop 4}}

If for $\theta \in \mathcal{N}_{\theta _{0}}$ the proxies satisfy the weak
proxies condition in equation (\ref{eq weak proxxx}), $T^{1/2}(\hat{\mu}%
_{T}-\mu _{0})$ is not asymptotically Gaussian because of the non-normality
of $T^{1/2}(\hat{\omega}_{T}-\omega _{0})$ established in Lemma \ref{Lemma
S.3}. We now show that also $T^{1/2}(\hat{\omega}_{T}^{\ast }-\hat{\omega}%
_{T})$, the bootstrap counterpart of $T^{1/2}(\hat{\omega}_{T}-\omega _{0})$%
, is not asymptotically Gaussian. In light of (\ref{relationship_boot}),
this suffices to claim that $T^{1/2}(\hat{\theta}_{T}^{\ast }-\hat{\theta}%
_{T})$ is not asymptotically Gaussian.

Notice that $\hat{\omega}_{T}^{\ast }=\omega (\hat{\sigma}_{T}^{+}{}^{\ast
}) $, the function $\omega (\cdot )$ being smooth. From Lemma \ref{Lemma S.1}%
(ii), $\hat{\sigma}_{T}^{+}{}^{\ast }-\hat{\sigma}_{T}^{+}{}\overset{p^{\ast
}}{\rightarrow }_{p}0$, in probability, hence also $\hat{\omega}_{T}^{\ast }-%
\hat{\omega}_{T}=o_{p}^{\ast }(1)$, in probability. The result holds
regardless of the strength of the instruments. Consider ($T$ times) the
quadratic expansion of $\hat{\omega}_{T}^{\ast }=\omega (\hat{\sigma}%
_{T}^{+}{}^{\ast })$ around $\hat{\sigma}_{T}^{+}$:%
\begin{equation}
T\left( \hat{\omega}_{T}^{\ast }-\hat{\omega}_{T}\right) =T^{1/2}J_{\sigma
^{+}}^{(1)}(\hat{\sigma}_{T}^{+})T^{1/2}(\hat{\sigma}_{T}^{+}{}^{\ast }-\hat{%
\sigma}_{T}^{+})+\tfrac{T}{2}R_{1,T}(\ddot{\sigma}_{T}^{+}{}^{\ast })
\label{eq exp for bootstrap omega weak instruments}
\end{equation}%
where $J_{\sigma ^{+}}^{(1)}(\hat{\sigma}_{T}^{+}):=\left. \frac{\partial
\omega }{\partial \sigma ^{+\prime }}\right\vert _{\sigma ^{+}=\hat{\sigma}%
_{T}^{+}}$ and the remainder term $R_{1,T}(\ddot{\sigma}_{T}^{+}{}^{\ast })$
has representation%
\begin{eqnarray*}
&&TR_{1,T}(\ddot{\sigma}_{T}^{+}{}^{\ast })\overset{}{:=}(I_{o_{1}}\otimes
T^{1/2}(\hat{\sigma}_{T}^{+}{}^{\ast }-\hat{\sigma}_{T}^{+})^{\prime
})H^{(1)}(\ddot{\sigma}_{T}^{+}{}^{\ast })T^{1/2}(\hat{\sigma}%
_{T}^{+}{}^{\ast }-\hat{\sigma}_{T}^{+})\text{,} \\
&&H^{(1)}(\ddot{\sigma}_{T}^{+}{}^{\ast })\overset{}{:=}\left. \frac{%
\partial }{\partial \sigma ^{+\prime }}{vec}\left( \frac{\partial \omega }{%
\partial \sigma ^{+\prime }}\right) ^{\prime }\right\vert _{\sigma ^{+}=%
\ddot{\sigma}_{T}^{+}{}^{\ast }},
\end{eqnarray*}%
$\ddot{\sigma}_{T}^{+}{}^{\ast }$ being an intermediate vector value between 
$\hat{\sigma}_{T}^{+}{}^{\ast }$ and $\hat{\sigma}_{T}^{+}$ (note that aside
from transposition, the matrix $H^{(1)}(\ddot{\sigma}_{T}^{+}{}^{\ast })$
above is the same as in (\ref{Hessian_partial_block})). We now show that the
cdf of $T\left( \hat{\omega}_{T}^{\ast }-\hat{\omega}_{T}\right) $,
conditionally on the data, converges in distribution (rather than converging
in probability) to a random cdf. That is, the (conditional) bootstrap
measure is random in the limit; see Cavaliere and Georgiev (2020).
Randomness essentially arises because of the limit behavior of the Jacobian $%
T^{1/2}J_{\sigma ^{+}}^{(1)}(\hat{\sigma}_{T}^{+})$: specifically, while in
the original non-bootstrap world it holds $T^{1/2}J_{\sigma
^{+}}^{(1)}(\sigma _{0}^{+})\rightarrow J^{(1)}$ (see the proof of Lemma \ref%
{Lemma S.3}), its analog in the bootstrap world, $T^{1/2}J_{\sigma
^{+}}^{(1)}(\hat{\sigma}_{T}^{+})$, does not converges to a constant.

First, from Lemma \ref{Lemma S.1}(ii), $T^{1/2}(\hat{\sigma}_{T}^{+}{}^{\ast
}-\hat{\sigma}_{T}^{+})\overset{d^{\ast }}{\rightarrow }_{p}\mathbb{G}%
_{\sigma ^{+}}^{\ast }\equiv N\left( 0,V_{\sigma ^{+}}\right) $. Moreover,
by continuity of second derivatives and by using the fact that $\hat{\sigma}%
_{T}^{+}=\sigma _{0}^{+}+o_{p}\left( 1\right) $, $H^{(1)}(\ddot{\sigma}%
_{T}^{+}{}^{\ast })\overset{p^{\ast }}{\rightarrow }_{p}H^{(1)}(\sigma
_{0}^{+})$ and hence 
\begin{equation*}
TR_{1,T}(\ddot{\sigma}_{T}^{+}{}^{\ast })\overset{d^{\ast }}{\rightarrow }%
_{p}(I_{o_{1}}\otimes \mathbb{G}_{\sigma ^{+}}^{\ast \prime })H_{\sigma
_{0}^{+}}^{(1)}\mathbb{G}_{\sigma ^{+}}^{\ast }\text{ }
\end{equation*}%
where $H_{\sigma _{0}^{+}}^{(1)}:=H^{(1)}(\sigma _{0}^{+})$. Consider now $%
T^{1/2}J_{\sigma ^{+}}^{(1)}(\hat{\sigma}_{T}^{+})$. By an expansion of ${vec%
}J_{\sigma ^{+}}^{(1)}(\hat{\sigma}_{T}^{+})$ around the true value ${vec}%
J_{\sigma ^{+}}^{(1)}(\sigma _{0}^{+})$, we obtain: 
\begin{equation*}
T^{1/2}{vec}J_{\sigma ^{+}}^{(1)}(\hat{\sigma}_{T}^{+})=T^{1/2}{vec}%
J_{\sigma ^{+}}^{(1)}(\sigma _{0}^{+})+H^{(1)}(\ddot{\sigma}_{T}^{+}{}^{\ast
})T^{1/2}(\hat{\sigma}_{T}^{+}-\sigma _{0}^{+}).
\end{equation*}%
From $\hat{\sigma}_{T}^{+}-\sigma _{0}^{+}=o_{p}\left( 1\right) $ and
continuity of the Hessian, $H^{(1)}(\ddot{\sigma}_{T}^{+}{}^{\ast
})\rightarrow H_{\sigma _{0}^{+}{}}^{(1)}$. This result, together with $%
T^{1/2}(\hat{\sigma}_{T}^{+}{}^{\ast }-\hat{\sigma}_{T}^{+})\overset{d^{\ast
}}{\rightarrow }_{p}N\left( 0,V_{\sigma ^{+}}\right) $ (Lemma \ref{Lemma S.1}%
(i)) and $T^{1/2}{vec}J_{\sigma ^{+}}^{(1)}(\sigma _{0}^{+})\rightarrow {vec}%
J^{(1)}$ (proof of Lemma \ref{Lemma S.2}), implies that\textbf{\ } 
\begin{equation*}
{vec}(\mathbb{G}_{J^{(1)}}):=T^{1/2}{vec}J_{\sigma ^{+}}^{(1)}(\hat{\sigma}%
_{T}^{+})\overset{d}{\rightarrow }N({vec}J^{(1)},\text{ }H_{\sigma
^{+}{}}^{(1)}V_{\sigma ^{+}}H_{\sigma ^{+}{}}^{(1)\prime })
\end{equation*}%
where $\mathbb{G}_{J^{(1)}}$ denotes a Gaussian matrix, implicitly defined.
Notice that albeit the covariance matrix $H_{\sigma ^{+}{}}^{(1)}V_{\sigma
^{+}}H_{\sigma ^{+}{}}^{(1)\prime }$is of reduced rank (see the proof of
Lemma \ref{Lemma S.3}), it is a not zero matrix. In summary, 
\begin{equation}
T\left( \hat{\omega}_{T}^{\ast }-\hat{\omega}_{T}\right) =\underset{\overset{%
d}{\rightarrow }\mathbb{G}_{J^{(1)}}}{\underbrace{T^{1/2}J_{\sigma
^{+}}^{(1)}(\hat{\sigma}_{T}^{+})}}\underset{\overset{d^{\ast }}{\rightarrow 
}_{p}\mathbb{G}_{\sigma ^{+}}^{\ast }}{\underbrace{T^{1/2}(\hat{\sigma}%
_{T}^{+}{}^{\ast }-\hat{\sigma}_{T}^{+})}}\text{ }+\underset{\overset{%
d^{\ast }}{\rightarrow }_{p}(I_{o_{1}}\otimes \mathbb{G}_{\sigma ^{+}}^{\ast
\prime })H_{\sigma _{0}^{+}}^{(1)}\mathbb{G}_{\sigma ^{+}}^{\ast }}{\tfrac{1%
}{2}\underbrace{R_{1,T}(\ddot{\sigma}_{T}^{+}{}^{\ast }).}}
\label{eq limits}
\end{equation}%
Because the term $T^{1/2}J_{\sigma ^{+}}^{(1)}(\hat{\sigma}_{T}^{+})$ does
not converge in probability to a constant but rather (in distribution) to a
random variable, the limit distribution of $T\left( \hat{\omega}_{T}^{\ast }-%
\hat{\omega}_{T}\right) $ is random in the limit. Specifically, the limit
can be described as a mixture of a Gaussian random variable $\mathbb{G}%
_{\sigma ^{+}}^{\ast }$ and the $\chi ^{2}$-type random variable $\left(
I_{o_{1}}\otimes \mathbb{G}_{\sigma ^{+}}^{\ast \prime }\right) H_{\sigma
_{0}^{+}}^{(1)}\mathbb{G}_{\sigma ^{+}}^{\ast }$, where the weight $\mathbb{G%
}_{\sigma ^{+}}^{\ast }$ is a random matrix (fixed across bootstrap
repetitions) and, precisely, distributed as $\mathbb{G}_{J^{(1)}}$. Put
differently, 
\begin{equation}
T\left( \hat{\omega}_{T}^{\ast }-\hat{\omega}_{T}\right) \overset{d^{\ast }}{%
\rightarrow }_{w}\mathbb{G}_{J^{(1)}}\mathbb{G}_{\sigma ^{+}}^{\ast }+\tfrac{%
1}{2}(I_{o_{1}}\otimes \mathbb{G}_{\sigma ^{+}}^{\ast \prime })H_{\sigma
_{0}^{+}}^{(1)}\mathbb{G}_{\sigma ^{+}}^{\ast }\big|\mathbb{G}_{J^{(1)}}
\label{eq weak convergence in distribution}
\end{equation}%
where the notation `$X_{T}^{\ast }\overset{d^{\ast }}{\rightarrow }_{w}X|G$'
indicates weak convergence in distribution; see, e.g., Appendix A in
Cavaliere and Georgiev (2020). The formal proof of (\ref{eq weak convergence
in distribution}) can be obtained by considering the bootstrap statistic $%
\mathbb{A}_{T}^{\ast }:=(T^{1/2}(\hat{\sigma}_{T}^{+}{}^{\ast }-\hat{\sigma}%
_{T}^{+})^{\prime },\tfrac{1}{2}TR_{T}(\ddot{\sigma}_{T}^{+}{}^{\ast
})^{\prime })^{\prime }$, such that $T\left( \hat{\omega}_{T}^{\ast }-\hat{%
\omega}_{T}\right) $ can be written as $h(T^{1/2}J_{\sigma ^{+}}^{(1)}(\hat{%
\sigma}_{T}^{+}),\mathbb{A}_{T}^{\ast })$ where $h$ is a continuous
transformation. We know from above that $T^{1/2}J_{\sigma ^{+}}^{(1)}(\hat{%
\sigma}_{T}^{+})\overset{d}{\rightarrow }\mathbb{G}_{J^{(1)}}$ and $\mathbb{A%
}_{T}^{\ast }\overset{d^{\ast }}{\rightarrow }_{p}\mathbb{A}:=(\mathbb{G}%
_{\sigma ^{+}}^{\ast \prime },((I_{o_{1}}\otimes \mathbb{G}_{\sigma
^{+}}^{\ast \prime })H_{\sigma _{0}^{+}}^{(1)}\mathbb{G}_{\sigma ^{+}}^{\ast
})^{\prime })^{\prime }$ (jointly). Then, we can consider a special
probability space where $\mathbb{G}_{J^{(1)}}$ and $\mathbb{A}$ are defined
and, for every sample size $T$, also the original and the bootstrap data can
be redefined, maintaining their distribution (we also maintain the
notation), such that (jointly) $T^{1/2}J_{\sigma ^{+}}^{(1)}(\hat{\sigma}%
_{T}^{+})\rightarrow _{a.s.}\mathbb{G}_{J^{(1)}}$ and $\mathbb{A}_{T}^{\ast }%
\overset{d^{\ast }}{\rightarrow }_{a.s.}\mathbb{A}$. Then, by Lemma A.3 in
Cavaliere and Georgiev (2020)\ we have that $(T^{1/2}J_{\sigma ^{+}}^{(1)}(%
\hat{\sigma}_{T}^{+}),\mathbb{A}_{T}^{\ast })\overset{d^{\ast }}{\rightarrow 
}_{a.s.}(\mathbb{G}_{J^{(1)}},\mathbb{A})|\mathbb{G}_{J^{(1)}}$ and, by a
continuous mapping theorem for a.s. weak convergence (Theorem 10 of
Sweeting, 1989), $T\left( \hat{\omega}_{T}^{\ast }-\hat{\omega}_{T}\right) 
\overset{w^{\ast }}{\rightarrow }_{a.s.}h(\mathbb{G}_{J^{(1)}},\mathbb{A})|%
\mathbb{G}_{J^{(1)}}$ on the special probability space. This implies the
desired result (\ref{eq weak convergence in distribution}) on a general
probability space. $\blacksquare $

\subsection{Proof of Proposition \protect\ref{prop 5}}

Given the distance defined in equation (\ref{eq def SNTB}), we consider the
following decomposition of $\tau _{T,N}^{\ast }(x)$: 
\begin{eqnarray}
&&N^{1/2}\hat{U}_{T}(x)^{-1/2}(\digamma _{T,N}^{\ast }(x)-\digamma _{%
\mathcal{G}}\left( x\right) )  \label{eq tau deco} \\
&&\hspace{1cm}\overset{}{=}N^{1/2}\hat{U}_{T}(x)^{-1/2}(\digamma
_{T,N}^{\ast }(x)-\digamma _{T}^{\ast }(x))  \notag \\
&&\hspace{1.3cm}\overset{}{+}N^{1/2}\hat{U}_{T}(x)^{-1/2}(\digamma
_{T}^{\ast }(x)-\digamma _{\mathcal{G}}\left( x\right) ).  \notag
\end{eqnarray}%
and provide the proof considering $\hat{U}_{T}(x):=\digamma _{T}^{\ast
}(x)(1-\digamma _{T}^{\ast }(x))$. First, the term $N^{1/2}\hat{U}%
_{T}(x)^{-1/2}(\digamma _{T,N}^{\ast }(x)-\digamma _{T}^{\ast }(x))$
converges to the normal distribution when $N\rightarrow \infty $ for any $%
T>1 $, since $\digamma _{T,N}^{\ast }(x)-\digamma _{T}^{\ast }(x)$ is the
(standardized)\ sum of (conditionally) i.i.d. indicators and hence, by the
Berry-Esseen bound, $\sup_{u\in \mathbb{R}}|P^{\ast }(N^{1/2}(\digamma
_{T,N}^{\ast }(x)-\digamma _{T}^{\ast }(x))\leq u)-\digamma _{\mathcal{G}%
}\left( u\right) |\leq CN^{-1/2}$ (a.s.), with $C$ a constant.

Second, since by assumption $\digamma _{T}^{\ast }(x)-\digamma _{\mathcal{G}%
}\left( x\right) $ admits a standard Edgeworth expansion such that $\digamma
_{T}^{\ast }(x)-\digamma _{\mathcal{G}}\left( x\right) $ $=O_{p}\left(
T^{-1/2}\right) $ uniformly in $x$, the second term on the right-hand side
in (\ref{eq tau deco}) is of order $O_{p}\left( N^{1/2}T^{-1/2}\right) $.
Hence, the desired result follows when $N,T\rightarrow \infty ,$ provided $%
N=o(T)$. $\blacksquare $

\subsection{Proof of Proposition \protect\ref{prop 6}}

Under the weak proxies condition, by Proposition \ref{prop 4}, ${plim}%
_{T\rightarrow \infty }\digamma _{T}^{\ast }(x)\neq \digamma _{\mathcal{G}%
}\left( x\right) $, which means that the second term on the right hand side
of (\ref{eq tau deco}) does not vanishes asymptotically, implying that $\tau
_{T,N}^{\ast }(x)$ diverges at the rate of $N^{1/2}$\textbf{\ }as\textbf{\ }$%
N,T\rightarrow \infty $. $\blacksquare $

\subsection{Proof of Proposition \protect\ref{prop 7}}

Let $\mathcal{D}_{T}$ denote the original data upon which the proxy-SVAR is
estimated, defined on the probability space $(\boldsymbol{%
\mathbb{Q}
},\mathcal{F},P)$. As is standard, the bootstrap (conditional)\ cdf $%
F_{T}^{\ast }(x):=P(\hat{\theta}_{T}^{\ast }\leq x|\mathcal{D}_{T})$ is a
function of the data only. Using $F_{T}^{\ast }(\cdot )$, we generate a set
of $N$ i.i.d. `bootstrap' random variables as follows. First, let $%
U_{b}^{\ast }$, $b=1,\ldots ,N$, be a sequence of i.i.d. $U[0,1]$ random
variables independent on the data (we implicitly extend the original
probability space such that it includes the $U_{b}^{\ast }$'s as well).
Then, the bootstrap random variables $\hat{\theta}_{T:b}^{\ast }$, $%
b=1,\ldots ,N$ that enter the argument of the statistic $\tau _{T,N}^{\ast
}:=\tau (\hat{\theta}_{T:1}^{\ast },\ldots ,\hat{\theta}_{T:N}^{\ast })$ are
defined as $\hat{\theta}_{T:b}^{\ast }:=F_{T}^{\ast -1}(U_{b}^{\ast })$, $%
b=1,\ldots ,N$, where $F_{T}^{\ast -1}(\cdot )$ is the generalized inverse
of $F_{T}^{\ast }(\cdot )$. Thus, we have 
\begin{equation*}
\tau _{T,N}^{\ast }=\tau (\hat{\theta}_{T:1}^{\ast },\ldots ,\hat{\theta}%
_{T:N}^{\ast })=\tau (F_{T}^{\ast -1}(U_{1}^{\ast }),\ldots ,F_{T}^{\ast
-1}(U_{N}^{\ast }))
\end{equation*}%
with cdf, conditional on $\mathcal{D}_{T}$, given by $\mathcal{H}%
_{T,N}(x)=P(\tau _{T,N}^{\ast }\leq x|\mathcal{D}_{T})$.

We now prove that $\rho _{T}$, where $\rho _{T}$ is a function of the
original data, and $\tau _{T,N}^{\ast }$ are independent asymptotically, in
the sense that for any $x_{1},x_{2}\in \mathbb{R}$, as $T,N\rightarrow
\infty $, the condition in equation (\ref{asymptotic_independence}), here
reported for convenience 
\begin{equation}
P(\{\rho _{T}\leq x_{1}\}\cap \{\tau _{T,N}^{\ast }\leq x_{2}\})-P(\rho
_{T}\leq x_{1})P(\tau _{T,N}^{\ast }\leq x_{2})\rightarrow 0
\label{eq independence}
\end{equation}%
holds. Observe that (\ref{eq independence}) trivially holds in the presence
of weak proxies because, by Proposition \ref{prop 4}, $\tau _{T,N}^{\ast }$
diverges for $N,T\rightarrow \infty $. In the presence of strong proxies,
Proposition \ref{prop 3}(i) ensures that as $T,N\rightarrow \infty $, $%
\mathcal{H}_{T,N}(x)\rightarrow _{p}\mathcal{H}(x)$, where $x\in \mathbb{R}$
and $\mathcal{H}(x)$ is a non-random cdf. By the law of iterated
expectations (and the fact that $P\left( X\in \mathcal{E}\right) =${${E}$}$(%
\mathbb{I}_{\{X\in \mathcal{E}\}})$), we have%
\begin{eqnarray*}
P(\{\rho _{T} &\leq &x_{1}\}\cap \{\tau _{T,N}^{\ast }\leq x_{2}\})={E}(%
\mathbb{I}_{\{\rho _{T}\leq x_{1}\}\cap \{\tau _{T,N}^{\ast }\leq x_{2}\}})={%
E}(\mathbb{I}_{\{\rho _{T}\leq x_{1}\}}\mathbb{I}_{\{\tau _{T,N}^{\ast }\leq
x_{2}\}}) \\
&=&{E}\left( {E}(\mathbb{I}_{\{\rho _{T}\leq x_{1}\}}\mathbb{I}_{\{\tau
_{T,N}^{\ast }\leq x_{2}\}}|\mathcal{D}_{T})\right) \\
&=&{E}\left( \mathbb{I}_{\{\rho _{T}\leq x_{1}\}}{E}(\mathbb{I}_{\{\tau
_{T,N}^{\ast }\leq x_{2}\}}|\mathcal{D}_{T})\right) \\
&=&{E}\left( \mathbb{I}_{\{\rho _{T}\leq x_{1}\}}\mathcal{H}%
_{T,N}(x_{2})\right) \\
&=&{E}\left( \mathbb{I}_{\{\rho _{T}\leq x_{1}\}}\mathcal{H}(x_{2})\right) +{%
E}\left( \mathbb{I}_{\{\rho _{T}\leq x_{1}\}}(\mathcal{H}_{T,N}(x_{2})-%
\mathcal{H}(x_{2}))\right) \\
&=&P(\rho _{T}\leq x_{1})\mathcal{H}(x_{2})+{E}\left( \mathbb{I}_{\{\rho
_{T}\leq x_{1}\}}(\mathcal{H}_{T,N}(x_{12})-\mathcal{H}(x_{2}))\right) .
\end{eqnarray*}%
For the last term, we have 
\begin{eqnarray*}
\left\vert {E}\left( \mathbb{I}_{\{\rho _{T}\leq x_{1}\}}(\mathcal{H}%
_{T,N}(x_{2})-\mathcal{H}(x_{2}))\right) \right\vert &\leq &{E}\left\vert 
\mathbb{I}_{\{\rho _{T}\leq x_{1}\}}(\mathcal{H}_{T,N}(x_{2})-\mathcal{H}%
(x_{2}))\right\vert \\
&\leq &{E}\left\vert (\mathcal{H}_{T,N}(x_{2})-\mathcal{H}%
(x_{2}))\right\vert .
\end{eqnarray*}%
Since we know that under strong proxies $\mathcal{H}_{T,N}(x_{2})\rightarrow
_{p}\mathcal{H}(x_{2})$, then {${E}$}$|\mathcal{H}_{T,N}(x_{2})-\mathcal{H}%
(x_{2})|\rightarrow 0$ provided $|\mathcal{H}_{T,N}(x_{2})-\mathcal{H}%
(x_{2})|$ is uniformly integrable. But $\mathcal{H}_{T,N}(x_{2})$ and $%
\mathcal{H}(x_{2})$ are cdfs, and hence they are both bounded and uniformly
integrable. Hence, as $T,N\rightarrow \infty $, 
\begin{equation*}
P(\{\rho _{T}\leq x_{1}\}\cap \{\tau _{T,N}^{\ast }\leq x_{2}\})-P(\rho
_{T}\leq x_{1})\mathcal{H}(x_{2})=o_{p}\left( 1\right) .
\end{equation*}%
Therefore,%
\begin{equation*}
P(\{\rho _{T}\leq x_{1}\}\cap \{\tau _{T,N}^{\ast }\leq x_{2}\})-P(\rho
_{T}\leq x_{1})P(\tau _{T,N}^{\ast }\leq x_{2})
\end{equation*}%
\begin{equation*}
=P(\{\rho _{T}\leq x_{1}\}\cap \{\tau _{T,N}^{\ast }\leq x_{2}\})-P(\rho
_{T}\leq x_{1})\mathcal{H}(x_{2})
\end{equation*}%
\begin{equation*}
\text{ \ \ \ \ \ \ \ \ \ \ \ \ \ \ \ \ \ \ }+P(\rho _{T}\leq x_{1})\left( 
\mathcal{H}(x_{2})-P(\tau _{T,N}^{\ast }\leq x_{2}\right) )
\end{equation*}%
\begin{equation*}
=P(\rho _{T}\leq x_{1})\left( \mathcal{H}(x_{2})-P(\tau _{T,N}^{\ast }\leq
x_{2}\right) )+o_{p}\left( 1\right) .
\end{equation*}%
Since $P(\rho _{T}\leq x_{1})\in \lbrack 0,1]$, we only need to prove that $%
P(\tau _{T,N}^{\ast }\leq x_{2})-\mathcal{H}(x_{2})$ vanishes
asymptotically. But this immediately follows from bootstrap consistency as 
\begin{eqnarray*}
P(\tau _{T,N}^{\ast } &\leq &x_{2})-\mathcal{H}(x_{2})={E}(\mathbb{I}%
_{\{\tau _{T,N}^{\ast }\leq x_{2}\}})-\mathcal{H}(x_{2}) \\
&=&{E}({E}(\mathbb{I}_{\{\tau _{T,N}^{\ast }\leq x_{2}\}}|\mathcal{D}_{T}))-%
\mathcal{H}(x_{2}) \\
&=&{E}\left( \mathcal{H}_{T,N}(x_{2})-\mathcal{H}(x_{2})\right) \rightarrow 0
\end{eqnarray*}%
by the uniform integrability of $\mathcal{H}_{T,N}(x_{2})$. $\blacksquare $
\ 

\section{Indirect-MD approach: identification restrictions on $B_{\bullet 1}$%
}

\label{Section_supplementary_restrictions_on_B1}Section \ref%
{Section_indirect_approach_MD} discusses the case in which in the multiple
shocks framework, $k>1$, the additional restrictions necessary for the
identification of the proxy-SVAR are placed on the parameters in the matrix $%
A_{1\bullet }$, see equation (\ref{structural_sub_A_bis}). In some cases,
however, the reference specification of the proxy-SVAR might be based on the
representation in equations (\ref{partition_B})-(\ref{moments_B_form}), and
the additional restrictions necessary to point-identify the model might
involve the parameters in the matrix $B_{\bullet 1}$, not those in $%
A_{1\bullet } $, i.e., the parameters $\alpha $. For instance, in Section %
\ref{Section_empirical_illustration_uncertainty}, the additional restriction 
$\beta _{F,M}=0$ is placed on $B_{\bullet 1}$. Recall that since $B_{\bullet
1}=\Sigma _{u}A_{1\bullet }^{\prime }$ (see (\ref%
{crucial_relationship_partial})), we can easily switch from one
representation to the other and map restrictions from parameters in $%
B_{\bullet 1}$ into parameters in $A_{1\bullet } $, and vice versa.

In this section we adapt the indirect-MD estimation approach discussed in
Section \ref{Section_indirect_approach_MD} to the case in which the
additional identifying restrictions involve the parameters in $B_{\bullet 1}$%
. These restrictions can be represented in the form: 
\begin{equation}
{vec}(B_{\bullet 1})=S_{B_{1}}\beta _{1}+s_{B_{1}}  \label{restrictions_B1}
\end{equation}%
where $\beta _{1}$ is the vector of (free) structural parameters in $%
B_{\bullet 1}$ and $S_{B_{1}}$ and $s_{B_{1}}$ have the same role as $%
S_{A_{1}}$and $s_{A_{1}}$\ in equation (\ref{restrictions_A1}),
respectively.\ Using (\ref{crucial_relationship_partial}), the moment
conditions in (\ref{moments_A1_variance2}) and (\ref{moments_A1}) can be
mapped into the expressions:%
\begin{equation}
B_{\bullet 1}^{\prime }\Sigma _{u}^{-1}B_{\bullet 1}=I_{k},
\label{monets_B1_variance}
\end{equation}%
\begin{equation}
B_{\bullet 1}^{\prime }\Omega _{u,w}=0_{k\times s}  \label{moments_B1}
\end{equation}%
where $\Omega _{u,w}:=\Sigma _{u}^{-1}\Sigma _{u,w}$. Under the restrictions
(\ref{restrictions_B1}), we can summarize the moment conditions (\ref%
{monets_B1_variance})-(\ref{moments_B1}) by the distance function:%
\begin{equation}
g^{o}(\omega ^{+},\beta _{1}):=\left( 
\begin{array}{c}
{vech}(B_{\bullet 1}^{\prime }\Sigma _{u}^{-1}B_{\bullet 1}-I_{k}) \\ 
{vec}(B_{\bullet 1}^{\prime }\Omega _{u,w})%
\end{array}%
\right)  \label{distance_mapped_to_B1}
\end{equation}%
where $\omega ^{+}:=({vech}(\Sigma _{u})^{\prime },{vec}(\Omega
_{u,w})^{\prime })^{\prime }$. Recall that $B_{\bullet 1}$ depends on $\beta
_{1}$ through (\ref{restrictions_B1}). Obviously, at the true parameter
values, $g^{o}(\omega ^{+},\beta _{1})=0_{m\times 1}$. The MD estimator of $%
\beta _{1}$ obtains from:%
\begin{equation}
\hat{\beta}_{1,T}:=\arg \min_{\beta _{1}\in \mathcal{P}_{\beta _{1}}}\hat{Q}%
_{T}^{o}(\beta _{1})\text{ , \ }\hat{Q}_{T}^{o}(\beta _{1}):=g_{T}^{o}(\hat{%
\omega}_{T}^{+},\beta _{1})^{\prime }\hat{V}_{gg}(\bar{\beta}%
_{1})^{-1}g_{T}^{o}(\hat{\omega}_{T}^{+},\beta _{1})
\label{MD_problem_mapped_B1}
\end{equation}%
where $g_{T}^{o}(\cdot ,\cdot )$ denotes the function $g^{o}(\cdot ,\cdot )$
once $\omega ^{+}$ is replaced with $\hat{\omega}_{T}^{+}$, $\mathcal{P}%
_{\beta _{1}}$ is the parameter space, $\hat{V}_{gg}(\bar{\beta}%
_{1}):=G_{\omega ^{+}}(\hat{\omega}_{T}^{+},\bar{\beta}_{1})\hat{V}_{\omega
^{+}}G_{\omega ^{+}}(\hat{\omega}_{T}^{+},\bar{\beta}_{1})^{\prime },$ $\hat{%
V}_{\omega ^{+}}$ is a consistent estimator of $V_{\omega ^{+}}$; finally, $%
G_{\omega ^{+}}(\omega ^{+},\beta _{1})$ is the $m\times m$ Jacobian matrix
defined by $G_{\omega ^{+}}(\omega ^{+},\beta _{1}):=\frac{\partial
g^{o}(\omega ^{+},\beta _{1})}{\partial \omega ^{+\prime }}$; $\bar{\beta}%
_{1}$ (interior point of $\mathcal{P}_{\beta _{1}}$) is some preliminary
estimate of $\beta _{1}$.

Under Assumptions 1-\ref{Assn 4}, the asymptotic properties of $\hat{\beta}%
_{1,T}$ are the same as those of the estimator $\hat{\alpha}_{T}$ discussed
in Section \ref{Section_indirect_approach_MD}. The IRFs of interest are
directly obtained from (\ref{IRF_j}). Given $\hat{\Sigma}_{u}$, the implied
estimate of $A_{1\bullet }$ follows from equation (\ref%
{crucial_relationship_partial}).

\section{Comparison with IV}

\label{Section_supplementary_IV_comparison}In this section we compare the MD
estimation approach presented in Section \ref{Section_indirect_approach_MD}
with its most natural frequentist alternative, represented by the IV
estimation method based on VAR residuals.

Assume that $k>1$ (multiple target shocks) and, for simplicity, that the
matrix $A_{1,1}$ in equation (\ref{structural_sub_A_bis})\ is nonsingular.
Note that this condition is not implied by Assumption 3; hence, the
nonsingularity of $A_{1,1}$ is not necessary in the MD approach. With $%
A_{1,1}$ nonsingular, one can write $A_{1\bullet }=A_{1,1}(I_{k}$ $,-\Psi )$%
, $\Psi :=-A_{1,1}^{-1}A_{1,2}$, and system (\ref{structural_sub_A_bis}) can
be represented as the multivariate regression model 
\begin{equation}
u_{1,t}=\Psi u_{2,t}+A_{1,1}^{-1}\varepsilon _{1,t}\text{, \ }t=1,\ldots ,T.
\label{system_structural_equations_A}
\end{equation}%
In some applications, (\ref{system_structural_equations_A}) can be
interpreted as a system of policy reaction functions; see, e.g., Caldara and
Kamps (2017) and Section \ref{Section_supplementary_fiscal_proxy-SVAR}.
Under Assumptions 1-2, the VAR disturbances $u_{1,t}$ and $u_{2,t}$ can be
replaced with the corresponding VAR residuals $\hat{u}_{1,t}$ and $\hat{u}%
_{2,t}$, $t=1,\ldots ,T$, and (\ref{system_structural_equations_A}) can be
written, for large $T$, as 
\begin{equation}
\hat{u}_{1,t}=\Psi \hat{u}_{2,t}+\xi _{t}\text{ \ \ , \ }t=1,\ldots ,T
\label{regression_IV_transformed}
\end{equation}%
where $\xi _{t}:=A_{1,1}^{-1}\varepsilon _{1,t}+o_{p}(1)$ is a disturbance
term with covariance matrix $\Theta =A_{1,1}^{-1}(A_{1,1}^{-1})^{\prime }$.

Consider the special case in which there exists proxies $w_{t}$ for all $s$
non-target shocks in $\varepsilon _{2,t}$, i.e. $\tilde{\varepsilon}%
_{2,t}\equiv \varepsilon _{2,t}$, $s=n-k$.\footnote{%
The IV estimation of system (\ref{system_structural_equations_A}) becomes
slightly more involving when $s<n-k$. With $s<n-k,$ it is necessary to
impose at least $n-k-s$ restrictions on the parameters in $\Psi $ in system (%
\ref{regression_IV_transformed}).} In this scenario, one can estimate the
parameters in the matrix $\Psi :=-A_{1,1}^{-1}A_{1,2}$ by IV using the
proxies $w_{t}$ as instruments for the (generated)\ regressors $\hat{u}%
_{2,t} $. This produces the IV estimator $\hat{\Psi}_{IV}$ and the IV
residuals $\hat{\xi}_{t}:=\hat{u}_{1,t}-\hat{\Psi}_{IV}\hat{u}_{2,t}$, $%
t=1,\ldots ,T$, which in turn can be used to estimate the covariance matrix $%
\Theta $: $\hat{\Theta}_{IV}=\frac{1}{T}\sum_{t=1}^{T}\hat{\xi}_{t}\hat{\xi}%
_{t}^{\prime }$. Given the IV estimators $\hat{\Psi}_{IV}$ and $\hat{\Theta}%
_{IV}$, the elements in $A_{1,1}$ and $A_{1,2}$ can be separately identified
only if $A_{1,1}$ is upper (lower) triangular, other than nonsingular. If $%
A_{1,1}$ is upper (lower) triangular, the estimated Choleski factor of $\hat{%
\Theta}_{IV}$ equals $\hat{A}_{11}^{-1}$. This implies the imposition of $%
\frac{1}{2}k(k-1)$ supplementary constraints on the proxy-SVAR parameters,
which guarantee exact point-identification.

The MD approach developed in Section \ref{Section_indirect_approach_MD} does
not involve `generated regressors' and is more flexible than the IV
approach: (i)\ the matrix $A_{1,1}$ in $A_{1\bullet }=(A_{1,1}$ $,$ $%
A_{1,2}) $ does not necessarily must be invertible or triangular;\ (ii)
point-identification is achieved under the general conditions of Proposition %
\ref{Prop 1}; hence, $A_{1,1}$ does not need to satisfy the requirements of
being upper or lower triangular.

\section{MBB algorithm}

\label{Section_Supplementary_MBB_algorithm}In this section we summarize Br%
\"{u}ggemann {\emph{et al.}} (2016)'s MBB\ algorithm. The reference model is
the proxy-SVAR represented in Section \ref{Section_Model}. The reference
proxy-SVAR\ model can be represented as in (\ref{large_B_SVAR}) and the
reduced form parameters of (\ref{large_B_SVAR}) are collected in the vector $%
\delta :=(\delta _{\psi }^{\prime },\delta _{\eta }^{\prime })^{\prime }$.\
Given (\ref{large_B_SVAR}), we consider the algorithm that follows.

\medskip\ 

\textsc{Algorithm (residual-based MBB)}

\begin{description}
\item[1.] Fit the reduced form VAR model in (\ref{large_B_SVAR}) to the data 
$W_{1},\ldots ,W_{T}$ and, given the estimates $\hat{\Psi}_{1},\ldots ,\hat{%
\Psi}_{l}$, compute the innovation residuals $\hat{\eta}_{t}=W_{t}-\hat{\Psi}%
_{1}W_{t-1}-\ldots -\hat{\Psi}_{l}W_{t-l}$ and the covariance matrix $\hat{%
\Sigma}_{\eta }:=\frac{1}{T}\sum_{t=1}^{T}\hat{\eta}_{t}\hat{\eta}%
_{t}^{\prime }$;

\item[2.] Choose a block of length $\ell <T$ and let $\mathcal{B}:=[T/\ell ]$
be the number of blocks such that $\mathcal{B}\ell \geq T$. Define the $%
\mathcal{M}\times \ell $ blocks $\mathcal{M}_{i,\ell }:=(\hat{\eta}%
_{i+1},\ldots ,\hat{\eta}_{i+\ell })$, $i=0,1,2,\ldots ,T-\ell $.

\item[3.] Let $i_{0},i_{1}$, ...,$i_{\mathcal{B}-1}$ be an i.i.d. random
sample of the elements of the set $\left\{ 0,1,2,\ldots ,T-\ell \right\} .$
Lay blocks $\mathcal{M}_{i_{0},\ell },\mathcal{M}_{i_{1},\ell },\ldots ,%
\mathcal{M}_{i_{\mathcal{B}-1},\ell }$ end-to-end and discard the last $%
\mathcal{B}\ell -T$ values, obtaining the residuals $\hat{\eta}_{1}^{\ast
},\ldots ,\hat{\eta}_{T}^{\ast };$

\item[4.] Center the residuals $\hat{\eta}_{1}^{\ast },\ldots ,\hat{\eta}%
_{T}^{\ast }$ according to the rule%
\begin{equation*}
e_{j\ell +e}^{\ast }:=\hat{\eta}_{j\ell +e}^{\ast }-{E}^{\ast }(\hat{\eta}%
_{j\ell +e}^{\ast })=\hat{\eta}_{j\ell +e}^{\ast }-\tfrac{1}{T-\ell +1}%
\sum\limits_{g=0}^{T-\ell }\hat{\eta}_{e+g}^{\ast }
\end{equation*}%
for $e=1,2,\ldots ,\ell $ and $j=0,1,2,\ldots ,\mathcal{B}-1,$ such that ${E}%
^{\ast }(e_{t}^{\ast })=0$ for $t=1,\ldots ,T;$

\item[5.] Generate the bootstrap sample $W_{1}^{\ast },W_{2}^{\ast },\ldots
,W_{T}^{\ast }$ recursively by solving, for $t=1,\ldots ,T$, the system 
\begin{equation}
W_{t}^{\ast }=\hat{\Psi}_{1}W_{t-1}^{\ast }+\ldots +\hat{\Psi}%
_{l}W_{t-l}^{\ast }+e_{t}^{\ast }  \label{system_boot}
\end{equation}%
with initial condition $W_{0}^{\ast },W_{-1}^{\ast },\ldots ,W_{1-p}^{\ast }$
set to the pre-fixed sample values $W_{0},W_{-1},\ldots ,W_{1-p}$;

\item[6.] Use the sample $W_{1}^{\ast },W_{2}^{\ast },\ldots ,W_{T}^{\ast }$
generated in the previous step to compute the bootstrap estimators of the
reduced form parameters $\hat{\delta}_{T}^{\ast }:=(\hat{\delta}_{\psi
,T}^{\ast \prime },\hat{\delta}_{\eta ,T}^{\ast \prime })^{\prime }$.
\end{description}

Once $\hat{\delta}_{T}^{\ast }$ is obtained from the algorithm above, the
bootstrap estimators $\hat{\mu}_{T}^{\ast }:=({vech}(\hat{\Omega}_{v}^{\ast
})^{\prime },{vec}(\hat{\Sigma}_{v,u}^{\ast })^{\prime })^{\prime }$
considered in the paper follow accordingly. See footnote 18 in the paper for
the practical rule we use to set the block length parameter $\ell $ in the
Monte Carlo experiments and the empirical illustrations considered in the
paper.

\section{Data generating process}

\label{Section_Supplement_DGP}In this section we summarize the DGP used for
the Monte Carlo experiments discussed in Section 6.3, and summarized in
Table 1 and Figure 1, respectively.

Data are generated from the following three-equation SVAR with one lag and
no deterministic component:%
\begin{equation}
Y_{t}=\Pi _{1}Y_{t-1}+u_{t}\text{\ , \ }t=1,\ldots ,T  \label{VAR1_DGP}
\end{equation}%
where:%
\begin{equation*}
\Pi _{1}:=\left( 
\begin{array}{ccc}
0.67 & -0.12 & 0.42 \\ 
0.03 & 0.43 & 0.08 \\ 
0.14 & 0.02 & 0.58%
\end{array}%
\right) \text{, \ }\lambda _{\max }(\Pi _{1})=0.86
\end{equation*}%
and $\lambda _{\max }(\cdot )$ denotes the largest eigenvalue (in absolute
value) of the matrix in the argument, and%
\begin{equation*}
\underset{u_{t}}{\underbrace{\left( 
\begin{array}{c}
u_{t}^{A} \\ 
u_{t}^{B} \\ 
u_{t}^{C}%
\end{array}%
\right) }}=\underset{B}{\underbrace{\left( 
\begin{array}{ccc}
0.196 & 0 & 0.19 \\ 
0.210 & 0.16 & -0.32 \\ 
0.017 & 0 & 0.09%
\end{array}%
\right) }}\underset{\varepsilon _{t}}{\underbrace{\left( 
\begin{array}{c}
\varepsilon _{t}^{A} \\ 
\varepsilon _{t}^{B} \\ 
\varepsilon _{t}^{C}%
\end{array}%
\right) }}\text{ }
\end{equation*}%
\begin{equation*}
\varepsilon _{t}:=\left( 
\begin{array}{c}
\varepsilon _{t}^{A}\equiv \varepsilon _{1,t} \\ 
\varepsilon _{t}^{B}\equiv \varepsilon _{2,t}^{1} \\ 
\varepsilon _{t}^{C}\equiv \tilde{\varepsilon}_{2,t}%
\end{array}%
\right) 
\begin{array}{l}
\text{{\small target shock}} \\ 
\text{{\small non-instrumented non-target shock}} \\ 
\text{{\small instrumented non-target shock}}%
\end{array}%
\text{ }
\end{equation*}%
\begin{equation*}
B_{\bullet 1}:=\left( 
\begin{array}{c}
0.196 \\ 
0.210 \\ 
0.017%
\end{array}%
\right)
\end{equation*}%
which imply 
\begin{equation*}
A_{1\bullet }=(\alpha _{1,1},\alpha _{1,2},\alpha _{1,3})=(6.246,0,-13.185).
\end{equation*}

\smallskip

\noindent \textsc{Results in Table 1. }The rejection frequencies of the test
of relevance reported in Table 1 are computed assuming that a proxy $w_{t}$
instruments the non-target shock $\varepsilon _{t}^{C}\equiv \tilde{%
\varepsilon}_{2,t}$ through the following linear measurement error model: 
\begin{equation}
w_{t}=\lambda \tilde{\varepsilon}_{2,t}+\omega _{w,t}\text{ , \ }\omega
_{w,t}:=\sigma _{w}er_{w,t}\text{, \ }er_{w,t}\perp \varepsilon _{t}\text{ \ 
}  \label{w_process}
\end{equation}%
where $er_{w,t}$ is a measurement error with zero mean and variance 1, and $%
\lambda $ is relevance parameter and is restricted as discussed below. By
defining $W_{t}:=(Y_{t}^{\prime },w_{t})^{\prime }$ and $\eta
_{t}:=(u_{t}^{\prime },w_{t})^{\prime }$, the analog of the proxy-SVAR
representation in (\ref{large_B_SVAR}) is given by 
\begin{equation*}
\underset{W_{t}}{\underbrace{\left( 
\begin{array}{c}
Y_{t} \\ 
w_{t}%
\end{array}%
\right) }}=\underset{\Psi _{1}}{\underbrace{\left( 
\begin{array}{cc}
\Pi _{1} & 0 \\ 
0 & 0%
\end{array}%
\right) }}\underset{W_{t-1}}{\underbrace{\left( 
\begin{array}{c}
Y_{t-1} \\ 
w_{t-1}%
\end{array}%
\right) }}+\underset{\eta _{t}}{\underbrace{\left( 
\begin{array}{c}
u_{t} \\ 
w_{t}%
\end{array}%
\right) }}\text{\ , \ }t=1,\ldots ,T
\end{equation*}%
with%
\begin{equation*}
\left( 
\begin{array}{c}
u_{t} \\ 
w_{t}%
\end{array}%
\right) =\left( 
\begin{array}{cc}
B & 0 \\ 
(0,0,\lambda ) & \sigma _{w}%
\end{array}%
\right) \left( 
\begin{array}{c}
\varepsilon _{t} \\ 
er_{w,t}%
\end{array}%
\right) .
\end{equation*}%
The term $\xi _{t}=(\varepsilon _{t}^{\prime }$, $er_{w,t})^{\prime }$ in
the expression above is generated as follows. In one case, $\xi _{t}\sim
iidN(0_{4\times 1},I_{4})$. In the other case, each component $\xi _{i,t}$ ($%
1\leq i\leq 4$) of $\xi _{t}$, with $\xi _{i,t}$ independent of $\xi _{j,s}$%
, all $i\neq j,t,s$, is a GARCH (1,1) process:%
\begin{eqnarray*}
\xi _{i,t} &=&\varsigma _{i,t}\xi _{i,t}^{0}\text{,}\ \ \ \xi _{i,t}^{0}\sim
iidN(0,1), \\
\varsigma _{i,t}^{2} &=&\varrho _{0}+\varrho _{1}\xi _{i,t-1}^{2}+\varrho
_{2}\varsigma _{i,t-1}^{2},\text{ }t=1,\ldots ,T
\end{eqnarray*}%
where $\varrho _{1}:=0.05$, $\varrho _{2}:=0.93$ and $\varrho
_{0}:=(1-\varrho _{1}-\varrho _{2})$.

In the `strong proxy' scenario considered in the upper panel of Table 1, the
relevance parameter $\lambda $ is set to the value $\lambda =0.8$ (and is
therefore independent on $T$); the implied correlation between the proxy and
the instrumented shocks is: 
\begin{equation*}
corr(w_{t},\tilde{\varepsilon}_{2,t})=\frac{\lambda }{(\lambda ^{2}+\sigma
_{w}^{2})^{1/2}}=0.588.
\end{equation*}%
In the `moderately weak proxy' scenario considered in the middle panel of
Table 1, $\lambda :=c/T^{1/2}$, with $c$ such that, for $T=250$,%
\begin{equation*}
corr(w_{t},\tilde{\varepsilon}_{2,t})=\frac{c/T^{1/2}}{(\frac{c^{2}}{T}%
+\sigma _{w}^{2})^{1/2}}=0.25;
\end{equation*}%
for $T=1,000$, the correlation becomes $0.13$. Finally, in the `weak proxy'
scenario (lower panel of Table 1), $c$ is such that for $T=250$, $corr(w_{t},%
\tilde{\varepsilon}_{2,t})=0.05$; for $T=1,000$ the correlation reduces to $%
0.03$.

~\smallskip

\noindent \textsc{Results in Figure 1}. Figure 1 in the paper plots actual
empirical coverage probabilities of 90\% confidence intervals built, in
samples of length $T=250$, for the response of the variable $Y_{3,t+h}$ to
the target shock $\varepsilon _{1,t}$, $h=0,1,\ldots ,12.$

In the indirect-MD approach, the dynamic causal effects produced by the
target shock $\varepsilon _{1,t}$ are recovered by estimating the structural
equation 
\begin{equation*}
A_{1\bullet }u_{t}=\alpha _{1,1}u_{1,t}+\alpha _{1,2}u_{2,t}+\alpha
_{1,3}u_{3,t}=\varepsilon _{1,t}
\end{equation*}%
using the proxy $w_{t}$ as instrument for the shock $\varepsilon
_{t}^{C}\equiv \tilde{\varepsilon}_{2,t}$ and the method discussed in
Section \ref{Section_indirect_approach_MD}. The restrictions $\alpha
_{1,2}=0 $ is correctly imposed by the econometrician in estimation.

In the `direct' approach, we consider a proxy $z_{t}$ for the target shock $%
\varepsilon _{t}^{A}\equiv \varepsilon _{1,t}$ that is `weak' in the sense
of equation (\ref{eq weak proxxx}). More precisely, the linear measurement
error model for $z_{t}$ is given by the equation%
\begin{equation*}
z_{t}=\tfrac{c}{T^{1/2}}\varepsilon _{1,t}+\omega _{z,t}\text{ \ , \ }\omega
_{z,t}:=\sigma _{z}er_{z,t}\text{, \ }\omega _{z,t}\perp \varepsilon _{t}
\end{equation*}%
with $\xi _{t}=(\varepsilon _{t}^{\prime }$, $er_{z,t})^{\prime }\sim
iidN(0_{4\times 1},I_{4})$. In this case, for $T=250$, $corr(z_{t},%
\varepsilon _{1,t})=0.045$. The dynamic causal effects and associated
weak-identification robust confidence intervals are inferred using Montiel
Olea {\emph{et al.}} (2021)'s weak-instrument robust approach.

\section{Another empirical illustration: US fiscal multipliers from a fiscal
proxy-SVAR}

\label{Section_supplementary_fiscal_proxy-SVAR}Fiscal multipliers are key
statistics for understanding how fiscal policy changes stimulate (or
contract) the economy. There is a large debate in the empirical literature
on the size of fiscal multipliers, especially the size and uncertainty
surrounding the tax multiplier, see Ramey (2019). This lack of consensus
also characterizes studies based on fiscal proxy-SVARs, see, e.g., Mertens
and Ravn (2014), Caldara and Kamps (2017) and Lewis (2021).

Using fiscal proxies for fiscal shocks, Mertens and Ravn (2014) uncover a
large tax multiplier on the period 1950-2006 and show that the tax
multiplier is larger than the fiscal spending multiplier. Conversely, using
non-fiscal proxies for non-fiscal shocks, Caldara and Kamps (2017) identify
fiscal multipliers through a Bayesian penalty function approach and the
estimation of fiscal reaction functions. Their analysis yields conflicting
outcomes relative to Mertens and Ravn (2014). Lewis (2021) exploits the
heteroskedasticity found in the data nonparametrically, and reports fiscal
multipliers only partially consistent with Mertens and Ravn (2014) and
Caldara and Kamps (2017).

In this section, we revisit the empirical evidence on fiscal multipliers
with our indirect-MD approach. This requires, as in Caldara and Kamps
(2017), the identification of a fiscal proxy-SVAR by using proxies for the
non-fiscal (non-target)\ shocks of the system.

We employ a VAR model for the variables $%
Y_{t}:=(TAX_{t},G_{t},GDP_{t},RR_{t})^{\prime }$ ($n=4$). Here, $TAX_{t}$
represents per capita real tax revenues, $G_{t}$ denotes per capita real
government spending, $GDP_{t}$ is per capita real output, and $RR_{t}$ is
the (ex-post) real interest rate, computed as $RR_{t}:=R_{t}-\pi _{t}$ where 
$R_{t}$ represents a short-term nominal interest rate and $\pi _{t}$ denotes
the inflation rate. The incorporation of the ex-post real interest rate in
the system allows us to concurrently capture both the short-term nominal
interest rate and the inflation rate, all while maintaining a manageable
dimensionality of the system. We use quarterly data spanning from 1950:Q1 to
2006:Q4 ($T=228$ quarterly observations), all sourced from Caldara and Kamps
(2017), where a more extensive explanation of the dataset can be found. The
time series are logarithmically transformed and linearly detrended. The
reduced form VAR includes $p=4$ lags and a constant term. Although we do not
report standard residual-based diagnostic tests here to save space, they
indicate that the VAR disturbances exhibit no serial correlation but display
conditional heteroskedasticity.

Let $\varepsilon _{1,t}:=(\varepsilon _{t}^{tax},\varepsilon
_{t}^{g})^{\prime }$ be the vector of target structural shocks ($k=2$),
where $\varepsilon _{t}^{tax}$ denotes the tax shock and $\varepsilon
_{t}^{g}$ the fiscal spending shock. The non-target shocks of the model are
collected in the vector $\varepsilon _{2,t}:=(\varepsilon
_{t}^{y},\varepsilon _{t}^{mp})^{\prime }$ ($n-k=2$), where $\varepsilon
_{t}^{y}$ denotes an output shock and $\varepsilon _{t}^{mp}$ can be
interpreted likewise a monetary policy shock. The analogue of the
representation in equation (\ref{partition_B}) is given by the system:%
\begin{equation}
\underset{u_{t}}{\underbrace{\left( 
\begin{array}{c}
u_{t}^{tax} \\ 
u_{t}^{g} \\ 
u_{t}^{y} \\ 
u_{t}^{rr}%
\end{array}%
\right) }}=\underset{B_{\bullet 1}}{\underbrace{\left( 
\begin{array}{ll}
\beta _{tax,tax} & \beta _{tax,g} \\ 
\beta _{g,tax} & \beta _{g,g} \\ 
\beta _{y,tax} & \beta _{y,g} \\ 
\beta _{rr,tax} & \beta _{rr,g}%
\end{array}%
\right) }}\underset{\varepsilon _{1,t}}{\underbrace{\left( 
\begin{array}{c}
\varepsilon _{t}^{tax} \\ 
\varepsilon _{t}^{g}%
\end{array}%
\right) }}+B_{\bullet 2}\underset{\varepsilon _{2,t}}{\underbrace{\left( 
\begin{array}{c}
\varepsilon _{t}^{y} \\ 
\varepsilon _{t}^{mp}%
\end{array}%
\right) }}  \label{B_partition_fiscal}
\end{equation}%
where $u_{t}$ is the vector of VAR innovations, and $\beta _{y,tax}$ and $%
\beta _{y,g}$ are the coefficients that capture the on-impact responses of
output to the tax shock and the fiscal spending shock, respectively. Since $%
k=2>1$, it is necessary to impose at least $\frac{1}{2}k(k-1)=1$ additional
restriction on the parameters to point-identify the model through external
instruments; see below. Once the parameters in $B_{\bullet 1}$ in (\ref%
{B_partition_fiscal}) are identified, fiscal multipliers follow from
properly scaling the responses of output to the identified fiscal shocks. In
particular, dynamic fiscal multipliers can be defined as\footnote{%
These definitions correspond to those used in, e.g., Angelini {\emph{et al.}}
(2023) and to the `alternative definition' considered in Caldara and Kamps
(2017), see their Section 5. Caldara and Kamps (2017) and Angelini {\emph{et
al.}} (2023) show that differences are not empirically relevant. Other
definitions, see, e.g., Ramey (2011), are equally possible.} 
\begin{equation}
\mathbb{M}_{h,tax}:=\frac{\beta _{y,tax}^{h}}{\beta _{tax,tax}}\times
Sc_{y,tax}\text{ \ , \ \ }\mathbb{M}_{h,g}:=\frac{\beta _{y,g}(h)}{\beta
_{g,g}}\times Sc_{y,g}\text{ \ },\text{ }h=0,1,\ldots
\label{impact_multipliers}
\end{equation}%
where $\beta _{y,tax}^{h}:=\beta _{y,tax}(h):=\frac{\partial GDP_{t+h}}{%
\partial \varepsilon _{t}^{tax}}$ is the dynamic response of tax revenues to
the tax shock after $h$ periods, $\beta _{tax,tax}\equiv \beta _{tax,tax}(0)$%
, $\beta _{y,g}^{h}:=\beta _{y,g}(h):=\frac{\partial GDP_{t+h}}{\partial
\varepsilon _{t}^{g}}$ and $\beta _{g,g}\equiv \beta _{g,g}(0)$ are defined
accordingly, and $Sc_{y,tax}$ and $Sc_{y,g}$ are scaling factors which serve
to convert the dynamic structural responses into US dollars.

In the next two sections we re-visit the direct approach to the
identification of fiscal multipliers (\ref%
{Section_supplementary_fiscal_direct}), and then explore the advantages of
the indirect-MD approach (\ref{Section_supplementary_fiscal_indirect}).

\subsection{Direct approach}

\label{Section_supplementary_fiscal_direct}The `direct' external variables
approach hinges on the availability of (at least) two proxies for the two
target shocks in $\varepsilon _{1,t}:=(\varepsilon _{t}^{tax},\varepsilon
_{t}^{g})^{\prime }$. We consider two proxies for the fiscal shocks ($r=k=2$%
) collected in the vector $z_{t}:=(z_{t}^{tax},z_{t}^{g})^{\prime }$ where,
as in Mertens and Ravn (2014), $z_{t}^{tax}$ is a time series of
unanticipated tax changes built upon Romer and Romer's (2010) narrative
records on tax policy decisions, and $z_{t}^{g}$ is Ramey's (2011) narrative
measure of expected exogenous changes in military spending. The counterpart
of the linear measurement system in equation (\ref{equation_link}) is given
by the system:%
\begin{equation}
\underset{z_{t}}{\underbrace{\left( 
\begin{array}{c}
z_{t}^{tax} \\ 
z_{t}^{g}%
\end{array}%
\right) }}=\underset{\Phi }{\underbrace{\left( 
\begin{array}{cc}
\varphi _{tax,tax} & 0 \\ 
0 & \varphi _{g,g}%
\end{array}%
\right) }}\underset{\varepsilon _{1,t}}{\underbrace{\left( 
\begin{array}{c}
\varepsilon _{t}^{tax} \\ 
\varepsilon _{t}^{g}%
\end{array}%
\right) }}+\underset{\omega _{t}}{\underbrace{\left( 
\begin{array}{c}
\omega _{t}^{tax} \\ 
\omega _{t}^{g}%
\end{array}%
\right) }}  \label{link_fiscal_direct}
\end{equation}%
where $\omega _{t}:=(\omega _{t}^{tax},\omega _{t}^{g})^{\prime }$ is a
vector of measurement errors uncorrelated with the structural shocks $%
\varepsilon _{t}$. The matrix $\Phi $ in (\ref{link_fiscal_direct}) is
specified diagonal, to capture the idea that the proxy $z_{t}^{tax}$ solely
instruments the tax shock (through the parameter $\varphi _{tax,tax}$), and
the proxy $z_{t}^{g}$ solely instruments the fiscal spending shock (through
the parameter $\varphi _{g,g}$). Notably, the diagonal structure assumed for 
$\Phi $ in (\ref{link_fiscal_direct})\ provides two restrictions on the
proxy-SVAR\ parameters that would in principle suffice to (over-)identify
the model if the proxies were strong in the sense of equation (\ref{eq
strong proxxx}); see Angelini and Fanelli (2019). Actually, below we show
that the zero restrictions on the off-diagonal terms of $\Phi $ are not
effectively exploited in the construction of weak-instrument robust
confidence sets for the fiscal multipliers using the proxies $%
z_{t}:=(z_{t}^{tax},z_{t}^{g})^{\prime }$; additional types of restrictions
are necessary to build weak-instrument robust confidence sets.

We proceed by assuming that the instruments in $z_{t}$ are potentially weak
proxies for the target structural shocks $\varepsilon _{1,t}$. Following
Montiel Olea {\emph{et al.}} (2021), we build weak-instrument confidence
sets for the simultaneous response of real output to the tax and fiscal
spending shocks, respectively. To simplify exposition and without loss of
generality, we now pretend that the VAR for $%
Y_{t}:=(TAX_{t},G_{t},GDP_{t},RR_{t})^{\prime }$ features only one lag,
which implies the VAR\ companion matrix coincides with the autoregressive
coefficients, i.e. $\mathcal{C}_{y}\equiv \Pi _{1}=\Pi $; the arguments that
follow can be easily extended to the case of our VAR model which features $%
p=4$ lags.

We consider the null hypothesis that at the horizon $h$, the simultaneous
response of real output to the fiscal shocks is equal to the values $\beta
_{y,tax}^{h}=\beta _{y,tax}(h)$ and $\beta _{y,g}^{h}=\beta _{y,g}(h)$ (see (%
\ref{impact_multipliers})), respectively, i.e., 
\begin{equation}
\gamma _{GDP,\varepsilon _{1,t}}(h):=\left( \frac{\partial GDP_{t+h}}{%
\partial \varepsilon _{t}^{tax}}\text{ , }\frac{\partial GDP_{t+h}}{\partial
\varepsilon _{t}^{g}}\right) =e_{4,3}^{\prime }(\Pi )^{h}B_{\bullet
1}=(\beta _{y,tax}^{h}\text{ , }\beta _{y,g}^{h})  \label{IRF_robust}
\end{equation}%
where $e_{4,3}^{\prime }:=(0,0,1,0)$ is the selection vector that picks out
the real output variable from the vector $Y_{t}$. Assuming constant scaling
factors $Sc_{y,tax}$ and $Sc_{y,g}$ , for given values $(\beta _{y,tax}^{h}$
, $\beta _{y,g}^{h})$ the multipliers $\mathbb{M}_{h,tax}$ and $\mathbb{M}%
_{h,g}$ can be easily computed from (\ref{impact_multipliers}). Moreover, by
post-multiplying both sides of equation (\ref{IRF_robust}) by $\Phi ^{\prime
}$ and using the covariance restriction $\Sigma _{u,z}=B_{\bullet 1}\Phi
^{\prime }$, we obtain the relationship 
\begin{equation}
e_{4,3}^{\prime }(\Pi )^{h}\Sigma _{u,z}-(\beta _{y,tax}^{h}\text{ , }\beta
_{y,g}^{h})\Phi ^{\prime }=(0,0)  \label{IRF_restrictions_robust}
\end{equation}%
which can be used to construct asymptotic valid confidence sets for $\beta
_{y,tax}^{h}$ and $\beta _{y,g}^{h}$ (hence, of their scaled counterparts, $%
\mathbb{M}_{h,tax}$ and $\mathbb{M}_{h,g}$)\ through test inversion.

To invert a test for the null hypothesis that the responses in (\ref%
{IRF_restrictions_robust}) are equal to the values $\beta _{y,tax}^{h}$ and $%
\beta _{y,g}^{h}$, consider the \emph{additional restrictions} $%
B_{1,1}=B_{1,1}^{0}$, where recall that $B_{1,1}$ is the $k\times k$ upper
block of the matrix of on-impact coefficients $B_{\bullet
1}=(B_{1,1}^{\prime },B_{2,1}^{\prime })^{\prime }$ (see the partition in
equation (\ref{partition_B})),\ and $B_{1,1}^{0}$ contains known values. The
restriction $B_{1,1}=B_{1,1}^{0}$ implies $k^{2}=4$ constraints on $%
B_{\bullet 1}$. Using $B_{1,1}=B_{1,1}^{0}$ and the representation (\ref%
{partition_B}), the proxy-SVAR\ moment conditions can be decomposed as: 
\begin{equation}
\left( 
\begin{array}{c}
\Sigma _{u_{1},z} \\ 
\Sigma _{u_{2},z}%
\end{array}%
\right) =\left( 
\begin{array}{c}
B_{1,1}^{0}\Phi ^{\prime } \\ 
B_{2,1}\Phi ^{\prime }%
\end{array}%
\right)  \label{moment_conditions_B_weak}
\end{equation}%
where it is seen that the reduced form covariance matrix $\Sigma _{u,z}$ has
been partitioned in the two blocks $\Sigma _{u_{1},z}$ and $\Sigma
_{u_{2},z} $, respectively, each of dimensions $2\times 2$. By solving the
first two equations in (\ref{moment_conditions_B_weak}) for $\Phi ^{\prime }$
gives:%
\begin{equation}
\Phi _{p}^{\prime }:=\left( B_{1,1}^{0}\right) ^{-1}\Sigma _{u_{1},z}\equiv
\left( B_{1,1}^{0}\right) ^{-1}(I_{k}\text{ },\text{ }0_{k\times
(n-k)})\Sigma _{u,z}  \label{Phi_p_second}
\end{equation}%
where the notation `$\Phi _{p}$' used in place of `$\Phi $' in (\ref%
{Phi_p_second}) simply\ remarks that the matrix of relevance parameters now
depends on the on-impact responses fixed in $B_{1,1}^{0}$. Expression (\ref%
{Phi_p_second}) suggests that a plug-in estimator of $\Phi _{p}^{\prime }$
is given by $\hat{\Phi}_{p}^{\prime }:=\left( B_{1,1}^{0}\right) ^{-1}(I_{k}$
$,$ $0_{k\times (n-k)})\hat{\Sigma}_{u,z}$. Hence, provided the restrictions 
$B_{1,1}=B_{1,1}^{0}$ hold in the DGP, the estimator $\hat{\Phi}_{p}$ is
consistent under the conditions of Lemma \ref{Lemma S.1}, regardless of the
strength of the proxies. Note that, as it stands, the estimator $\hat{\Phi}%
_{p}^{\prime }:=\left( B_{1,1}^{0}\right) ^{-1}(I_{k}$ $,$ $0_{k\times
(n-k)})\hat{\Sigma}_{u,z}$\ does not explicitly incorporate the diagonal
structure postulated for $\Phi $ in (\ref{link_fiscal_direct}).

Let $\kappa :=({vec}(\Pi )^{\prime },{vec}(\Sigma _{u,z})^{\prime })^{\prime
}$ be the vector containing the reduced form proxy-SVAR\ parameters; let $%
\kappa _{0}$ be the corresponding true value and $\hat{\kappa}_{T}$ the
estimator of $\kappa $; $\kappa $ is a function of the parameters $\delta $,
see Section \ref{Section_supplementary_Lemmas}. Then, by Lemma \ref{Lemma
S.1}, under Assumptions 1--2, $T^{1/2}(\hat{\kappa}_{T}-\kappa _{0})\overset{%
d}{\rightarrow }N(0,V_{\kappa })$, where $V_{\kappa }$ follows from a
delta-method argument. This result is valid regardless of the strength of
the proxies. Using the expression in (\ref{Phi_p_second}) for $\Phi
_{p}^{\prime }$, and taking the ${vec}$ of both terms in equation (\ref%
{IRF_restrictions_robust}), the null hypothesis that $\beta _{y,tax}^{h}$
and $\beta _{y,g}^{h}$ are the true responses at horizon $h$ can be
re-stated as 
\begin{equation*}
S(\kappa _{0},\beta _{y,tax}^{h}\text{,}\beta _{y,g}^{h},B_{1,1}^{0})={vec}%
\left\{ e_{4,3}^{\prime }(\Pi )^{h}\Sigma _{u,z}-(\beta _{y,tax}^{h}\text{ , 
}\beta _{y,g}^{h})\Phi _{p}^{\prime }\right\} =0_{2\times 1}.
\end{equation*}%
Then, by a simple delta-method argument it follows that:%
\begin{equation*}
T^{1/2}S(\hat{\kappa}_{T},\beta _{y,tax}^{h}\text{ , }\beta
_{y,g}^{h},B_{1,1}^{0})\overset{d}{\rightarrow }N(0_{2\times 1},V_{S})
\end{equation*}%
where $V_{S}$ is a covariance matrix that depends on $V_{\kappa }$. A valid $%
\nu $-level test for the null hypothesis that $(\beta _{y,tax}^{h}$ , $\beta
_{y,g}^{h})$ are the true responses rejects whenever 
\begin{equation}
T\times S(\hat{\kappa}_{T},\beta _{y,tax}^{h}\text{ , }\beta
_{y,g}^{h},B_{1,1}^{0})^{\prime }\hat{V}_{S}^{-1}S(\hat{\kappa}_{T},\beta
_{y,tax}^{h}\text{ , }\beta _{y,g}^{h},B_{1,1}^{0})>\chi _{2,1-\nu }^{2},
\label{Wald_S_region}
\end{equation}%
where $\hat{V}_{S}$ is a consistent estimator of $V_{S}$\ and $\chi
_{2,1-\nu }^{2}$ is the $(1-\nu )100\%$ quantile of the $\chi ^{2}$
distribution with $2$ degrees of freedom. An asymptotically valid
weak-instrument robust confidence set for $\beta _{y,tax}^{h}$ and $\beta
_{y,g}^{h}$ with asymptotic coverage $1-\nu $ will contain all postulated
responses $(\beta _{y,tax}^{h}$,$\beta _{y,g}^{h})$ that are not rejected by
the Wald test. Confidence intervals for the tax and fiscal spending shocks
can be obtained by the projection method.

Before moving to the empirical results, two considerations are in order.
First, we remark that we need at least $k^{2}=4$ restrictions on $B_{\bullet
1} $, given by $B_{1,1}=B_{1,1}^{0}$, to derive the asymptotic normality
result and the rejection region in (\ref{Wald_S_region}). It should be noted
that the two zero restrictions that lead to the diagonal structure of the
relevance parameter matrix $\Phi $, see (\ref{link_fiscal_direct}), have not
been taken into account in our analysis. In order for $\Phi _{p}^{\prime }$
to be diagonal in (\ref{Phi_p_second}), a sufficient condition is that both $%
B_{1,1}^{0}$ and $\Sigma _{u_{1},z}$ are diagonal. The diagonal structure of 
$\Sigma _{u_{1},z}$ can be easily tested using standard methods, see below.
Second, the computation burden necessary to invert the test through (\ref%
{Wald_S_region})\ simplifies when the investigator has a strong confidence
on the credibility and validity of the restrictions $B_{1,1}=B_{1,1}^{0}$.
However, this assumption may not be realistic in many empirical
applications. To reduce the computation burden, hereafter we consider the
hypothesis%
\begin{equation}
B_{1,1}\equiv \left( 
\begin{array}{ll}
\beta _{tax,tax} & \beta _{tax,g} \\ 
\beta _{g,tax} & \beta _{g,g}%
\end{array}%
\right) =B_{1,1}^{0}:=\left( 
\begin{array}{ll}
1 & 0 \\ 
0 & 1%
\end{array}%
\right) \text{ \ }  \label{B11_fixed}
\end{equation}%
which amounts to imposing the `unit the effect responses' $\beta
_{tax,tax}=1 $ and $\beta _{g,g}=1$. Unit effect responses imply that the
size of the tax and fiscal spending shocks is of a magnitude that makes the
on-impact responses of GDP to these shocks equal to 1; moreover, (\ref%
{B11_fixed}) features two zero contemporaneous restrictions, i.e., that
fiscal spending does not react instantaneously to an exogenous tax shock ($%
\beta _{g,tax}=0$) and that tax revenues do not react instantaneously to an
exogenous fiscal spending shock ($\beta _{tax,g}=0)$. These two zero
restrictions are extensively debated in the empirical fiscal proxy-SVAR
literature; a detailed discussion on this topic is deferred to Angelini 
\emph{et al.} (2023).

Moving to the data, our bootstrap pre-test for the relevance of the proxies $%
z_{t}:=(z_{t}^{tax},z_{t}^{g})^{\prime }$ rejects the null of strong proxies
with a p-value of 0.003. We ignore temporarily the outcome of the test and
proceed by estimating the dynamic multipliers in (\ref{impact_multipliers})
pretending that the vector $z_{t}$ is a relevant proxy for the fiscal shocks 
$\varepsilon _{1,t}$. The impact and peak tax and fiscal spending
multipliers are summarized in the left column of Table S1.\footnote{%
We normalize the signs of the responses of output consistently with a fiscal
expansions induced by exogenous tax cuts on the one hand, and increases in
fiscal spending on the other hand. Estimates are obtained by the CMD
estimation approach developed in Angelini and Fanelli (2019).} The estimated
peak fiscal spending multiplier is 1.52 (after three quarters) with 68\% MBB
confidence interval given by (-0.73, 3.38); the estimated peak tax
multiplier is 2.46 (after three quarters) with 68\% MBB confidence interval
given by (-0.91, 9.76). Figure S1 plots the estimated dynamic fiscal
multipliers over an horizon of $h_{\max }=$40 quarters with associated 68\%
MBB confidence intervals. The graph confirms that by assuming strong proxy
asymptotics, the fiscal multipliers estimated by the direct approach exhibit
substantial uncertainty, a somewhat expected result in light of the outcome
of our pre-test of relevance of $z_{t}:=(z_{t}^{tax},z_{t}^{g})^{\prime }$.
Table S1 also reports the estimated elasticity of tax revenues and fiscal
spending to output, two crucial parameters in the fiscal multipliers
literature, see Mertens and Ravn (2014), Caldara and Kamps (2017) and Lewis
(2021). The estimated elasticity of fiscal spending to output is close to
zero, while the estimated elasticity of tax revenues to output is almost
3.5, a value comparable to that reported in Mertens and Ravn (2014).
Consistent with the uncertainty surrounding the fiscal multipliers, the
estimation of the elasticity of tax revenues to output is also characterized
by a relatively wide 68\% MBB confidence interval.

We robustify the inference on the fiscal multipliers by computing
weak-instrument confidence sets. To do so, we impose the four restrictions
in (\ref{B11_fixed}) on the parameters $B_{1,1}$ and, for $h=0,1,\ldots
,h_{\max }=40$, invert the Wald-type test in (\ref{Wald_S_region}), yielding
68\% Anderson-Rubin weak-instrument robust confidence sets for $\beta
_{gdp,tax}^{h}$and $\beta _{gdp,g}^{h}$.\footnote{%
To construct economically reasonable grid of values for $(\beta
_{gdp,tax}^{h}$, $\beta _{gdp,g}^{h})$, we exploit both economic
considerations and the survey in Ramey (2019) on the size of fiscal
multipliers. For each horizon $h$, we consider values of the tax multiplier
ranging from 0 to 6, and values of the fiscal spending multiplier ranging
from 0 to 3, respectively.} Assuming constant scaling factors $Sc_{y,tax}$
and $Sc_{y,g}$ in (\ref{impact_multipliers}), the confidence sets for $\beta
_{gdp,tax}^{h}$ and $\beta _{gdp,g}^{h}$ can be easily mapped to the fiscal
multipliers $\mathbb{M}_{h,tax}$ and $\mathbb{M}_{h,g}$, respectively. Part
of our results are summarized in the central column of Table S1. It can be
noticed that the projected 68\% weak-instrument robust confidence set for
the peak fiscal spending multiplier is $(0,3)$, with associated
Hodges-Lehmann estimate of 1.06 (after three quarters); the projected 68\%
weak-instrument robust confidence set for the peak tax multiplier is $%
(0.37,6)$ and the associated Hodges-Lehmann estimate is 2.55 (after three
quarters).\footnote{%
The Hodges-Lehmann point estimate corresponds to the multiplier within the
confidence set that has the highest associated p-value. We refer to point
estimates to facilitate a comparison of results with the point estimates
obtained through the indirect-MD approach, as discussed in the subsequent
section.}

\footnote{%
To infer whether the diagonal structure assumed for $\Phi $ in (\ref%
{link_fiscal_direct})\ is not rejected by the data when $%
B_{11}=B_{11}^{0}:=I_{4}$, we compute a Wald-type test for the hypothesis
that the covariance matrix $\Sigma _{u_{1},z}$ is diagonal; see the
expression of $\Phi _{p}^{\prime }$ in (\ref{Phi_p_second}). The test
delivers a p-value of 0.34.}

\subsection{Indirect MD-approach}

\label{Section_supplementary_fiscal_indirect}The analogue of the proxy-SVAR
representation (\ref{structural_sub_A_bis}) is given by the system: 
\begin{equation}
\text{ \ \ \ \ \ \ }\underset{A_{1,1}}{\underbrace{\left( 
\begin{array}{cc}
\alpha _{tax,tax} & \alpha _{tax,g} \\ 
\alpha _{g,tax} & \alpha _{g,g}%
\end{array}%
\right) }}\underset{u_{1,t}}{\underbrace{\left( 
\begin{array}{c}
u_{t}^{tax} \\ 
u_{t}^{g}%
\end{array}%
\right) }}+\underset{A_{1,2}}{\underbrace{\left( 
\begin{array}{cc}
\alpha _{tax,y} & \alpha _{tax,rr} \\ 
\alpha _{g,y} & \alpha _{g,rr}%
\end{array}%
\right) }}\underset{u_{2,t}}{\underbrace{\left( 
\begin{array}{c}
u_{t}^{y} \\ 
u_{t}^{rr}%
\end{array}%
\right) }}=\underset{\varepsilon _{1,t}}{\underbrace{\left( 
\begin{array}{c}
\varepsilon _{t}^{tax} \\ 
\varepsilon _{t}^{g}%
\end{array}%
\right) }}  \label{counterpart_A_fiscal}
\end{equation}%
and can be interpreted, under the identification conditions we discuss
below, as a model comprising two fiscal reaction functions, whose innovation
components coincide with the two target fiscal shocks $\varepsilon
_{1,t}:=(\varepsilon _{t}^{tax},\varepsilon _{t}^{g})^{\prime }$. The
crucial assumption here is Assumption 4, which postulates there are
available proxies for the non-target shocks in $\varepsilon
_{2,t}:=(\varepsilon _{t}^{y},\varepsilon _{t}^{mp})^{\prime }$ where,
recall, $\varepsilon _{t}^{y}$ is an output shock and $\varepsilon _{t}^{mp}$
a monetary policy shock. In this framework $n-k=2$ and $s\leq n-k$, where $s$
is the dimension of the vector of instruments $w_{t}$ for the non-target
shocks. If the proxies $w_{t}$ for the non-target shocks are chosen such
that Proposition \ref{Prop 1} holds, asymptotic inference on the fiscal
multipliers is of standard type, see Proposition \ref{prop 2}.\ 

We consider the following vector of instruments: $%
w_{t}:=(w_{t}^{tfp},w_{t}^{rr})^{\prime }$, $s=(n-k)=2$, where as in Caldara
and Kamps (2017), $w_{t}^{tfp}$ is Fernald's (2014) measure of TFP, used as
an instrument for the output shock, $\varepsilon _{t}^{y}$, and $w_{t}^{rr}$
is Romer and Romer's (2004) narrative series of monetary policy shocks, used
as an instrument for the monetary policy shock, $\varepsilon _{t}^{mp}$.
Hence, $\varepsilon _{2,t}:=(\varepsilon _{t}^{y},\varepsilon
_{t}^{mp})^{\prime }\equiv $ $\tilde{\varepsilon}_{2,t}$. The associated
linear measurement error model can be written in the form:%
\begin{equation}
\underset{w_{t}}{\underbrace{\left( 
\begin{array}{c}
w_{t}^{tfp} \\ 
w_{t}^{rr}%
\end{array}%
\right) }}=\Lambda \underset{\varepsilon _{2,t}}{\underbrace{\left( 
\begin{array}{c}
\varepsilon _{t}^{y} \\ 
\varepsilon _{t}^{mp}%
\end{array}%
\right) }}+\underset{\omega _{t}}{\underbrace{\left( 
\begin{array}{c}
\omega _{t}^{tfp} \\ 
\omega _{t}^{rr}%
\end{array}%
\right) }}  \label{link_non_fiscal_indirect}
\end{equation}%
where $\omega _{t}:=(\omega _{t}^{tfp},\omega _{t}^{rr})^{\prime }$ is a
measurement error term assumed uncorrelated with the structural shocks.
Since $k>1$, it is necessary to complement the two instruments used for the
two non-target shocks with at least one additional restriction on the
parameters in $A_{1\bullet }:=(A_{1,1},$ $A_{1,2})$; see Proposition \ref%
{Prop 1}. Based on previous contributions, we postulate that fiscal spending
does not react instantaneously to output, i.e. we set $\alpha _{g,y}=0$ in (%
\ref{counterpart_A_fiscal}). Equations (\ref{moments_A1_variance2})-(\ref%
{moments_A1}) provide $m=\frac{1}{2}k(k+1)+ks=7$ moment conditions that can
be used to estimate the $7$ structural parameters in the vector $\alpha $ by
the MD approach, i.e. the free structural parameters in $A_{1\bullet
}:=(A_{1,1},$ $A_{1,2})$.

The proxy $w_{t}^{rr}$ is available from 1969Q1, hence we consider the
common sample period 1969Q1--2006Q4 for estimation (based on $T=152$
quarterly observations). The bootstrap pre-test for the relevance of the
chosen proxies $w_{t}$ does not reject the null hypothesis with a p-value of 
$0.88$.\footnote{%
Formally, the test is computed as DH multivariate normality test computed on
the sequence $\{\hat{\beta}_{2,T:1}^{\ast },...,\hat{\beta}_{2,T:N}^{\ast
}\} $\ of MBB replications, with $N=[T^{1/2}]$=12. See Section \ref%
{Section_testing} for details.} The impact and peak fiscal multipliers are
summarized in the right column of Table S1. The estimated peak fiscal
spending multiplier is $1.54$ (after two quarters), with 68\% MBB confidence
interval equal to $(0.64,1.76)$; the estimated peak tax multiplier is 0.96
(after four quarters), with 68\% MBB confidence interval equal to $%
(0.18,1.44)$. The estimated elasticity of tax revenues to output is 2.06, a
value surprisingly close to the value 2.08 calibrated by Blanchard and
Perotti (2002); the 68\% MBB confidence interval for this parameter is $%
(1.6,2.5)$.

Figure S1 displays the dynamic fiscal multipliers estimated using the
indirect-MD approach (red dots) for a horizon of $h_{\max }=$40 quarters.
The associated 68\% MBB confidence intervals correspond to the red shaded
areas. For the purpose of comparison, the graph also includes the dynamic
fiscal multipliers estimated by the direct approach, assuming that the
proxies $z_{t}:=(z_{t}^{tax},z_{t}^{g})^{\prime }$ are strong for the target
fiscal shocks. These estimates are represented by blue dots, and the
corresponding 68\% MBB confidence intervals are represented as blue shaded
areas.

In her recent review of the theoretical and empirical literature on fiscal
multipliers, Ramey (2019) highlights the significant lack of consensus
regarding the magnitude and uncertainty of fiscal multipliers, particularly
concerning the uncertainty surrounding the tax multiplier.\textbf{\ }Our
empirical findings suggest that one possible explanation for the lack of
consensus on the tax multiplier could be attributed to the challenges
associated with finding `sufficiently strong' proxies for the tax shock. The
suggested estimation and testing strategy offer a potential solution to this
issue.

\section{ References}

\begin{description}
\item Angelini, G. and Fanelli, L. (2019), Exogenous uncertainty and the
identification of Structural Vector Autoregressions with external
instruments, \emph{Journal of Applied Econometrics} 34, 951-971.

\item Angelini, G., Caggiano, G., Castelnuovo, E. and Fanelli, L. (2023),
Are fiscal multipliers estimated with proxy-SVARs robust? \emph{Oxford
Bulletin of Economics and Statistics} 85, 95-122.

\item Basawa, I., Mallik, A., McCormick, W., Reeves, J.\ and Taylor, R.
(1991), Bootstrapping unstable first-order autoregressive processes,\ \emph{%
The Annals of Statistics} 19, 1098--1101.

\item Blanchard, O. and Perotti, R. (2002), An empirical characterization of
the dynamic effects of changes in government spending and taxes on output, 
\emph{Quarterly Journal of Economics} 117, 1329--1368.

\item Br\"{u}ggemann, R., Jentsch, C. and Trenkler, C. (2016), Inference in
VARs with conditional volatility of unknown form, \emph{Journal of
Econometrics} 191, 69-85.

\item Boubacar Mainassara, Y. and Francq, C. (2011), Estimating structural
VARMA models with uncorrelated but non-independent error terms, \emph{%
Journal of Multivariate Analysis} 102, 496-505.

\item Caldara, D. and Kamps, C. (2017), The analytics of SVARs: A unified
framework to measure fiscal multipliers, \emph{Review of Economic Studies}
84, 1015-1040.

\item Cavaliere, G.\ and Georgiev, I. (2020), Inference under random limit
bootstrap measures, \emph{Econometrica} 88, 2547-2974.

\item Fernald, J. (2014), A quarterly, utilization-adjusted series on Total
Factor Productivity, \emph{Federal Reserve Bank of San Francisco Working
Paper} No. 2012-19.

\item Francq, C. and Ra\"{\i}ssi, H. (2006), Multivariate portmanteau test
for autoregressive models with uncorrelated but nonindependent errors, \emph{%
Journal of Time Series Analysis} 28, 454-470.

\item Jentsch, C. and Lunsford, K.C. (2019), The dynamic effects of personal
and corporate income tax changes in the United States: Comment, \emph{%
American Economic Review} 109, 2655-2678.

\item Jentsch, C. and Lunsford, K.C. (2022), Asymptotic valid bootstrap
inference for Proxy SVARs, \emph{Journal of Business and Economic Statistics}
40, 1876--1891.\emph{\ }

\item Kallenberg, O.\ (1997), \emph{Foundations of Modern Probability}. New
York: Springer.

\item Lewis, D.J. (2021), Identifying shocks via time-varying volatility, 
\emph{Review of Economic Studies} 88, 3086-3124.

\item Magnus, J.R. and Neudecker, H. (1999), \emph{Matrix differential
calculus with applications in Statistics and Econometrics}, Wiley \& Sons,
2nd edition.

\item Mertens, K. and Ravn, M. (2013), The dynamic effects of personal and
corporate income tax changes in the United States, \emph{American Economic
Review} 103, 1212-1247.

\item Mertens, K. and Ravn, M. (2014), A reconciliation of SVAR and
narrative estimates of tax multipliers, \emph{Journal of Monetary Economics}
68, S1-S19.

\item Montiel Olea, J.L., Stock, J.H. and Watson, M.W. (2021), Inference in
SVARs identified with an external instrument, \emph{Journal of Econometrics}
225, 74-87.

\item Ramey, V. (2011), Identifying government spending shocks: It's all in
the timing, \emph{Quarterly Journal of Economics} 126, 1--50. 2016.

\item Ramey, V. (2019), Ten years after the Financial Crisis: What have we
learned from the renaissance in fiscal research? \emph{Journal of Economic
Perspectives}, 33, 89-114.

\item Romer, C., and Romer, D. (2004), A new measure of monetary policy
shocks: Derivation and implications, \emph{American Economic Review} 94,
1055--1084.

\item Romer, C. and Romer, D. (2010), The macroeconomic effects of tax
changes: Estimates of based on a new measure of fiscal shocks, \emph{%
American Economic Review} 100, 763-801.

\item Sweeting, T.J. (1989), On ConditionalWeak Convergence, \emph{Journal
of Theoretical Probability} 2, 461--474.
\end{description}

\clearpage
\newpage

\begin{table}[h]
\centering
\begin{tabular}{ccc}
\multicolumn{3}{c}{\textbf{Fiscal proxy-SVARs}} \\ \hline\hline
&  &  \\ 
\multicolumn{1}{c}{Direct ``Plug-in''} & \multicolumn{1}{c}{Direct A\&R} & 
\multicolumn{1}{c}{Indirect-MD} \\ 
&  &  \\ 
$%
\begin{array}{c}
\mathbb{M}_{0,g} = \underset{( -0.6359; 2.3364)}{1.0809 } \\ 
\mathbb{M}_{0,tr} = \underset{( -1.0294; 7.5788)}{1.8394 } \\ 
\mathbb{M}_{3,g} = \underset{( -0.7307; 3.3828)}{1.5214 [3]} \\ 
\mathbb{M}_{3,tr} = \underset{( -0.9058; 9.7567)}{2.4598[3] } \\ 
\psi_y^{tr} = \underset{( 0.0608;4.8160 )}{ 3.4814 } \\ 
-0.7365;0.5616 ){\ -0.0553 }%
\end{array}%
$ & $%
\begin{array}{c}
\mathbb{M}_{0,g} = \underset{( 0.0000; 3.000)}{0.7440 } \\ 
\mathbb{M}_{0,tr} = \underset{( 0.2162; 6.000)}{1.9072} \\ 
\mathbb{M}_{3,g} = \underset{( 0.0000; 3.000)}{1.0639 [3]} \\ 
\mathbb{M}_{3,tr} = \underset{( 0.3661; 6.000)}{2.5513 [3]} \\ 
\\ 
0.0608;4.8160 ){\ 3.4814 } \\ 
-0.7365;0.5616 ){\ -0.0553 }%
\end{array}%
$ & $%
\begin{array}{c}
\mathbb{M}_{0,tr} = \underset{( 0.9009; 1.5594)}{1.4662 } \\ 
\mathbb{M}_{0,tr} = \underset{( 0.0431; 0.9313)}{0.6382 } \\ 
\mathbb{M}_{2,g} = \underset{( 0.6411; 1.7603)}{1.5365 [2]} \\ 
\mathbb{M}_{4,tr} = \underset{( 0.1800; 1.4418)}{0.9553[4] } \\ 
\psi_y^{tr} = \underset{( 1.6419;2.4932 )}{ 2.0673 } \\ 
\end{array}%
$ \\ 
&  &  \\ 
\multicolumn{2}{c}{$p$-value $DH_{\theta=B_{\bullet 1} } = 0.0031$} & 
\multicolumn{1}{c}{$p$-value $DH_{\theta=\widetilde{B}_{\bullet 2} } =
0.8224 $} \\ 
&  &  \\ \hline\hline
\end{tabular}%
\caption{US fiscal Multipliers and pretests of relevance. \newline
\newline
{\protect\small {Notes: Results are based on U.S. quarterly data, period
1950:Q1-2006:Q4. Estimated multipliers and elasticities with associated 68\%
MBB confidence intervals; quarters of the peak effects in brackets. $p$%
-values of the diagnostic tests are based on $N:=[T^{1/2}]$ bootstrap
replications of the CMD estimator (see, Section 5). $DH_{\protect\theta %
=B_{\bullet 1} }$ ($DH_{\protect\theta =\tilde{B}_{\bullet 2} }$) is Doornik
and Hansen's (2008) multivariate normality test computed with respect to the
vector of on-impact coefficients in $B_{\bullet 1}$ ($\tilde{B}_{\bullet 2} $%
).}}}
\label{table:emp_app_spending}
\end{table}

\newpage
\clearpage 

\begin{figure}[h]
\centering
\begin{tabular}{c}
\hline
\hline
\\
\includegraphics[width=12cm,keepaspectratio]{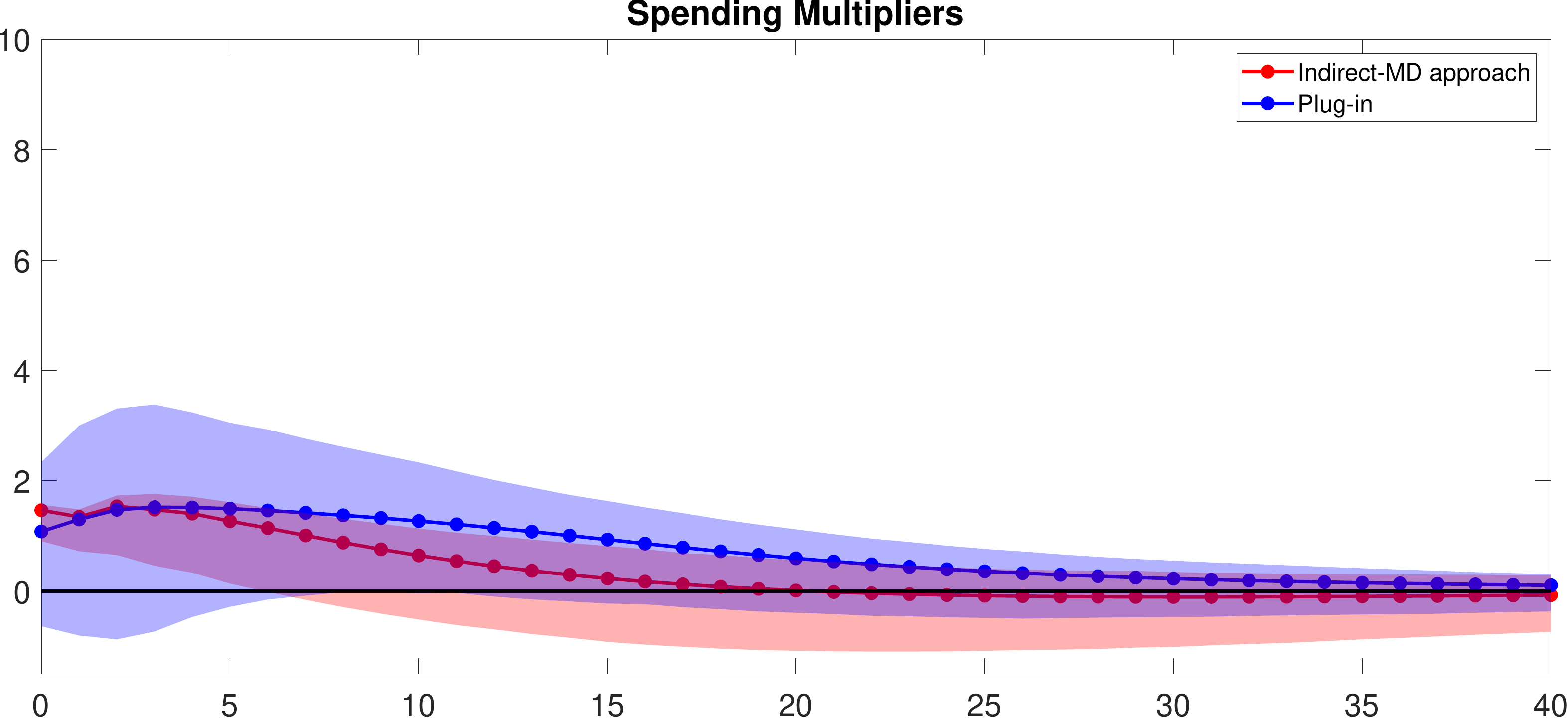} \\
\\
\includegraphics[width=12cm,keepaspectratio]{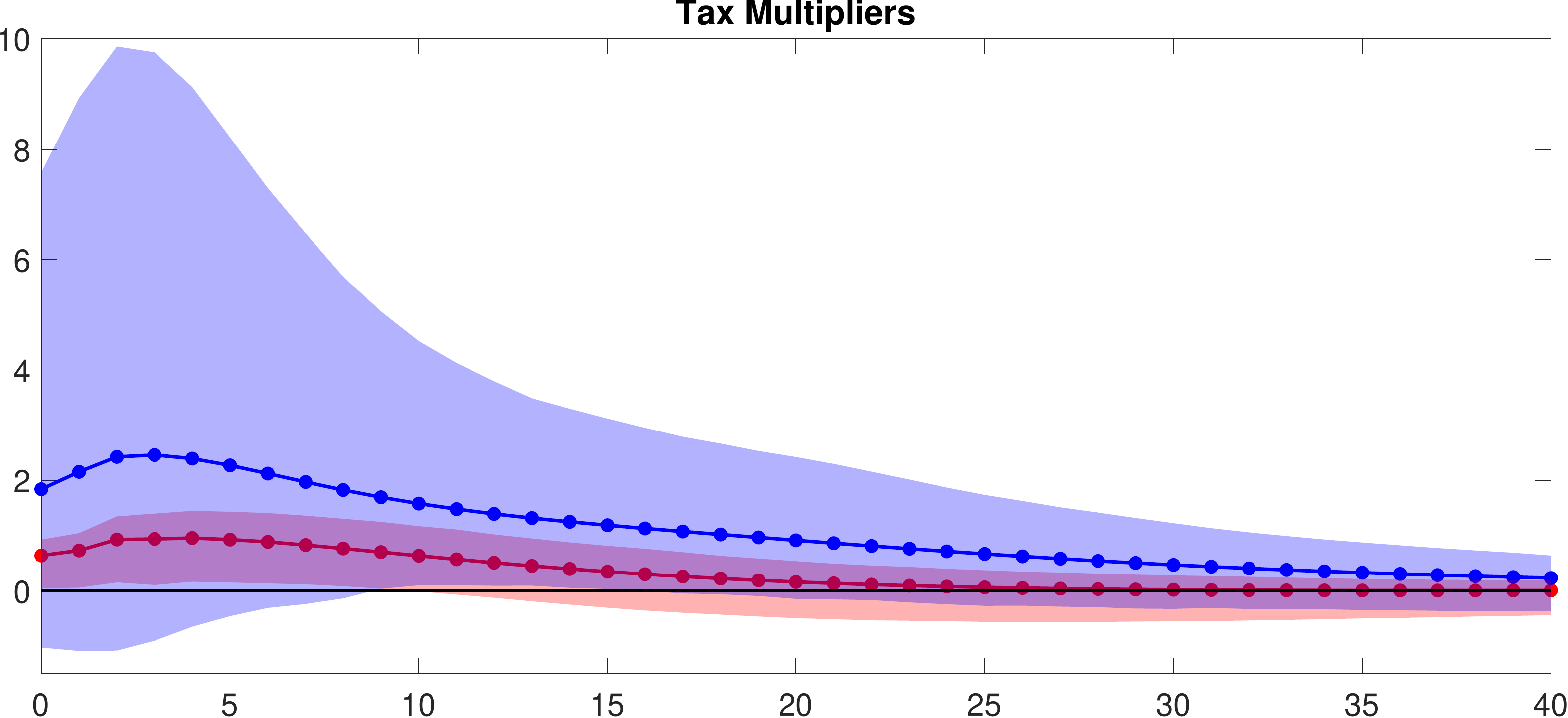} \\
\\
\hline
\hline
\end{tabular}
\caption{Fiscal multipliers. Red dotted lines correspond to the multipliers estimated with our indirect-MD approach; red shaded areas are the corresponding 68\% MBB confidence intervals; blue dotted lines correspond to the Plug-in multipliers obtained pretending that the proxies $z_t^{tax}$ and $z_t^{g}$ (direct approach)
are strong for the tax and spending shocks; blue shaded areas are the corresponding 68\%  Plug-in confidence intervals.}
\label{figure:Fiscal_shocks}
\end{figure}

\end{document}